\documentclass[12pt]{article}
\pdfoutput=1
\usepackage{jheppub}
\usepackage{amsmath}
\usepackage{amsfonts}
\usepackage{amssymb}
\usepackage{latexsym}
\usepackage{simplewick}
\usepackage{colonequals}
\usepackage{float}
\usepackage{tikz}
\usetikzlibrary{patterns}
\usepackage{simpler-wick}
\usepackage{pgfplots}
\usetikzlibrary{positioning, arrows.meta}

\RequirePackage{amssymb,amsmath,amsthm,amsfonts}
\RequirePackage{bbm}
\numberwithin{equation}{section}

\usepackage{mathtools}
\usepackage{simpler-wick}

\usepackage{extarrows}

\usepackage{epsf}
\usepackage{color}
\usepackage{graphicx}
\usepackage{dsfont}
\usepackage{subcaption}
\usepackage[export]{adjustbox}
\usepackage{hyperref}
\definecolor{darkred}{rgb}{0.8,0.1,0.1}
\hypersetup{colorlinks=true, linkcolor=darkred, citecolor=blue, linktoc=page}

\graphicspath{{figures/}}
\usepackage[utf8]{inputenc} 

\def\junk#1{}
\def\details#1{}

\pgfplotsset{compat=1.18}
\AtBeginDocument{}

\usepackage{comment}

\title{Solvable Models of Heat Transport in Quantum Mechanics}

\author[a]{R Loganayagam}
\author[b]{, Prithvi Narayan}
\author[b]{and Swathi T S}
\emailAdd{nayagam@icts.res.in}
\emailAdd{prithvi.narayan@gmail.com}
\emailAdd{swathisubramanian3@gmail.com}
\affiliation[a]{International Centre for Theoretical Sciences (ICTS-TIFR),\\
 Tata Institute of Fundamental Research, Shivakote, Hesaraghatta, Bengaluru 560089, India.}
\affiliation[\,b]{
\it{Department of Physics, Indian Institute of Technology, Palakkad 678557, India.}}

\abstract{
We investigate solvable models of heat transport between a pair of quantum mechanical systems initialized at two different temperatures. At time $t=0$, a weak interaction is switched on  between the systems, and we study the resulting energy transport. Focusing on the heat current as the primary observable, we analyze both the transient dynamics and the long-time behavior of the system. We demonstrate that simple toy models - including Random Matrix Theory like models ({\it RMT models})  and Schwarzian like models ({\it conformal models}) - can capture many generic features of heat transport, such as  transient current peaks and the emergence of  non-equilibrium steady state (NESS).  For these models, we derive a variety of exact results characterizing the short time transients, long time approach to NESS and thermal conductivity. Finally, we show how these features appear in  a more realistic solvable model, the Double-Scaled SYK (DSSYK) model. We demonstrate that the DSSYK model interpolates between the seemingly distinct toy models discussed earlier, with the toy models in turn providing a useful lens through which to understand the rich features of DSSYK.}

\begin{document}

\setcounter{footnote}{0}

\maketitle
\setcounter{equation}{0}
\setcounter{footnote}{0}

\section{Introduction}

Transport is a well studied topic in physics, with relevance across a wide range of disciplines including statistical mechanics, condensed matter physics, fluid dynamics, and high energy physics. Measurements of transport properties—such as thermal and electrical conductivities—provide critical insights into the underlying structure and dynamics of physical systems.  From a theoretical standpoint, transport constitutes one of the simplest and most accessible non-equilibrium processes, offering a natural  bridge between theoretical predictions and experimental observations. As we will explore, key theoretical constructs such as two-point functions are directly related to experimentally measurable currents\cite{zwanzig2001nonequilibrium, Kapusta:2006pm, kamenev2011field,balakrishnan2020elements}. For all these reasons, transport studies have become an indispensable tool in physics. 

In this work, we focus on one of the simplest scenario of heat transport : a pair of systems initialized at different temperatures are suddenly brought into contact, initiating an exchange of energy and setting up a heat current\cite{Bernard:2012je,Almheiri:2019jqq}.  The evolution typically proceeds through three stages: an initial transient regime at short times, an intermediate-time window characterized by a non-equilibrium steady state (NESS), and eventually at long times, thermalization to a common intermediate temperature.

Theoretical models that allow this entire transport process to be studied analytically are of fundamental importance. They serve not only as conceptual tools for testing physical intuition and identifying the mechanisms underlying observed phenomena, but also serve as essential testing grounds for broader ideas in non-equilibrium statistical mechanics. However, identifying exactly solvable models that capture all three stages of heat transport—transients, steady-state NESS, and thermalization poses a challenge. For instance, integrable systems are solvable, but do not thermalize in the standard sense \cite{Bernard:2016nci,Essler:2014qza}.

As for the NESS, in some exactly solvable theories, the exact NESS density matrix can be found \cite{2012PhRvE..85a1126D,Doyon:2012bg} and exact non-equilibrium heat currents can be computed in spin-chain models\cite{PhysRevB.88.195129,  PhysRevB.90.161101, PhysRevB.93.205121, Biella:2019pxg}. In weakly coupled theories, the heat  transport can be captured by the Landauer formalism applied to quasiparticles (see \cite{kamenev2011field,RevModPhys.93.025003} for a review). However, even in these systems, it is hard to precisely describe the transients as well as the \emph{approach} to NESS. Of course, in strongly coupled systems, there are no quasiparticles, and special methods like holography become necessary\cite{hartnoll2018holographic}. 

There is hence a need for solvable models that can access both transient and NESS regimes without relying on weak coupling or quasiparticles.  In this work, we study a class of solvable models\footnote{By solvable models, as will be clearer later in specific examples, we mean models where the two point function (and more generally, all correlation functions) are determined perhaps up to an integral. In all the examples we deal with in this work this turns out to be the case, since spectrum and matrix elements of operators are known.} that provide full control in these regimes.  

In this work, we consider  a pair of systems brought into thermal contact via operator couplings (for details see section \ref{sec:Heat Transport setup}) and study the heat current that flows from one to the other as a function of time.  In the regime where the coupling between the two systems is weak, the heat current depends solely on the two-point function of the operator which couples the two systems, and we consider diverse systems in which the two point function is known explicitly.

We begin by analyzing transport between a pair of quantum systems which  are described by simple toy models that allow full analytic control in the regimes of interest. We will consider three sets of toy models namely
\begin{itemize}
    \item \textbf{Random matrix theory (RMT) model} (in section \ref{subsec:RMT model example}) : Here the system is described by a random matrix theory with 
    following the well known Wigner semicircle.  
    \item \textbf{Conformal model}(in section \ref{subsec:Conformal model example}) : Here the system is taken to have two point function  motivated from conformal symmetry (and also arising in some explicit models). While we do not have an explicit density of states here, we will later see that exponential density of states can give rise to these models. 
    \item  \textbf{Gaussian  model}(in section \ref{subsec:Gaussian model example}) :  Here the density of states is gaussian. This model  can also arise as a special limit of the  conformal model. 
\end{itemize}
Despite their simplicity, one of the key results of our work is that these toy models already capture key qualitative features of the dynamics  
- a transient regime with a peak in the energy current  followed by a relaxation into a non-equilibrium steady state obeying Fourier's law where heat current is proportional to temperature difference.  In each of the above models, we systematically characterize the transient regime (including the time at which heat current peaks and its height), the approach to the steady state (which may follow power law, exponential  or gaussian behavior depending on the model) and the  heat conductivity at  NESS regime particularly its temperature dependence. 

We then turn  to a more realistic model the Sachdev-Ye-Kitaev (SYK) model  in its double scaling limit, namely the Double Scaled SYK (DSSYK) model. The SYK model\cite{KitaevTalk1,PhysRevLett.70.3339,PhysRevD.94.106002} has recently emerged as a paradigmatic example of a solvable, strongly interacting quantum systems. The model consists of $N$ fermions with {\it disordered} $p$ fermion interactions and becomes analytically tractable in the large-$N$ and at low energies\cite{PhysRevD.94.106002,Polchinski2016TheSI}. In this regime, the model exhibits many intriguing features that have attracted broad attention. The two-point functions become conformal, reflecting an emergent reparameterization symmetry\cite{PhysRevD.94.106002,Polchinski2016TheSI}. The system also displays strong quantum chaos, with a Lyapunov exponent saturating the universal bound\cite{Sachdev:2010uj,PhysRevD.94.106002,Maldacena:2015waa}, a hallmark of maximal chaos effectively captured by a low-energy Schwarzian theory. Furthermore, the model is believed to admit a holographic dual description, with the Schwarzian action corresponding to the boundary dynamics of nearly AdS$_2$ gravity\cite{Maldacena:2016upp,Engelsoy:2016xyb}, making it a powerful toy model for exploring aspects of quantum gravity and holography. Motivated by the remarkable features outlined above, numerous studies have explored SYK variants, including extensions with higher symmetries \cite{ PhysRevX.5.041025,Gross:2016kjj,Yoon:2017nig,Narayan2018SupersymmetricSM,Gu2020NotesOT,Bhattacharya2017SYKMC,Liu:2019niv} and supersymmetry \cite{Wenbo,Yoon2017SupersymmetricSM,Peng:2017spg,Narayan2018SupersymmetricSM}, among others. It is thus natural to ask what happens if the SYK model were used as the quantum system in our transport setup. This direction has been pursued in a number of works \cite{Larzul:2022kri,Larzul:2022yss,Almheiri:2019jqq}, where heat transport between coupled SYK systems was studied in various contexts. In \cite{Larzul:2022kri}, two SYK systems coupled via a channel were considered, and the resulting transport properties were analyzed. The authors found that the conductivity at intermediate temperatures follows a power-law dependence on temperature, with a distinct change in the exponent at a critical temperature. In \cite{Almheiri:2019jqq}, the authors derive general bounds on the rate of heat transfer in quantum systems, with implications for holography. For coupled SYK systems, they also numerically compute the energy flow and demonstrate eventual thermalization.

While the SYK model provides analytical control in the low-temperature and intermediate-time regime, at finite temperatures or during early-time transients the results are only numerical. To analytically study transport at finite temperatures, including the transient regime, it is necessary to consider a model that is solvable beyond these limits. To access these regimes, one considers the double scaling limit of the SYK model\cite{Erds2014PhaseTI,Cotler:2016fpe}, defined by $N \to \infty$, $p \to \infty$, with $\lambda \equiv \frac{2p^2}{N}$ held fixed. The resulting DSSYK model \cite{Cotler:2016fpe,2018JHEPNarayan,Berkooz:2018jqr} is solvable at all temperatures and retains key features of the original SYK model in an appropriate limit. There has been a significant body of work on the DSSYK model, including the computation of correlation functions\cite{2018JHEPNarayan,Berkooz:2018jqr}, the analysis of multi-trace operator structures\cite{Berkooz:2020fvm}, and investigations into its implications for emergent gravitational dynamics\cite{Lin:2022rbf,Berkooz:2022mfk,Mertens:2022irh,Susskind:2021esx,Susskind:2022bia,Susskind:2022dfz,Blommaert:2024ydx}. In particular, its correlation functions can be given as an integral (or equivalently as an infinite sum) of certain special functions. As mentioned above, the conventional SYK physics is recovered in a particular limit termed \textit{triple-scaling limit} where $\lambda \rightarrow 0$ at low temperatures with $T \sim \lambda^{\frac{3}{2}}$\cite{Cotler:2016fpe}. The parameter $\lambda$ plays  a role somewhat analogous to the t'hooft coupling : for small $\lambda$, the system is essentially free (although, at sufficiently low temperature, with $T \sim \lambda^{\frac{3}{2}}$  the physics is nontrivial as we mentioned above) and at large $\lambda$, the theory is strongly coupled. As with the SYK model, numerous variants of the DSSYK model have also been investigated, including those with higher symmetries \cite{Narayan:2023wlk,Berkooz2020ComplexSM} and supersymmetric versions \cite{Berkooz:2020xne}.

With this background in place, we now turn to the next focus of this work: transport between two quantum systems, each described by a DSSYK model. The results obtained in earlier studies of coupled SYK systems is reproduced in appropriate regimes. For instance the power-law scaling of conductivity with respect to temperature observed in \cite{Larzul:2022kri} is observed here as well; moreover our analytic results allow is to reliably calculate the prefactors that were previously computed only numerically. General bounds on energy transfer rates \cite{Almheiri:2019jqq} are also explicitly verified. An added advantage of the DSSYK model is the presence of tunable parameters such as $\lambda$ and the operator dimension  $m$. 

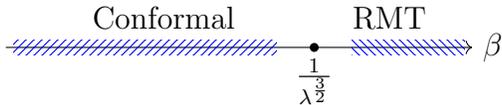
\begin{figure}
    \centering
    \begin{tikzpicture}
    \draw[->] (-0.1,0) -- (6.1,0) node[right] {$\beta$};

   \node[below] at (4,0) {$\frac{1}{\lambda^{3\over 2}}$};
   \node at (4,0)[circle,fill,inner sep=1.2pt]{};

    \fill[pattern=north west lines, pattern color=blue] (4.5,-0.1) rectangle (6,0.1);
    \node[align=center] at (5,0.4) {RMT};

    \fill[pattern=north east lines, pattern color=blue] (0,-0.1) rectangle (3.5,0.1);
    \node[align=center] at (2,0.4) {Conformal};
\end{tikzpicture}
\caption{Schematic depiction of different regimes in the DSSYK model.}
    \label{fig:cartoon beta regimes}
\end{figure}

The DSSYK model stands out due to the availability of explicit results,  we get additional insights :  
We find that the DSSYK system, even in isolation, is accurately described by simpler toy models introduced earlier in many regimes.
\begin{itemize}
\item We identify a characteriztic time scale, $\tau_s$, referred to as the \textit{scrambling scale}, after which the dynamics matches that of the RMT model. 
\item We further see that DSSYK model reduces to the \textit{RMT}, \textit{Conformal}, and \textit{Gaussian} models in appropriate limits of $\beta$, $\lambda$, and $q$, as summarized in fig(\ref{fig:Cartoon relations the DSSYK model and other toy models}). 
\begin{figure}
\begin{tikzpicture}[
 box/.style = {draw=red!60, fill=red!5, rectangle, minimum width=2.5cm, minimum height=1cm, align=center},
  boxg/.style = {draw=blue!60, fill=blue!5, rectangle, minimum width=2.5cm, minimum height=1cm, align=center},
  ->, >=Stealth,
]

\node[box] (dssyk) {DSSYK\\ $\tilde q = q^m,q=e^{-\lambda}, J, \beta$};
\node[boxg, below left=2cm and 2.5cm of dssyk] (rmt) {RMT\\ $f_0, \Delta,\beta$};
\node[boxg, below right=2cm and 3.5cm of dssyk] (conformal) {Conformal\\$m,\phi_0$,$\beta $};
\node[boxg, below =5cm  of dssyk] (gaussian) {Gaussian\\$f_0,\tilde q,J$};

\draw[->] (dssyk) -- (rmt) node[midway, left]{$\beta \gg (1-q)^{-{3\over 2}}$};
\draw[->] (dssyk) -- (conformal) node[pos=0.3, right, xshift=20pt, align=left]{$\lambda \rightarrow 0, \beta\ll \lambda^{-{3\over 2}}$\\ $\cos \phi_0 = {2 \over \beta J \sqrt \lambda}\phi_0$};
\draw[->] (dssyk) -- (gaussian) node[pos=0.5, right, xshift=0.5pt, align=left]{$q=1$} ;
\draw[->] (conformal) -- (gaussian) node[pos=0.7, right, xshift=25pt, align=left]{$m\rightarrow \infty,\phi_0\sqrt m$ held fixed\\$J(1-\tilde q)= {2\phi_0 \sqrt m \over \beta}$} ;
\end{tikzpicture}
\caption{Relation between the DSSYK model and other toy models. Parameters of all the models and their relations are also given (see eq(\ref{eq:DSSYK long time effective f and Delta}) for RMT and $f_0=1$ for Gaussian model). See section \ref{sec:heat transport in toy models} for the description of toy models and section \ref{subsec:DSSYK review} for the description of the DSSYK model.}
\label{fig:Cartoon relations the DSSYK model and other toy models}
\end{figure}
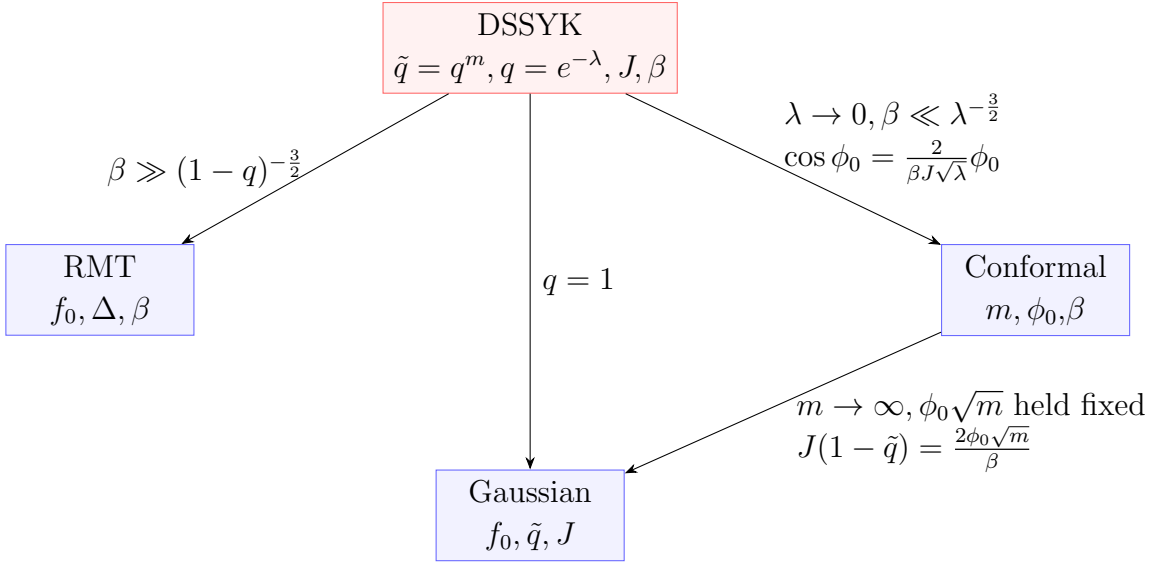
\item The transition between conformal and RMT regime can also be realized  by changing the temperature. In fact, loosely speaking, DSSYK exhibits a renormalization group (RG) flow  between a UV theory  described by a conformal model to an IR theory described by RMT model. 
\item The above point  is most transparent in the $\lambda \rightarrow 0$ regime,  where we have the most analytic control and we observe the RMT to conformal crossover around $\beta\sim \lambda^{-3/ 2}$ as shown in fig(\ref{fig:cartoon beta regimes}). At finite $\lambda$, this separation is less sharp, but a qualitative version of the flow persists : RMT region expands and we numerically\footnote{We emphasize here that these numerics are not very intensive unlike simulating the full system in the generic model or solving the Schwinger Dyson equations in the case of SYK model.} observe that there is still a conformal regime which shrinks as $\lambda$ increases.  The scenario we outline above is consistent with the $\lambda \to \infty$ limit where the model reduces exactly to the RMT model. 
\item As we discuss later, the above statements translate directly into statements about transport. For the conductivity of the DSSYK model, we observe a smooth transition from the conformal to the RMT results as the temperature is varied, as shown in fig(\ref{fig:rationew}), consistent with the RG flow picture discussed above.
\end{itemize}


In section \ref{sec:Heat Transport setup}, we set up the computation of heat current and describe its properties. In subsections \ref{subsec:RMT model example} and \ref{subsec:Conformal model example}, we discuss the transport in the case of two toy models of interest namely the RMT model and conformal model. 
In subsection \ref{subsec:DSSYK review}, we quickly review the DSSYK model with emphasis on those properties which are relevant to our work. In the remaining part of section \ref{sec: Heat Transport in DSSYK}, we discuss the transport in the DSSYK systems.  In section \ref{sec:Conlusions and Outlook}, we conclude with some possible future directions.

\newpage
\section{Setup for the heat transport and some toy models}\label{sec:Heat Transport setup}

In this section, we describe the setup used to study heat transport in detail. We consider two distinct systems, referred to as the hot and cold systems, initially held at temperatures $T_c$ and $T_h$ respectively with $T_h > T_c$.  These systems are coupled at time $t=0$ via operator interaction. The total Hamiltonian is given by\footnote{The equation below does not fully capture the setup for the DSSYK systems discussed in section \ref{sec: Heat Transport in DSSYK}, where the coupling $\epsilon$ also includes disorder. See footnote \ref{footnote reference} in Appendix \ref{sec: Appendix energy transport} for details. However, as shown there, the results for the heat current remain identical to those obtained using the simplified form presented here.}
\begin{eqnarray}\label{eq:Setup of the model}
    {\cal H} = H_c + H_h + \theta(t) \epsilon \ O_c O_h \ ,
\end{eqnarray}
where $O_c,O_h$ are operators in the cold and the hot system respectively. The entire system evolves under the full Hamiltonian ${\cal H}$ and the energies of the cold and hot system denoted by $E_c(t), E_h(t)$ change with time. The goal of this section is to establish some general formulae that can be used to compute  the heat current $\dot E_c(t)\equiv{dE_c \over dt}$ to leading order in $\epsilon$, the strength of the coupling between the systems. In addition, we will derive expressions for the thermal conductivity from the long-time behavior of the heat current $\dot E_c(t)$.

\subsection{Description of the isolated quantum system}\label{subsec:Description of the isolated quantum system}

We begin by describing the properties of individual quantum systems. The Hamiltonian is denoted by $H$ and we restrict ourselves to systems whose energy spectrum is confined to the interval  $(-E_0, E_0)$ with $E_0>0$\footnote{In some cases, we will consider the limit  $E_0 \rightarrow \infty$ - see the conformal regime in section \ref{subsec:DSSYK conformal regime}}. We assume that the spectrum is symmetric about  $E=0$, such that the density of states $\rho(E)$ satisfies 
\begin{equation}\label{eq:DOS property}
    \rho(E ) = 0 \quad \forall |E|\ge E_0, \qquad \rho(E) = \rho(-E)\ .
\end{equation}
The partition function of the system at inverse temperature $\beta$ is given by 
\begin{equation}\label{eq:Def Z}
    Z(\beta) \equiv  \mbox{ Tr }(e^{-\beta H}) = \int_{-E_0}^{E_0} dE \rho(E) e^{-\beta E}\ .
\end{equation}
Note that since $\rho(E)$ is an even function of $E$, $Z(\beta)$ is an even function of $\beta$. The two point function of any operator $O$ in the system is given by
\begin{equation}\label{eq:Def G Two point function}
    G(t) \equiv {\mbox{Tr}(O(t) O e^{-\beta H}) \over Z(\beta)}  = {1\over Z(\beta)} \int \int dE_1 dE_2 \rho(E_1) \rho(E_2) e^{-\beta E_2+i t(E_2-E_1)} f(E_1,E_2)\ ,
    \end{equation}
where $f(E_1,E_2) = \overline{| \langle E_1 | O | E_2 \rangle |^2}$ represents the average over the energy eigen states of the square of matrix element of operator $O$. By definition, $f(E_1,E_2)$ is a positive and symmetric function i.e., $f(E_1,E_2) = f(E_2,E_1)>0$. 

Without loss of generality, we assume that the one-point function of $O$ vanishes, i.e $\mbox{Tr}(Oe^{-\beta H}) = 0 $. In addition, we impose the symmetry  $f(E_1,E_2) = f(-E_1,-E_2)$, in analogy with the behavior of the density of states. It will also be useful to write the two point function in the frequency domain as
\footnote{Our convention for Fourier transform is,
\[
    \tilde f(\omega)= \int\limits_{-\infty}^{\infty} dt\  e^{i\omega t} f(t) , \qquad 
    f(t)=  \int\limits_{-\infty}^{\infty} {d\omega \over 2\pi}\ e^{-i\omega t} \tilde f(\omega) 
\] 
}
\begin{equation}\begin{split}\label{eq:Def tilde G}
    \tilde G(\omega) &=  {2\pi e^{\beta \omega \over 2} \over Z(\beta)} \int_{-E_0+{|\omega| \over 2}}^{E_0-{|\omega| \over 2}} dE \rho(E+{\omega \over 2})\rho(E-{\omega \over 2})  \ f(E+{\omega\over2 },E-{\omega \over 2}) \ e^{-\beta E}\ .
    \end{split}
\end{equation}
From the above definition, it is clear that $\tilde G(\omega)$ vanishes for $|\omega| > 2E_0 $ since that is the maximum energy the system can absorb. It is useful to note the following properties of the two point function 
\begin{itemize}
\item \textbf{Reality condition}: The two point function satisfies
\begin{equation}
G^*(-t) = G(t), \qquad \tilde G^*(\omega)      =  \tilde G(\omega) \ .
\end{equation}
\item \textbf{KMS relation}: Two point function satisfies the KMS relation 
\begin{equation}\label{eq:KMS definition}
    G(t) = G(-t - i \beta) , \qquad \tilde G(-\omega) = e^{-\beta \omega} \tilde G(\omega)\ .
\end{equation}
\item \textbf{Behavior near $t=0$}: We will normalize the two point function so that $G(t=0)>0$ is $\beta$ independent\footnote{This is an unusual choice, but it happens to be true for the DSSYK model as we will see explicitly later}. Additionally, note that $\mbox{Im}\dot G(0)<0$, since  
\begin{equation*}
Z(\beta)\ \mbox{Im}[ \dot G(0) ] = \int dE_1 dE_2\  \rho(E_1)\rho(E_2)\ [E_2-E_1]\ [e^{-\beta E_2}-e^{-\beta E_1}]\ , 
\end{equation*}
which is explicitly negative.
\end{itemize}

\subsection{Heat transport between quantum systems}

To study the heat transport, we consider two individual quantum systems as described in the previous subsection, initially held at different temperatures $T_c$ and $T_h$ with $T_c < T_h$ (the subscript refer to cold and hot systems respectively). These systems are coupled at times $t\ge 0$, via operator coupling  $O_c O_h$.  The Hamiltonian for $t>0$ is given by eq(\ref{eq:Setup of the model}). 

Our goal is to compute the energy change in each of the two systems as a function of time $t$ to leading order in $\epsilon$.   The energy current/heat current of the cold system is given by
\begin{equation}
    \dot E_c(t) \equiv {1 \over Z(\beta_c)Z(\beta_h) } \mbox{Tr}\left[ \dot H_c(t) \rho(t=0) \right] ,\qquad \rho(t=0)= e^{-\beta H_c - \beta H_h}\ .
\end{equation}
Here $H_c(t)$ denotes the cold system Hamiltonian evolved using the full Hamiltonian ${\cal H}$ given in eq(\ref{eq:Setup of the model}).

\begin{figure}
    \centering
    \includegraphics[width=0.75\linewidth]{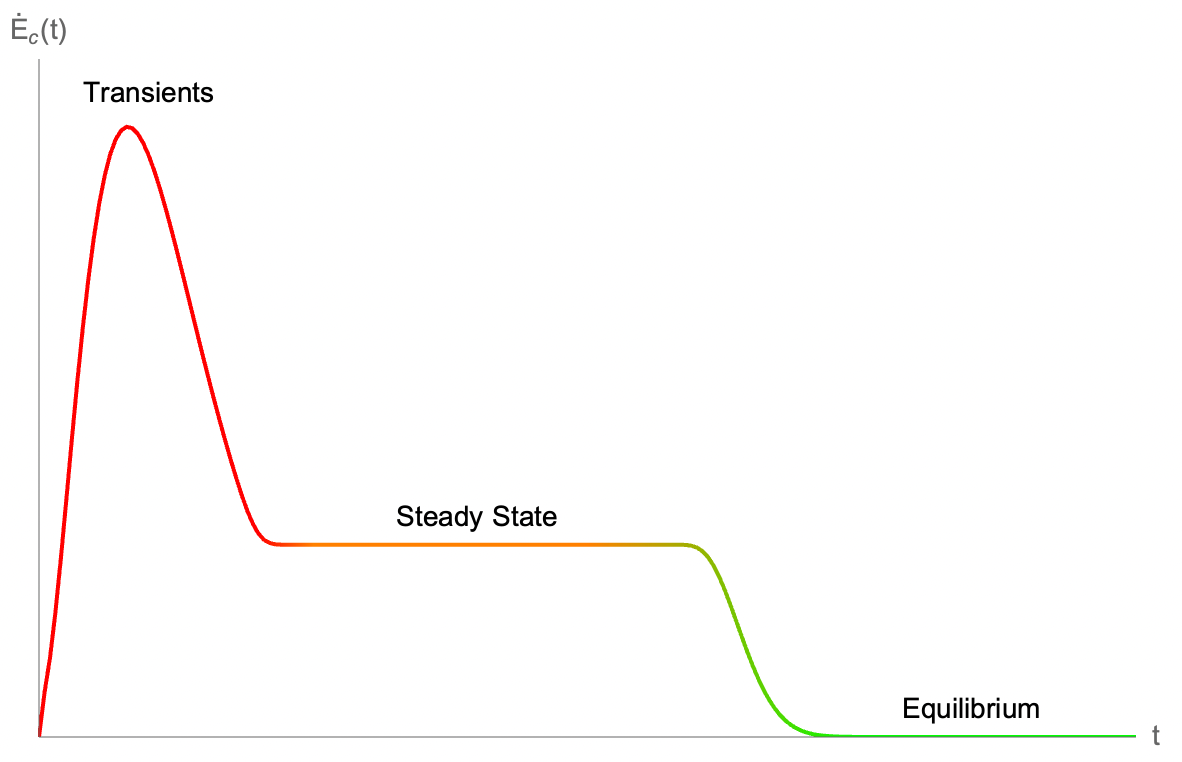}
    \caption{Typical heat current when two systems at different temperatures are coupled, adapted from fig(1) of \cite{Almheiri:2019jqq}}
    \label{fig: typical heat current}
\end{figure}

In figure(\ref{fig: typical heat current}), we give a typical expected behavior of the heat current $\dot E_c(t)$ computed this way (see \cite{Almheiri:2019jqq} for instance).
As shown in the figure, the heat current typically starts from zero, reaches a peak at early times, then decreases and settles to an approximately constant value at intermediate times. This plateau corresponds to a non-equilibrium steady state(NESS). At late times, the heat current eventually vanishes as the whole system thermalizes to an intermediate temperature. If the initial systems are very large, the NESS survives for a very long time.  

To leading order in coupling $\epsilon$, the heat current can be expressed in terms of the two point function of the operator $O$ in the isolated quantum system. Denoting the two point function of  $O_c, O_h$ by $G_c(t), G_h(t)$ respectively (see eq(\ref{eq:Def G Two point function})),  the heat current to leading order in $\epsilon$ is given by (see Appendix \ref{sec: Appendix energy transport} for a detailed derivation)
\begin{equation}\begin{split}\label{eq:Def Heat current}
\dot E_c(t) &= - \epsilon^2 \mbox{ Im } \left\{ G_c(t) G_h(t) \right\} + 
\underbrace{\epsilon^2 \mbox{ Im } \int_0^t d\tilde t \left\{G_c(\tilde t) \dot  G_h(\tilde t) - \dot G_c(\tilde t)  G_h(\tilde t)    \right\}}_{\equiv \dot E_- } \ .
\end{split}
\end{equation}
Here we have defined the energy current between the systems $\dot E_-(t)$  for use in later discussions. It is useful to note that the first term in the expression for the heat current dominates during the transient regime — i.e., at early times—after which it decays and becomes negligible. At longer times, the second term, namely $\dot E_-$ becomes relevant.

In this work, we compute the heat current for various toy models as well as the DSSYK model focusing on its behavior up to intermediate times where the non-equilibrium steady state (NESS) emerges. Specifically, we will analyze and characterize the following aspects
\begin{itemize}
\item \textbf{Transients} : At early times, the heat current is sensitive to the details of the quantum systems. In all the models analyzed in this work, we observe a linear increase in heat current (both $\dot E_c(t),\dot E_h(t)$) at small times, followed by a peak at a characteriztic peak time $t_p$, and then an eventual decrease  of the heat current to a constant value.\footnote{The fall need not be monotonic, there could be  oscillations too as we will see.} This is consistent with findings of \cite{Almheiri:2019jqq},  where it was also proven that the integrated energy flux 
\begin{equation*}
F_\kappa \equiv  \int_0^\infty dt e^{-\kappa t} \dot E_h\ .
\end{equation*}
obeys the the following inequality
\begin{equation}
    F_\kappa \ge 0 , \qquad  \forall \kappa \ge {2 \over \beta_h}\ .
\end{equation}
We verify that this result holds for all the models considered in this work.

The linear rise of $\dot E_c(t)$ at early times can be understood analytically since   
\[\dot E_c(t) = -2 t\   G_h(0) \mbox{Im} [\dot G_c(0)]+{\cal O}(t^2)\ ,\]
which is manifestly positive since $G(0)>0,\  \mbox{Im} G(0)<0$, as noted before. 

\item \textbf{Approach to steady state} : At intermediate times, the two-point functions decay, and the heat current, $\dot  E_c$ approaches a positive constant. The nature of this approach depends on the decay profile of the two-point function—whether it be power-law, exponential, or Gaussian. All these behaviors are observed in the  models we study. It is useful to note that for systems with long-lived quasiparticles, an exponential decay is typically expected. 

\item \textbf{Steady state} :   As mentioned above, at later times, the system settles to a steady state characterized by a constant heat current denoted by $\dot E_c(\infty)$ which can be obtained as the $t \rightarrow \infty$ limit of eq(\ref{eq:Def Heat current})  neglecting the first term, \footnote{At large enough times, the ${\cal O}(\epsilon)^2$  the approximation  we are working with eventually breaks down. Here, by "large times," we mean $t$ much greater than the decay timescale of the two-point function.}  which vanishes at large $t$. In the Fourier representation, the time integral yields a Dirac delta enforcing $\tilde \omega=- \omega$, and we obtain
\begin{equation}\label{eq:Heat current saturation formulae}
\begin{split}
    \dot E_c(\infty) &=  -\epsilon^2\int_{-\infty}^{\infty} \frac{d\omega}{2\pi} \omega\ \tilde G_c(\omega) \tilde G_h(-\omega)  \\
    &=- \epsilon^2 \int_0^\infty {d\omega\over 2\pi} \omega\ \tilde G_c(\omega) \tilde G_h(\omega) \left( e^{-\beta_c \omega} - e^{-\beta_h \omega} \right) .
\end{split}\end{equation}
Here we have used the KMS relation in frequency space eq(\ref{eq:KMS definition}). 

The above expression captures the steady-state energy flow from the hot system to the cold one, as indicated by the fact that $\dot E_c(\infty) = -\dot E_h(\infty)$ and that $\dot E_c(\infty)>0$.   In the limit where $T_h$ is very close to $T_c$, the heat current satisfies Fourier’s law, viz.,
\[\dot E_c(\infty) = \sigma (T_h- T_c) + {\cal O}  (T_h-T_c)^2\ ,\]
where  $\sigma$ is the thermal conductivity. This can be computed as
\begin{equation}\begin{split}\label{eq:Conductivity formulae general}
    \sigma & = \epsilon^2 \beta^2 \ \mbox{Im} \int_0^\infty dt \left( \partial_{\beta} G(t) \dot G(t) - \partial_\beta \dot G(t) G(t)  \right) \\
    & = \epsilon^2 \beta^2 \int_{0}^\infty {d\omega\over 2\pi} \omega^2\ \tilde G(\omega)^2 e^{-\beta \omega}\ , \\
\end{split}
\end{equation}
where $\beta$ is the average inverse temperature. In going to the second line, we have used the KMS relation eq(\ref{eq:KMS definition}) and linearized the temperature dependence around the average $\beta$.
\end{itemize}

In the next subsections, we analyze three toy models, in which  we observe all three key features of heat transport discussed earlier—namely, the transient rise in heat current, the approach to a steady state, and the establishment of a non-equilibrium steady-state (NESS) with a constant heat current.

\section{Heat transport in toy models}\label{sec:heat transport in toy models}

In this section, we study the heat transport between two quantum systems as outlined in the previous section when the quantum systems involved are described by certain toy models. In what follows, we  introduce three toy models, namely the {\it Random Matrix Theory model} in  subsection \ref{subsec:RMT model example}, the {\it conformal model} in subsection \ref{subsec:Conformal model example} and the {\it gaussian model} in the subsection \ref{subsec:Gaussian model example} \footnote{We  will see that the gaussian model can be thought of as a special case  of conformal model}. We systematically characterize both the transient dynamics and the steady-state behavior of the heat current. 

\subsection{Toy Model 1 : Random Matrix Theory (RMT)}\label{subsec:RMT model example} 
In the Random Matrix Theory (RMT) model, the energy spectrum is taken to be the well-known Wigner semicircle law:
\begin{equation}\label{eq:rho RMT semicircle}
\rho(E)= {2{\cal N}_r\sqrt{E_0^2-E^2} \over \pi E_0^2}\ .    
\end{equation}
Here ${\cal N}_r$ is  a normalization factor, which we will keep arbitrary for now. 

Now we move to modeling the matrix element square of a generic operator in the energy basis. The simplest choice is 
perhaps to take  the matrix element square average  $f(E_1,E_2)$  to be a constant independent of the energies $E_1$ and $E_2$. However, we want to allow for a suppression of matrix elements between energetically distant states compared to nearby ones. To this end, we consider a form with a mild energy dependence\footnote{ In general, $f(E_1,E_2)$ can be expressed as a systematic power series in $E_1,E_2$ (expanded around say $E = - E_0$)  consistent with the symmetries. In fact, for the realistic model of DSSYK as we will see in eq(\ref{eq:Def f in DSSYK result}), $f(E_1,E_2)$ takes precisely such a form. While the truncated form we use here may or may not appear natural for generic operators in RMT, it is justified in the regime of long times or low temperatures. In these regimes, the dominant contribution to the dynamics arises from states near the ground state (or near the highest excited state, due to symmetry).  Truncating to ${\cal O}((E_1-E_0)(E_2-E_0))$ and imposing the symmetries results in  the expression below.}
\begin{equation}\label{eq:f for RMT}
   f(E_1,E_2) =  f_0 \ \left(1+{\Delta\over E_0^2}\   E_1E_2 \right) \ ,
\end{equation}
for some constants $f_0,\Delta$ where $1>\Delta>-1$ since $f(E_1,E_2)>0$. The above form means that, for positive $\Delta$, the average matrix element square  $ f(E_1,E_2)<f_0$ when $E_1$ and $E_2$ are of opposite sign, i.e., when they are on either side of the maximum energy, When
$E_1$ and $E_2$ are of the same sign, $ f(E_1,E_2)>f_0$.
The combination ${1-\Delta \over 1+\Delta}$ can then be interpreted as characterizing the suppression of matrix elements between energetically distant states compared to nearby ones. We take the above formulae for $\rho(E)$ and  $ f(E_1,E_2)$ to be the definition of our toy model 1. Phrased this way, 
our model depends on only on two dimensionless parameters: $\beta E_0$ and $\Delta$ (the parameters $f_0,{\cal N}_r$ appear only as overall factors). 

The above model may seem somewhat crude and ad hoc - after all, we are not giving some detailed picture of where the underlying spectrum and matrix elements come from. But, as will be seen momentarily, this already captures lot physics behind both the heat transients as well as of approach to NESS. The very simplicity of the model allows us to do exact computations. Furthermore, as we shall see, in an appropriate regime, DSSYK model reduces to this model, thus giving a concrete realization in terms of disordered fermions.

\begin{figure}
    \centering
    \includegraphics[width=0.7\linewidth]{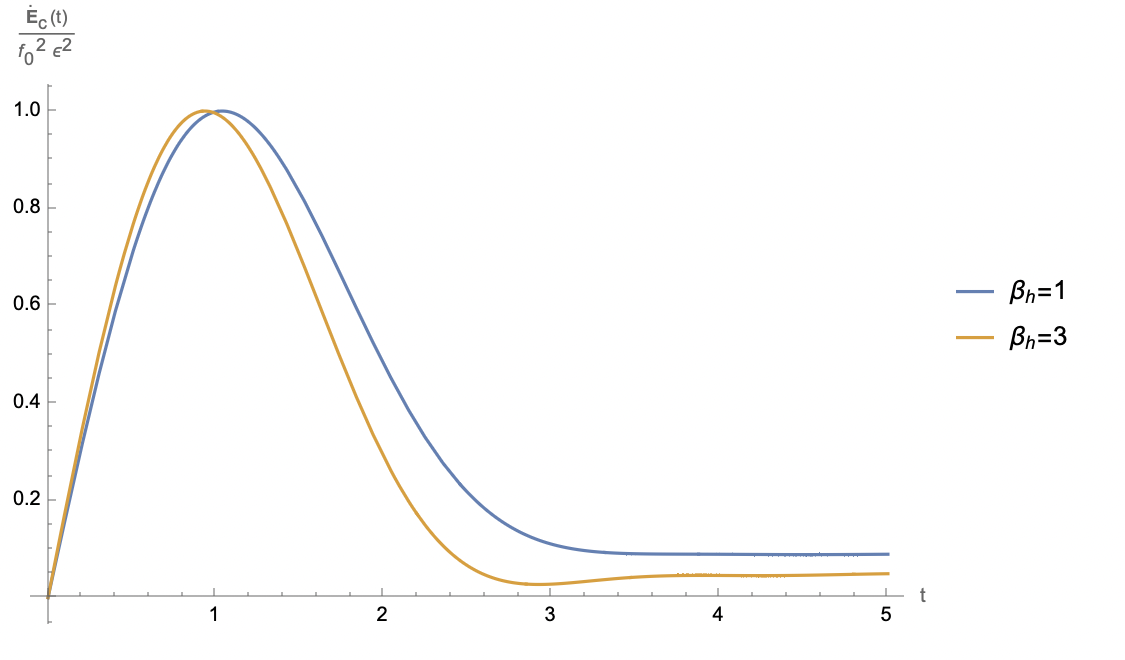}
    \caption{behavior of heat current for systems at different temperatures with $\Delta=0.1,\ E_0=1,\ \beta_c=1.1\beta_h,\ {\cal N}_r=1$, normalized so that peak height is 1}
    \label{fig:J RMT}
\end{figure}

The partition function  eq(\ref{eq:Def Z}) can be calculated  analytically and is given by 
\begin{equation*}
Z(\beta) = {2 {\cal N}_r I_1(\beta E_0) \over \beta E_0}\ ,
\end{equation*}
where $I_1$
is the modified Bessel function of the first kind. It will be convenient to define the quantity
\begin{equation}\label{eq:def tilde Z RMT}
    \tilde Z(\beta) \equiv -{1 \over E_0} \int_{-E_0}^{E_0} dE \rho(E) e^{-\beta E}E  = {2 {\cal N}_rI_2(\beta E_0) \over \beta E_0}\ ,
\end{equation}
which is the (negative of the) internal energy of the system in $E_0$ units. 
For the two point function, we get  from eq(\ref{eq:Def G Two point function}) the following result:
\begin{equation}\begin{split}\label{eq:Two point function RMT}
    G(t) &= f_0 { Z(it) Z(\beta - i t) \over Z(\beta)} +  f_0 \Delta{ \tilde Z(it) \tilde Z(\beta - i t) \over Z(\beta)} \ .
\end{split}
\end{equation}
In figure(\ref{fig:J RMT}), we plot the  the heat current $\dot E_c(t)$ at two different temperatures, computed using  eq(\ref{eq:Def Heat current}). The plot clearly displays both the transient regime at early times and the emergence of a steady-state regime at intermediate times. In the following discussion, we provide a detailed characterization of these two regimes.

\paragraph{Transients}
To characterize the transient behavior, we identify the peak time $t_p$ —the time at which the heat current $\dot E_c(t)$ reaches its maximum.
\begin{itemize}
    \item As shown in figure(\ref{fig:rmtpeakvsbeta}), $t_p$  is observed to be largely independent of $\beta$, regardless of the value of 
$\Delta$.
 \item  In figure(\ref{fig:rmtmaximavsbeta}), we also plot the peak height $\dot E_c(t_p)$ as a function of $\beta$. This plot shows how the magnitude of the initial energy surge depends on temperature: we see the peak height vary linearly with $\beta$ at high temperatures. At low temperatures, the peak height is pretty much independent of temperature.
 \begin{figure}
    \centering
      \includegraphics[width=0.65\linewidth]{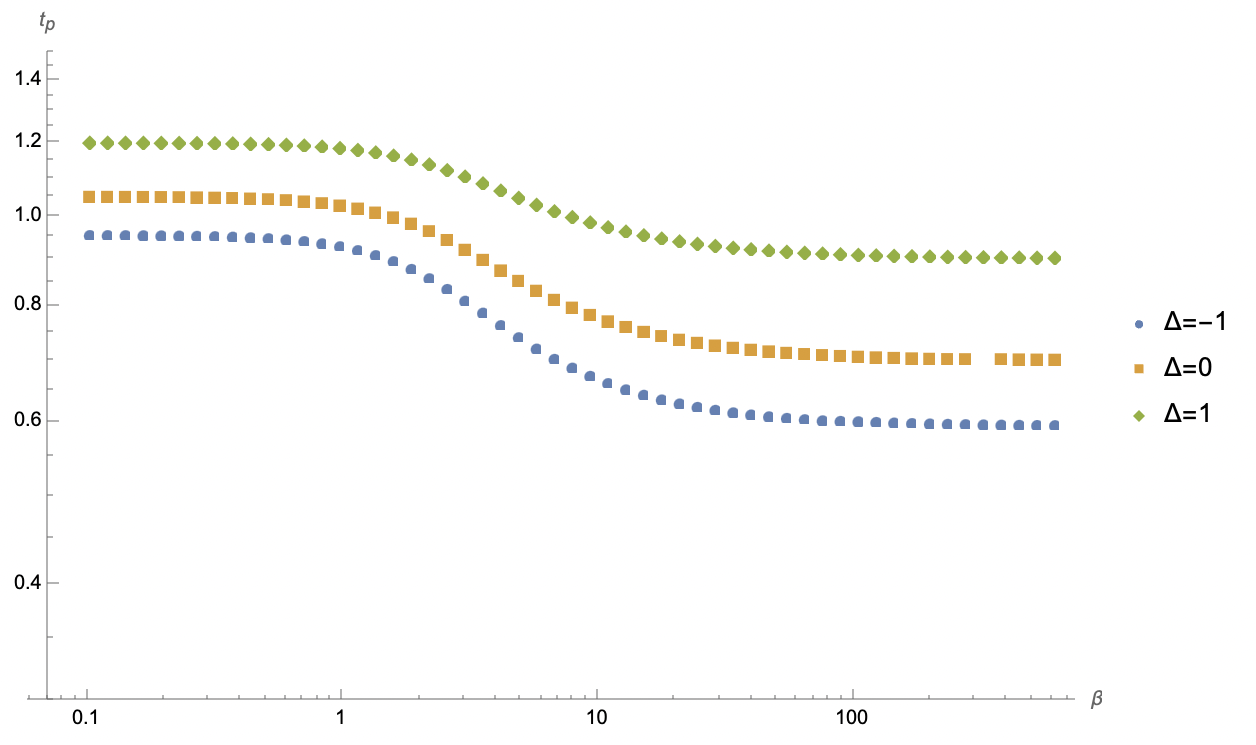}
    \caption{Peak time vs temperatures for systems with $E_0=1,\ \beta_c=1.1\beta_h$ and different values of $\Delta$.}
    \label{fig:rmtpeakvsbeta}
\end{figure}
\begin{figure}
    \centering
        \includegraphics[width=0.65\linewidth]{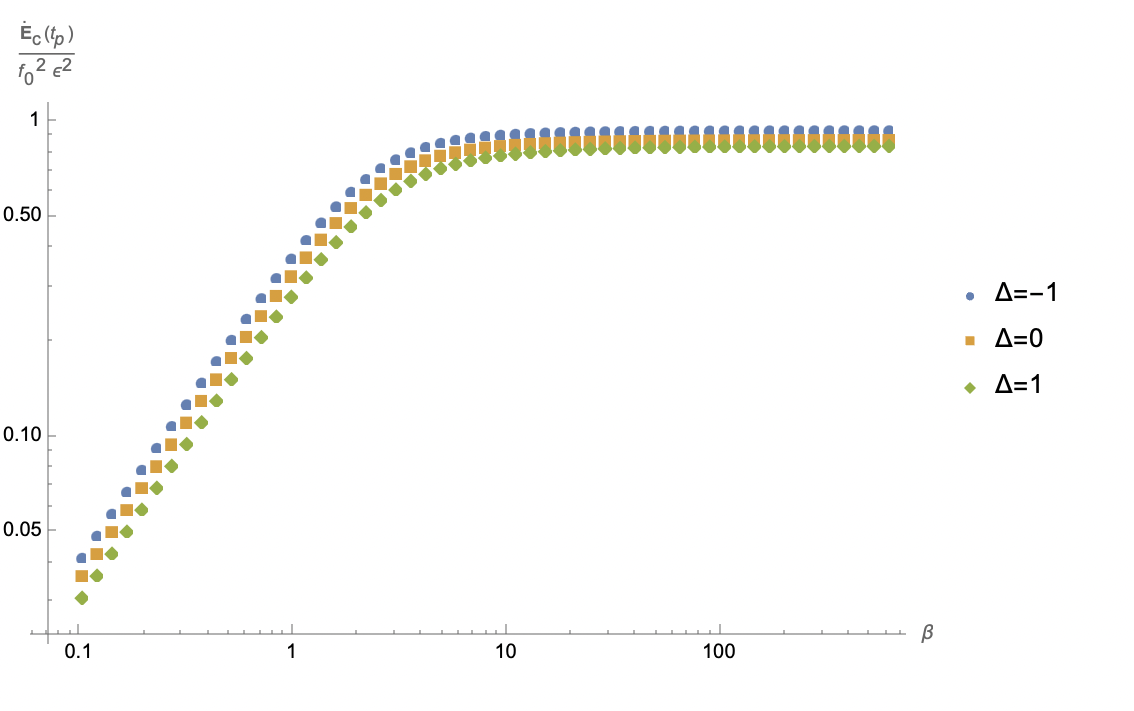}
    \caption{Peak height vs temperatures for systems with $E_0=1,\ \beta_c=1.1\beta_h,\ {\cal N}_r=1$ and different values of $\Delta$. The peak height fits to $a \beta$ for small $\beta$ with the slope $a$ values  $\{0.31,0.28,0.24\}$ corresponding to $\Delta$ values $\{-1,0,1\}$. }
    \label{fig:rmtmaximavsbeta}
\end{figure}
\item The integrated energy flux $F_\kappa$ is positive for all $\kappa \ge {2\over \beta_h}$ as can be seen in figure(\ref{fig:RMT Fk}).
\begin{figure}
    \centering
    \includegraphics[width=0.65\linewidth]{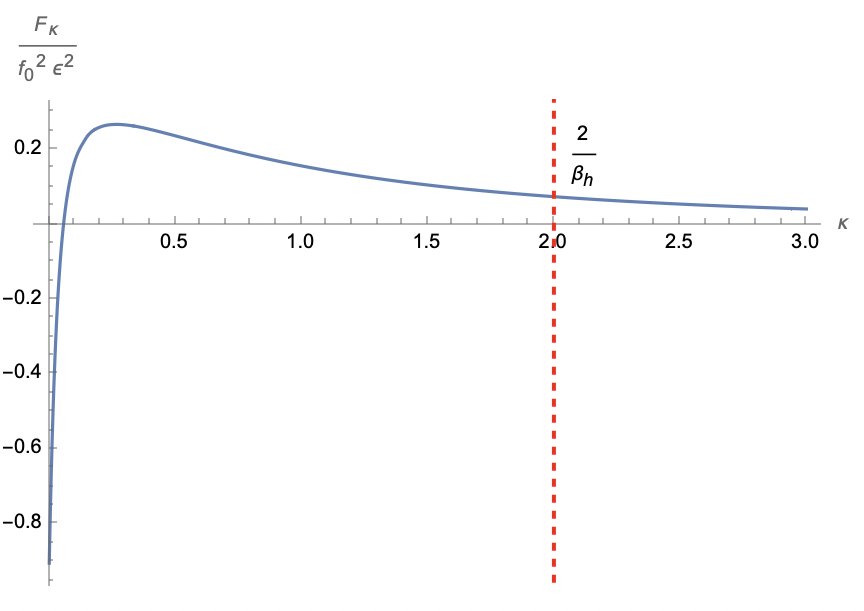}
    \caption{$F_\kappa$ vs $\kappa$ for a sytem with $E_0=1,\ \beta_c=1.1,\ \beta_h=1,\  \Delta=0.1,\ {\cal N}_r=1$.}
    \label{fig:RMT Fk}
\end{figure}
\end{itemize}

\paragraph{Approach to the steady state and thermal conductivity}
Given the formulae given earlier, we can easily analyze the long-time behavior of the heat current using Bessel asymptotics. In the limit $t E_0 \gg 1$ (with no restrictions on $\beta$), the two-point function given in  eq(\ref{eq:Two point function RMT}) simplifies considerably using the asymptotic form of Bessel functions (see appendix \ref{sec:Asymptotics} for details). We get 
\begin{equation}\label{eq:Two point function RMT long time}
G(t) =  \frac{4 (1-\Delta )f_0 {\cal N}_r^2 \  }{\pi E_0^3  Z(\beta)}\frac{   \left[\  \frac{1+\Delta}{1-\Delta} \cosh (\beta E_0  ) -  \sin (2 E_0 t+i \beta E_0  )\  \right]}{[t(t+i\beta)]^{3\over 2}} \left[ 1 +  {\cal O}((t^2+\beta^2)^{-{1\over 2}}) \right]\ .
\end{equation}
This is an interesting expression exhibiting power law at late times, modulated by oscillations.
We also see that  the parameter $1-\Delta$ governs the oscillatory behavior in time.  

Substituting the above result into the heat current expression given in eq(\ref{eq:Def Heat current}) we have the following asymptotic form\footnote{Interestingly, for the case $\Delta =1$ case, the leading contribution to the heat current vanishes, and the behavior of $\dot E_c(t)$ is suppressed as  $ {\cal O}(t^{-7})$ at long times.
}
\begin{equation}\begin{split}
    \dot E_c(t) & = \dot E_c(\infty) -{ 32 f_0^2 {\cal N}_r^4 (1-\Delta)^2 \over \pi^2 Z(\beta_c) Z(\beta_h) (E_0t)^5 } \left[ \frac{ \sinh \left(E_0 \left(\beta _c-\beta
   _h\right)\right)}{5} \right. \\
   & \quad \left. +\frac{ \sin  (4 E_0 t) \sinh  \left(E_0 \left(\beta _c+\beta _h\right)\right)-4 \frac{1+\Delta}{1-\Delta} \cos  (2 E_0 t) \sinh  \left(E_0 \beta _c\right) \cosh  \left(E_0 \beta _h\right)}{4 E_0 t} \right] \\
   & \qquad +{\cal O}(t^{-7})\ .
   \end{split}\label{eq:RMT Long time heat current}
\end{equation}
We thus observe a power law $1/t^5$ type approach to the steady state. Given the simplicity of this derivation, we suspect that this power law approach to steady state is universal across a large class of models which look like random matrix theory.

\begin{figure}
    \centering
\includegraphics[width=0.75\linewidth]{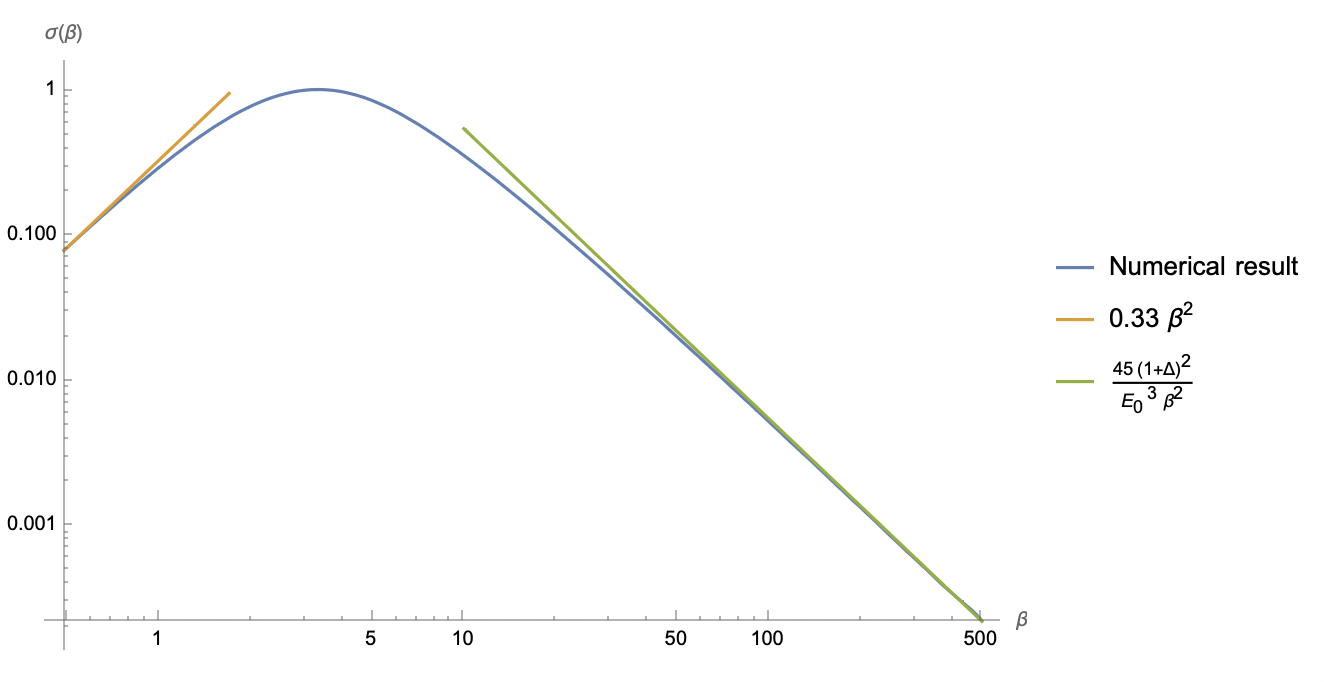}
    \caption{Conductivity as a function of $\beta$ for $E_0=1  ,\ \Delta=0.1,\  {\cal N}_r=1$.}
    \label{fig:conductivityRMT}
\end{figure}

Finally, we turn to the steady state itself: 
the conductivity eq(\ref{eq:Conductivity formulae general}) can be numerically evaluated and the result is shown in figure(\ref{fig:conductivityRMT}). As seen in the plot, the conductivity exhibits power-law fall off at both the high- and low-temperature regimes, and it shows a maxima at an intermediate temperature. The low-temperature (i.e., large $\beta$) scaling seen in the figure can be understood via a low-temperature analysis, to be outlined in the subsection \ref{subsec:Cold RMT model example} (See  eq(\ref{eq: cold RMT conductivity})).

We will now like to explore whether even simpler effective models can capture the physics of the RMT model in the long-time, low-temperature regime. In this limit, we expect the two-point function—and consequently, the heat current to be governed by the singularities in the density of states $\rho(E)$, and the  matrix element averages $f(E_1,E_2)$, in the vicinity of these singularities. Since the edges of the density of states exhibit a square-root dependence on energy i.e., $\rho(E) \propto \sqrt{E-E_0}$,  it is natural to ask whether this feature alone is sufficient to reproduce the observed RMT model behavior. In the next subsection, \ref{subsec:Cold RMT model example}, we answer this question affirmatively. 

We introduce in next subsection a  simplified model, which we refer to as \textit{Cold RMT model}, that retains only the relevant features near the  edge of the spectrum, and allows for explicit analytic expressions for the long time observables like conductivity and approach to NESS regimes. This toy model is particularly relevant to the DSSYK model since, as we will see later, at sufficiently low temperatures and long times (for $\beta,t \gg  (1-q)^{-{3 \over 2} }$), DSSYK also reduces to the Cold RMT model.

\subsubsection{Cold Random Matrix Theory}\label{subsec:Cold RMT model example}

In this subsection, we introduce the Cold Random Matrix Theory (Cold RMT) model which effectively captures the long time low temperature behavior of the RMT model described in the previous section \ref{subsec:RMT model example}. As noted earlier, near the lower edge of the spectrum (i.e., close to the ground state), the RMT density of states takes the form 
\begin{equation*}
\rho(E) = {2 \sqrt 2  {\cal N}_r \over \pi E_0^{3\over 2}} \sqrt{E_0+E}\ . 
\end{equation*}
While we could work with just this model, we consider slightly more general class of models parametrized by $\alpha$ to capture the effect of singularity structure of density of states as follows
\begin{equation}\label{eq:rho cold RMT}
    \rho(E) = {\sqrt{2}{\mathcal{N}_r} \over  \sqrt{\pi}\Gamma(1+\alpha) E_0^{\alpha+1} }\ (E_0-|E|)^\alpha, \qquad |E|<E_0\ ,
\end{equation}
for some constants $\alpha >0$. The model above matches the RMT model near the edges for $\alpha = {1 \over 2}$. We are interested in the long time low temperature limit of this model where the results are expected to match RMT model results for $\alpha={1 \over 2}$. The partition function  at low temperatures, i.e., large $\beta E_0$ becomes  
\begin{equation}\begin{split}\label{eq:Z for cold RMT}
    Z(\beta) 
    \approx   {\sqrt{2}{\mathcal{N}_r} e^{\beta E_0} \over \sqrt{\pi}(\beta E_0)^{\alpha +1}}\ .
    \end{split}
\end{equation}
\begin{itemize}
\item The partition function of the RMT model at low temperatures (i.e., large $\beta E_0$),  agrees with the above result for $\alpha={1\over 2}$, confirming that the Cold RMT correctly captures the low temperature physics of RMT (this also justifies the name). 
\item Further, in the low temperature, long time limit of the two point function in  the cold RMT matches the result in the RMT case given in eq(\ref{eq:Two point function RMT long time}). 
\item This in turn implies that the power law approach of the heat current to its NESS in the cold RMT case is $t^{-4\alpha -3}$ which for the $\alpha = {1 \over 2}$  is $t^{-5}$ which is the same as the RMT case seen before in eq(\ref{eq:RMT Long time heat current}). 
\end{itemize}

We now turn to obtaining analytical expression for conductivity in the cold RMT model. To do this, we first  evaluate  the two-point function in frequency space using eq(\ref{eq:Def tilde G}) in the low-temperature limit. By extending the upper limit of the time integral to $\infty$, valid for $\beta E_0 \gg 1$, we obtain: 
\begin{equation}\begin{split}\label{eq:tilde G for cold RMT}
    \tilde G(\omega) 
    & \approx \frac{2\sqrt{2}f {\cal N}_r   (1+\Delta)}{E_0 \Gamma (\alpha +1)}   e^{\frac{\beta  \omega }{2}}\ (|\omega|/E_0)^\alpha \  \sqrt{\beta |\omega|} K_{\alpha +\frac{1}{2}}\left(\frac{\beta  | \omega | }{2}\right),
    \end{split}
\end{equation}
where we have kept $\omega \beta$ fixed. The resulting expression explicitly satisfies KMS. We can obtain conductivity from eq(\ref{eq:Conductivity formulae general}) 
\begin{equation}\label{eq: Cold RMT conductivity}
\begin{split}
    \sigma &  \approx  {   2^{2\alpha+3} \epsilon^2 \mathcal{N}_r^2 f_0^2 (1+\Delta)^2     \over E_0 } \ \frac{(\alpha +1) \Gamma \left(2 \alpha +\frac{5}{2}\right)}{\Gamma (\alpha +1) \Gamma \left(\alpha +\frac{5}{2}\right)}\ (\beta E_0)^{-1-2\alpha}.
    \end{split}
\end{equation}
For $\alpha={1\over 2}$ we get,
\begin{equation}\label{eq: cold RMT conductivity}
    \sigma = {45\epsilon^2 f_0^2 \mathcal{N}_r^2 (1+\Delta)^2 \over  E_0^3 \beta^2}.
\end{equation}
This is consistent with  low temperature features seen before in figure(\ref{fig:conductivityRMT}). This line of analysis shows how the low temperature power laws in RMT model directly follow from the behavior of the spectrum near the edges.

\subsection{Toy Model 2 : Conformal  model }\label{subsec:Conformal model example}
    
We will now describe a second toy model for heat transport which exhibits a qualitatively different behavior than the RMT model above. We will see later that, like the RMT model, this model can also be realized as a limit of DSSYK in an appropriate regime. 

We will call this second model as \emph{conformal model}, since it captures features of conformal quantum mechanical systems at finite temperature. The conformal regime is characterized by the zero temperature two point function $G(t)$ scaling as $t^{-2m}$ for some exponent $m$. The corresponding finite temperature result can then be obtained by a simple conformal transformation. This motivates a finite temperature two point Wightman function of the form
\begin{equation}\label{eq:G for SYK}
    G(t) \sim  \sinh^{-2m}\left( {\pi (t-i\delta) \over \beta}\right) ,
\end{equation}
where $m, \delta$ are some parameters. Here $\delta$ is a UV cutoff which regulates the conformal behavior for small times $t\ll \delta$.

A two point function of the above form arises for instance in the low  temperature limit of 
SYK models, in an emergent scale invariant regime\cite{Maldacena:2016hyu}.
The exponent $m$ is related to the conformal dimension of the operator in the IR, whereas $\delta$
models the sensible UV behavior that cuts off the conformal regime in this model. In the SYK model $\delta$ is fixed by UV physics - see \cite{Larzul:2022kri} for corresponding analytical statements and \cite{Almheiri:2019jqq} for numerical results. We will see in the next section that, in the DSSYK model as well, an analogous SYK-like limit leads to a finite value of $\delta$.\footnote{For instance, in frequency space, DSSYK two point function is given in eq(\ref{eq: Two point function omega conformal})  where $\phi_0$ is related to $\delta$ via $\phi_0={\pi\over 2} -\delta$. }

We would like a two point function of the above form that also obeys  the KMS 
condition $G(t) = G(-t - i \beta)$ (which the above form does not as it currently stands). This is achieved by the following ansatz for the two point function
\begin{equation}\label{eq:G for conformal model}
    G(t) = { f_0 \sinh^{2m}(- i \theta_0) \over \sinh^{2m}({2\phi_0 t \over \beta} - i \theta_0 )},
\end{equation}
where $f_0$ is some overall normalization, $m$ is the scaling exponent, and 
$\phi_0$ and $\theta_0$ are parameters related by  $\theta_0+ \phi_0 ={\pi \over 2}$ so that KMS is preserved.   In some cases, it is convenient to make the KMS condition explicit by expressing 
$\theta_0$ in terms of $\phi_0$, yielding:
\begin{equation}\label{eq:G for conformal model preserving KMS}
    G(t) = { f_0 \cos^{2m}(\phi_0) \over \cos^{2m}({\phi_0\over \beta}(\beta - 2it) )}.
\end{equation}

Note that, in specifying the conformal model, we have not described the density of states $\rho(E)$ and the averaged squared matrix element of operators $f(E_1,E_2)$ as we did for the RMT model. Rather we find it convenient to directly write down the two point function which is after all what is needed to study heat transport.\footnote{We will see later two cases where we get explicit expressions for $\rho(E),f(E_1,E_2)$ which can give rise to such a two point function. First one is the Gaussian model in subsection \ref{subsec:Gaussian model example}, which can be thought of as a heavy operator (i.e., $m\gg 1$) limit of the conformal model. The second instance is a specific conformal regime of DSSYK (see section \ref{subsec:DSSYK conformal regime}).
}

Having specified the model via the above two point function, let us move to describe its consequences. One striking difference between the RMT vs. the conformal models is their large time behavior. At large times, the two-point function of conformal model exhibits exponential decay, i.e.,  \begin{equation*}
G(t) = f_0 \left( 2e^{-i\phi_0}\cos \phi_0 \right)^{2m}\ e^{-{4m t \phi_0 \over \beta}},
\end{equation*}
in sharp  contrast to the power-law falloff seen in the RMT models.  The two point function in frequency domain can be determined to be\footnote{ We have used the  result \cite{chakrabarti2012remarkable}
\[
   2^{2m} \int\limits_{-\infty}^{\infty} \frac{d\omega}{2\pi} e^{-i\tau \omega}\ \frac{\Gamma(m + i\omega)\Gamma(m - i\omega)}{\Gamma(2m)} = \cosh^{-2 m}{\tau\over 2},\qquad \mbox{Re}(m)>0\ .
 \]
 Note that since $\lim_{\omega \rightarrow \pm \infty} \Gamma(m + i\omega)\Gamma(m - i\omega) = 2 \pi |\omega|^{2m-1} e^{-\pi |\omega|}$, the integral converges for $-\pi < \mbox{Im}(\tau) < \pi$. The result in the main text follows by setting $\tau = {4 \phi_0 t \over \beta} + 2i \phi_0$. The convergence criterion then translates to the condition $-{\pi \over 2} < \phi_0 < {\pi \over 2}$.},
\begin{equation}\label{eq: Conformal two point function in frequency}
    \tilde G(\omega)= {f_0 \beta \over 4\phi_0 }{(2\cos{\phi_0})^{2m}\over \Gamma(2m)} e^{\beta \omega \over 2} \Gamma(m+{i\beta\omega\over4\phi_0 }) \Gamma(m-{i\beta\omega\over4 \phi_0 })
\end{equation}
The function $\tilde G(\omega)$ has isolated poles at $\beta \omega =\pm 4 i (m+n) \phi_0  ,\ \forall n \ge 0$. For the SYK models at low temperatures, these quasinormal modes were observed in \cite{Dodelson:2024atp}. This should be contrasted with the case of Cold RMT where $\tilde G(\omega)$ had branch cut singularities at least at low temperatures - see eq(\ref{eq:tilde G for cold RMT}). This difference is perhaps one reason why the large time asymptotics are so different in these two models.

\begin{figure}
    \centering
    \includegraphics[width=0.7\linewidth]{ 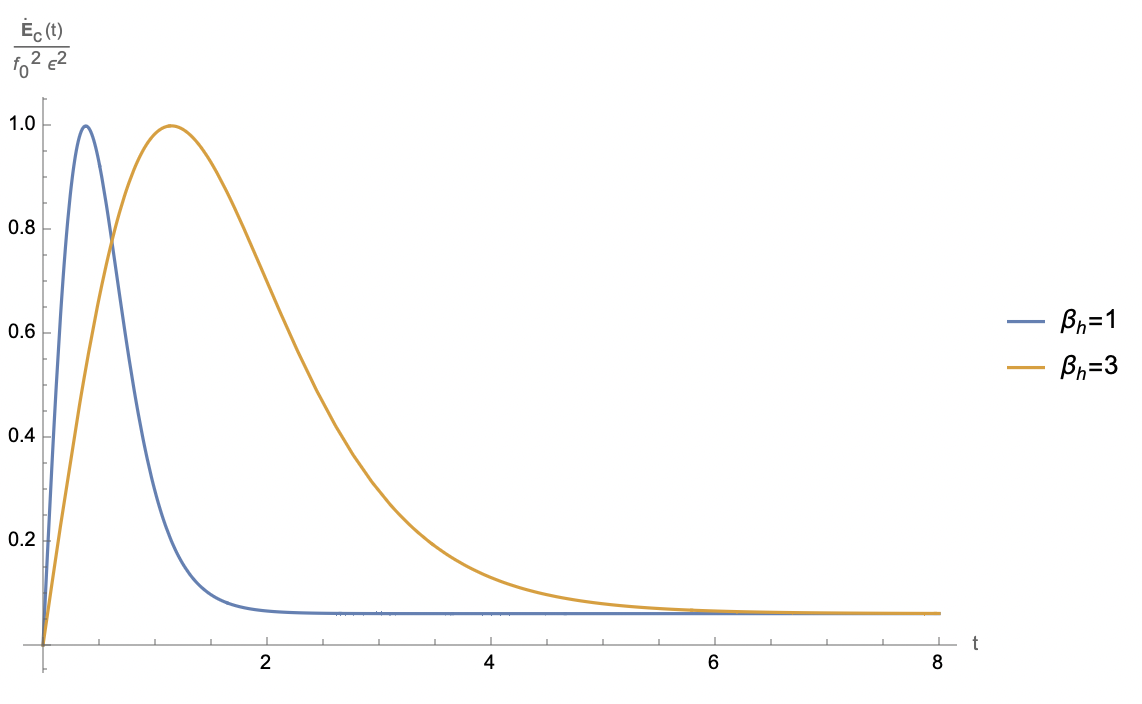}
    \caption{Behavior of $\dot E_c$ for systems at different temperatures, $\theta_0={\pi\over3}$ and $m=1$,\ $\phi_0={\pi\over 6}$ with peak heights normalized to 1.}
    \label{fig:J Schwarzian general}
\end{figure}

\begin{figure}
    \centering    \includegraphics[width=0.65\linewidth]{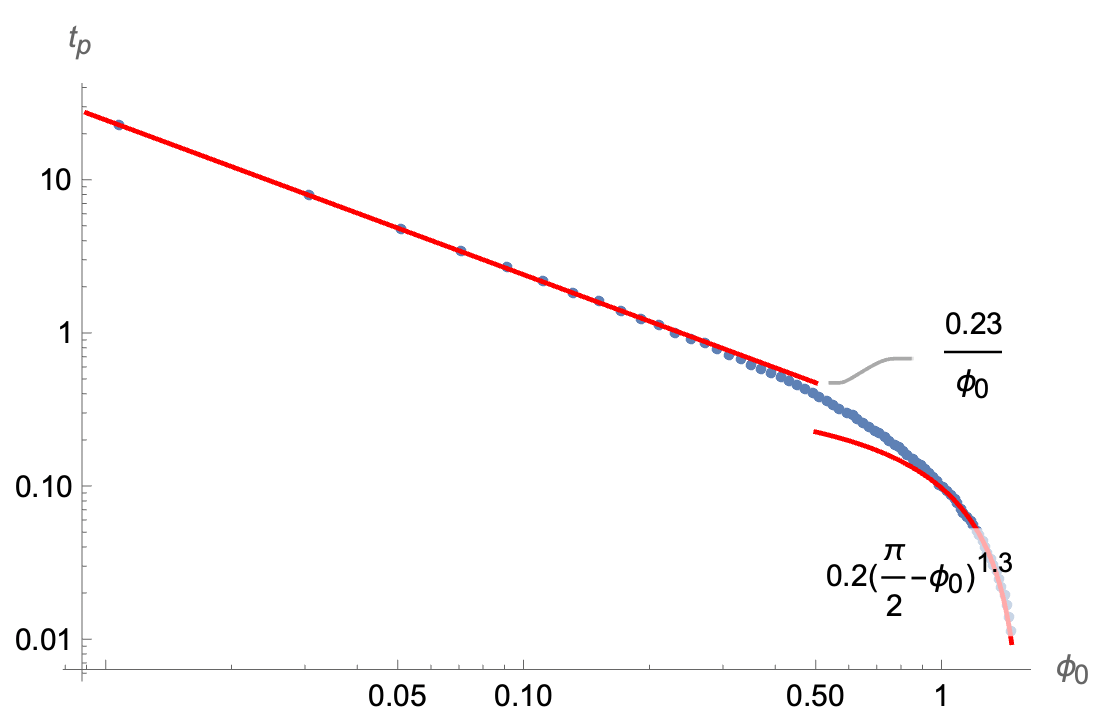}
   \caption{Peak time as a function of $\phi_0$ for conformal model  with parameters $m=1,\ \beta_c=1.1\beta_h$ in $\beta_h=1$ units.}
    \label{fig:Conformal tp}
\end{figure}

\begin{figure}
    \centering    \includegraphics[width=0.65\linewidth]{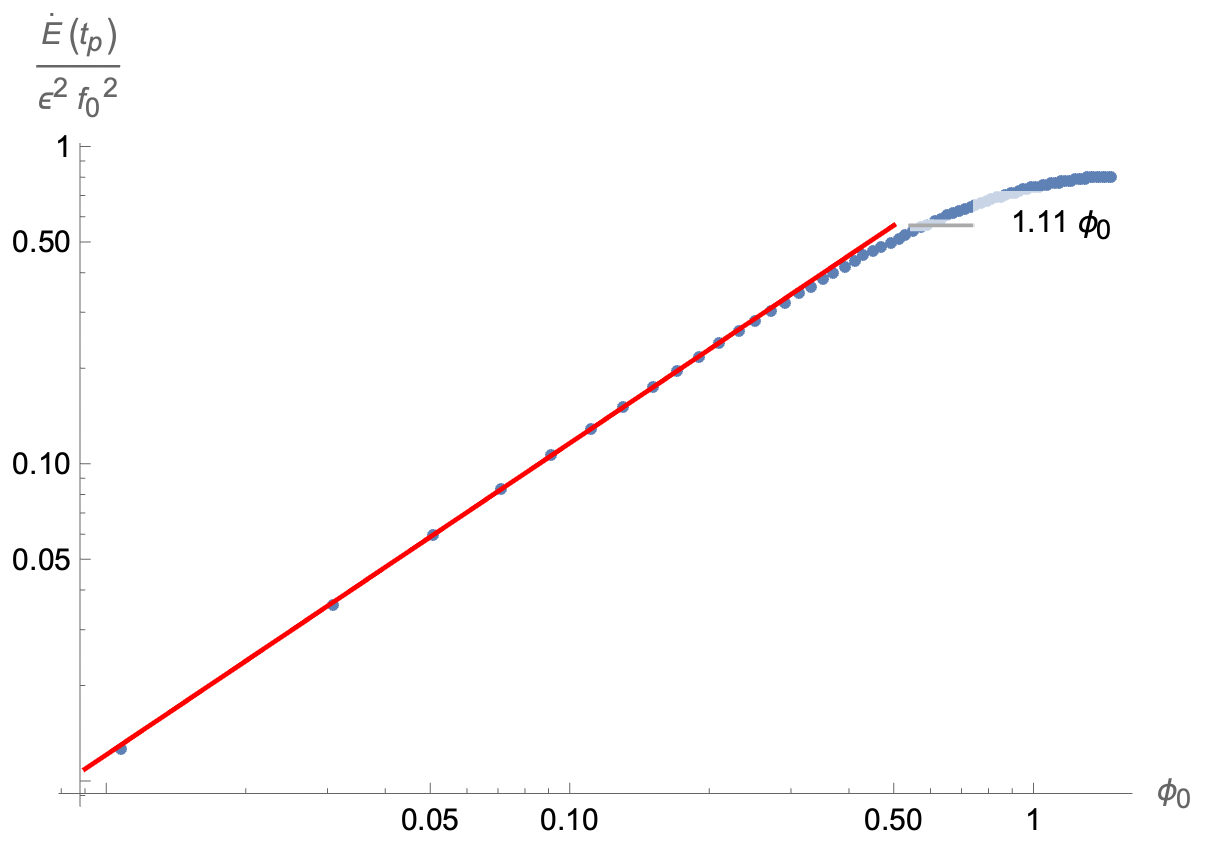}
     \caption{Peak height as a function of $\phi_0$ for conformal model  with parameters  $m=1,\ \beta_c=1.1\beta_h$ in $\beta_h=1$ units.}
    \label{fig:Conformal maxima}
\end{figure}

\begin{figure}
    \centering
    \includegraphics[width=0.65\linewidth]{ 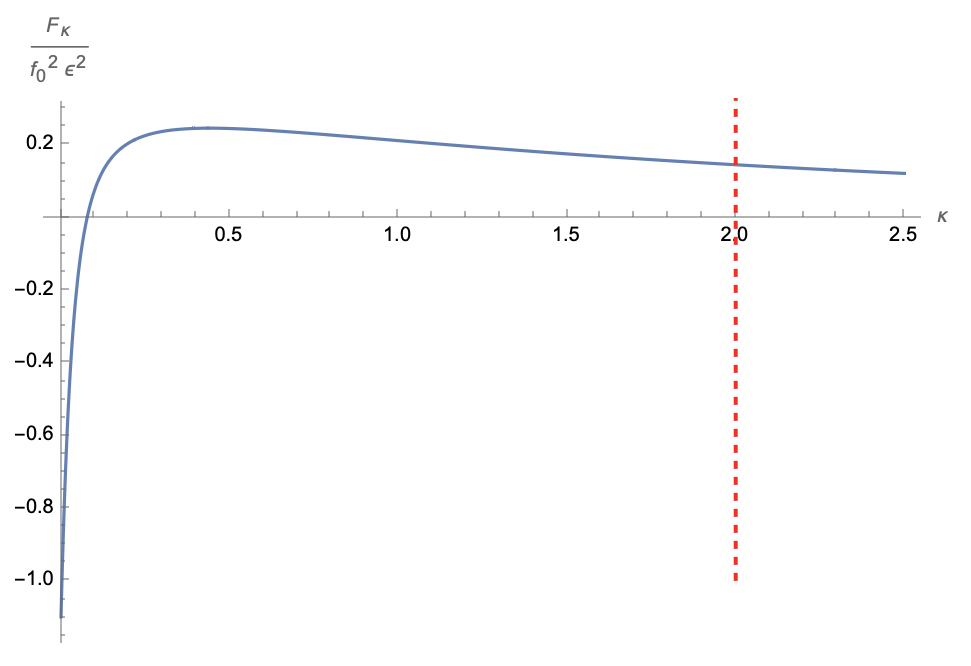}
    \caption{$F_\kappa$ for system at $\beta_h=0.1, \  \ \beta_c=0.11$ and $\phi_0={\pi\over 6},\ m=1$}
    \label{fig:Conformal Fk}
\end{figure}

We can  compute the heat current by plugging the above expressions into eq(\ref{eq:Def Heat current}) and plot $\dot E_c(t)$ in   the figure(\ref{fig:J Schwarzian general}) for two different temperatures. The plot clearly shows all the three features : transients, approach to steady state and steady state. 
\paragraph{Transients:}
As in the RMT model, we see a heat surge with a peak. However, the scaling of peak times $t_p$ and peak height are qualitatively different. We  give below the peak position $t_p$ and peak height for typical values of parameters $\beta_c,\beta_h,m$ as a function of $\phi_0$ in figures (\ref{fig:Conformal tp},\ref{fig:Conformal maxima}). We see that the peak in the heat current occurs earlier and earlier as we increase the parameter $\phi_0$, with $t_p$ scaling inversely with $\phi_0$ for small $\phi_0$. As to the peak height, it grows linearly with $\phi_0$ and saturates eventually to a constant value as $\phi_0 \rightarrow {\pi \over 2}$. To summarize, we get smaller and earlier peaks at small $\phi_0$ whereas larger $\phi_0$s result in larger peaks, occurring later. The integrated energy flux $F_\kappa$ is positive for all $\kappa \ge {2\over \beta_h}$ as can be seen in figure(\ref{fig:Conformal Fk}).

\paragraph{Approach to steady state:} At long times, as we mentioned before  $G(t)\sim e^{-{4\phi_0  t m \over \beta} }$ and the  long time behavior of heat current can be computed  to be\footnote{up to factors which are time independent, but depending on $\phi_0,m, \beta$.},
\begin{equation}
   \dot E_c(t) -  \dot E_c(\infty) \ \propto  \ e^{-4\phi_0 t m(\beta_c^{-1}+\beta_h^{-1}) } .
\end{equation}
In contrast to the RMT results eq(\ref{eq:RMT Long time heat current}) where we saw a power law approach, here we see an exponential approach to the steady state.

\paragraph{Steady state:} The conductivity can be calculated using the two point function eq(\ref{eq: Conformal two point function in frequency}) as, 
\begin{equation}\label{eq:Conductivity conformal model}
\sigma   = c_m f_0^2 \epsilon ^2 \ \phi_0 \beta \cos ^{4 m}(\phi_0)\ ,
\end{equation}
where $c_m\equiv {2^{4m +1} \over \pi \Gamma(2m)^2} \int\limits_0^\infty d\omega \omega^2 |\Gamma(m+i \omega)|^4 =2\sqrt\pi m^2 {\Gamma(2m)\over \Gamma(2m+{3\over 2})}$.
For a fixed $\phi_0$, we see that the conductivity varies linearly with $\beta$. At a fixed temperature, $\sigma$ exhibits a maximum at an intermediate $\phi_0$.
     
 As a brief aside, we present the expression for thermal conductivity in a more general scenario where the coupled operators $O_c$ and $O_h$ (see eq(\ref{eq:Setup of the model})) have different scaling dimensions, denoted by  $m_c$ and $m_h$. In this case, the thermal conductivity takes the form
     \begin{equation}\label{conductivity conformal mc mh}
    \begin{split}
        \sigma  & = {\epsilon^2 \beta^2 \over 2 \pi } \ \int_0^\infty G_c(\omega) G_h(\omega) e^{-\beta \omega}
    \omega^2 d\omega 
    = c_{m_c,m_h}  \epsilon^2 f_0^2 \beta \phi_0 (\cos \phi_0)^{2m_c+2m_h},
    \end{split}
\end{equation}
where we have defined the prefactor
\[ c_{m_c,m_h}\equiv 2^{2m_c+2m_h +1}  \int\limits_0^\infty \frac{d\omega}{\pi} \ \omega^2 \frac{|\Gamma(m_c+i \omega)|^2 |\Gamma(m_h+i \omega)|^2}{\Gamma(2m_c) \Gamma(2m_h)}\ .\]

\paragraph{Heavy operator limit } 
There is a special limit in which the conformal toy model results are particularly simple. In the limit of heavy operator, $m \rightarrow \infty, \phi_0  \rightarrow 0$ limit, with the parameter $J_e \equiv {2\phi_0\sqrt m \over \beta }$ held fixed, the two point function given in eq(\ref{eq:G for conformal model preserving KMS}) becomes 
\begin{equation}\label{eq:G(t) in Gaussian model}
G(t) = f_0 e^{-{t(t+i \beta)J_e^2} }.
\end{equation}
As we describe in the next subsection, such a two point function can arise from another class of toy models which we term {\it gaussian models}.

\subsection{Toy Model 3 : Gaussian model}\label{subsec:Gaussian model example}
We will now describe a third simple toy model for heat transport. As indicated before, this model can be thought of as a limit of conformal model, but we think it worthwhile to give the third model an independent description ab initio. 

Consider then the {\it gaussian model}  with a  density of states 
 \begin{equation*}
 \rho(E) = {1 \over J \sqrt{2\pi} } e^{-{ E^2  \over 2 J^2} }.
 \end{equation*}
Here $J$ sets the energy scale within which most states occur. For example, we expect such a density of states if our system is made of many identical, non-interacting subsystems each having a somewhat random spectrum. By central limit theorem, the distribution of the total energy automatically takes the above form.

The partition function corresponding  to the above density of states is also a Gaussian given by 
 \begin{equation*}
 Z(\beta) = e^{\beta^2 J^2 \over 2}.
 \end{equation*}
We  take the averaged squared operator matrix elements to also have a Gaussian dependence on energies, i.e.,
\begin{equation}\label{eq:f in gaussian model}
    f(E_1,E_2) = { f_0 \over \sqrt{1-\tilde q^2} }  e^{ {\tilde q \over 2(1-\tilde q^2)J^2} \left[ 2 E_1 E_2 - \tilde q (E_1^2+E_2^2) \right]} .
\end{equation} 
Here $f_0$ is an  overall normalization constant. We have also introduced another dimensionless parameter $\tilde{q}$ in a way that is convenient for computations below. The two point function can be obtained from eq(\ref{eq:Def G Two point function}) to be
\begin{equation}
    G(t) = f_0 e^{-{t(t+i \beta) J^2 (1-\tilde q) }} \ .
\end{equation}
We note that this exactly matches the heavy operator limit of the conformal toy model given in  eq(\ref{eq:G(t) in Gaussian model}) with $J_e \equiv  {J  \sqrt{1-\tilde q} }$. The Fourier transform is then given by another Gaussian 
\begin{equation}\label{eq:Two point function as function of omega at q=1 arbit tilde q}
    \tilde G(\omega) = { \sqrt \pi \ f_0 \over J_e} e^{-\frac{1}{4}\left\{{\beta J_e} - {\omega \over J_e}\right\}^2}\ . 
\end{equation}
This result can also be understood as a limiting case of the conformal model’s frequency-space two-point function $\tilde G(\omega)$ eq(\ref{eq: Conformal two point function in frequency}) using the asymptotic behavior of the $\Gamma$ function. 

\begin{figure}
    \centering
    \includegraphics[width=0.75\linewidth]{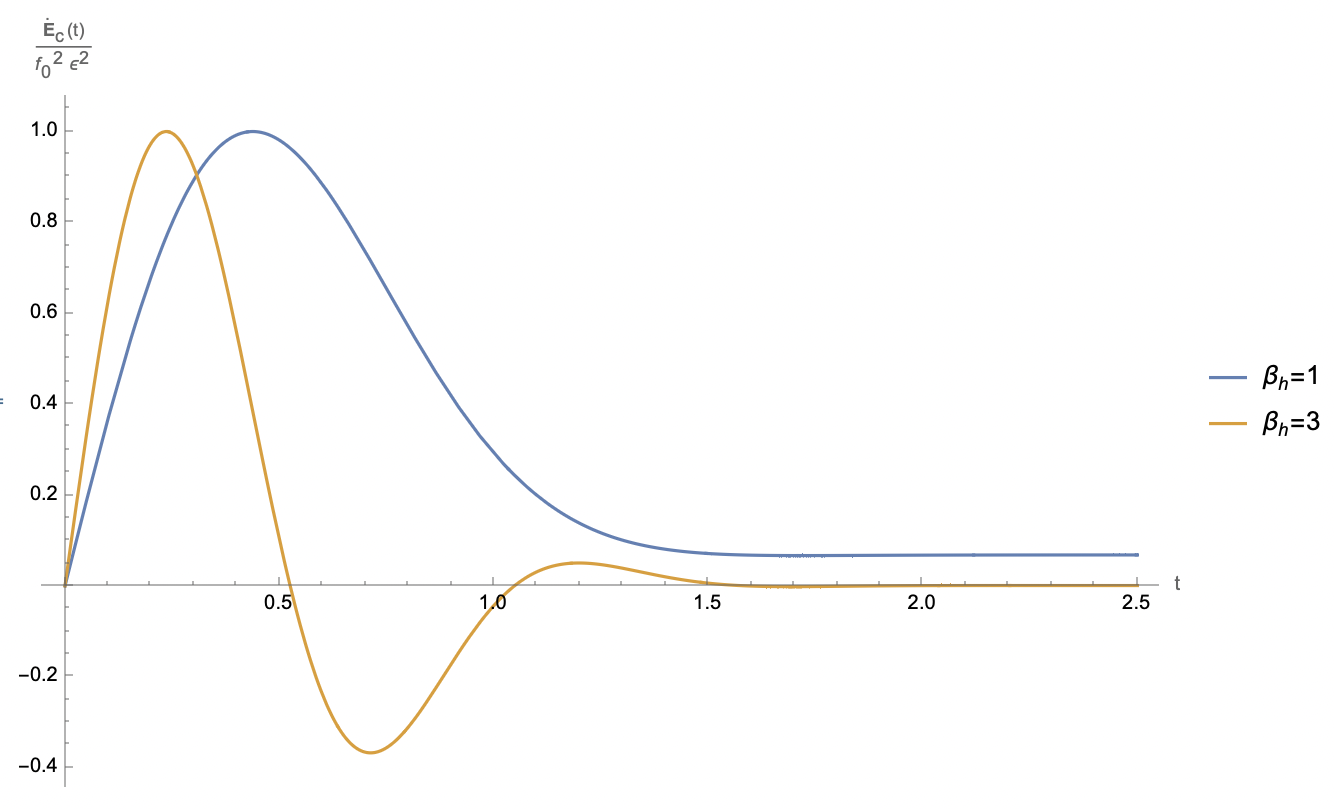}
    \caption{behavior of heat current for systems with $J_e=1,\  \Delta=0.1$. The peak heights are normalized to 1.}
    \label{fig:Gaussian Ec}
\end{figure}

Now that we have the two point function, we can compute the heat current 
 eq(\ref{eq:Def Heat current}) for this model - the result is plotted in figure(\ref{fig:Gaussian Ec}).
We see the same qualitative behavior as the other two models: a heat surge and an approach to steady state. Given that all the integrals involved are Gaussian, many exact analytical expressions can be obtained.

\begin{figure}
    \centering
    \includegraphics[width=0.65\linewidth]{ 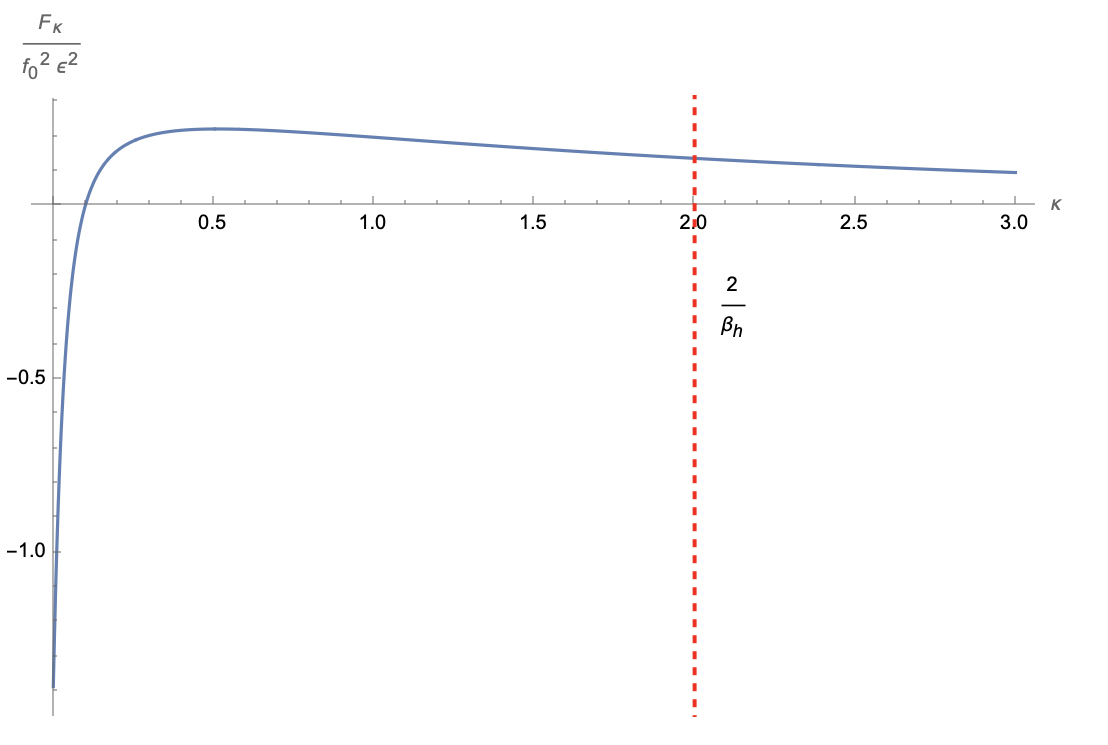}
    \caption{$F_\kappa$ for Gaussian model with $J_e=1$ }
    \label{fig:GaussianFk}
\end{figure}

\paragraph{Transients: } We find the peak time  $t_{p}$  by solving a transcendental equation 
\begin{equation}
{2t_{p}  \over \beta_c} \tan {[t_pJ_e^2(\beta_c + \beta_h) ]} = 1\ .
\end{equation}
This maps the $t_p$ computation to computing ground state energy in a finite square well potential (and the subsequent peaks/troughs amount to computing the higher excited states of even parity in the well). The answer to the finite square well problem is well-known: as the well becomes broader, the ground state energy goes down. Here, similarly, we see that, at a fixed temperature, the heat peak moves to earlier times as we increase $J_e$. In the well problem, we get more bound states as the well gets deeper: here we find more peaks and troughs as $\beta$ increases. Additionally, the integrated energy flux $F_\kappa$ is positive for all $\kappa \ge {2\over \beta_h}$ as illustrated in fig(\ref{fig:GaussianFk}). 

\paragraph{Approach to Steady state: }  We can also explicitly compute the approach to the steady state at large times in this model. In this limit, we obtain
\begin{equation}\begin{split}
    {\dot E_c(t)\over \epsilon^2 f_0^2}
    & =e^{-2 t^2 J^2_e} \sin \left(t J_e^2 (\beta _c+\beta _h)\right) \\
    & \quad + {(\beta_c - \beta_h) \over 4} \left( \sqrt{2\pi}\ J_e e^{-{J_e^2(\beta_c + \beta_h)^2 \over 8 }}    - {e^{-{2t^2 J_e^2}} \cos(t(\beta_c + \beta_h)J_e^2)    \over t}\left[ 1+ {\cal O}(t^{-1}) \right] \right).
    \end{split}
\end{equation} 
What we see here is that the steady state is reached very quickly governed by a gaussian falloff in time. We see that this is even more faster approach to steady state than the exponential approach in the conformal model, and is in stark contrast to power law approach in the RMT model.

\paragraph{Steady state: } It is also equally straightforward to characterize the steady state in this model. Using eq(\ref{eq:Conductivity formulae general}), we can  evaluate the conductivity to be,
\begin{equation}
\sigma =   { \sqrt{\pi \over 2} }{\epsilon^2 \beta^2 f_0^2 J_e \over 4 } e^{-{\beta^2 J_e^2\over 2 }}.
\end{equation}
Note that for a fixed $J_e$, conductivity falls off both at small and large temperature, with a maximum at some intermediate temperature.

\section{Heat transport in DSSYK model}\label{sec: Heat Transport in DSSYK}
In the previous section, we looked at heat transfer in a variety of toy models. encountering broad similarities but also many qualitative differences. The models were simple, but to some extent pulled out of the hat: we never wrote down an explicit Hamiltonian to solve. We would now like to fix this by formulating a full microscopic model which can be solved. It would also be nice to have a single model realise all the behaviors we described in the last section so that we can see what kind of microscopic regimes lead to different kind of heat transfers, To this end, we now move on to the more realistic case of the transport in the Double Scaled Sachdev Ye Kitaev (DSSYK) models. 

As in the case of our toy models, the input we need from the DSSYK model to compute the transport is the two point function. Hence we begin in  section \ref{subsec:DSSYK review} where we  quickly summarize the relevant results of DSSYK model.   In section \ref{subsec:Regimes of DSSYK} identify the regimes where the DSSYK model simplifies and reduces to one of the toy models described in the previous section. Finally,  we turn to the question of addressing the transport in the DSSYK models in section \ref{subsec:Transport in DSSYK}.

\subsection{Review of the double scaled SYK Model}\label{subsec:DSSYK review}
The SYK model \cite{KitaevTalk1,PhysRevLett.70.3339,PhysRevD.94.106002} is a quantum mechanical model of $N$ fermions with disordered interactions. The Hamiltonian is given by 
\begin{equation}\label{eq: DSSYK Hamiltonian}
H = \sum_{I} i^{p \over 2} J_{I} X_{I}.
\end{equation}
Here $I$ denotes the ordered index set $\{i_1,i_2,\dots i_{p}\}$ (with $i_1<i_2<\dots i_N$) where each of the indices $i_k \in \{1,\dots N \}$, $X_I$ denotes a fermion product $X_I = \chi_{i_1} \dots \chi_{i_{p}}$ and $p$ is even. 
Here $\chi_i$ are Majorana fermions satisfying $\{\chi_i , \chi_j \} = 2 \delta_{ij}$. The couplings $J_I$ are disorder and are taken from an ensemble satisfying
\begin{equation}\label{eq:disorder average J}
\langle \langle  J_I J_J  \rangle \rangle  = {\delta_{IJ} \over \binom{N}{p} } \ J^2\ .
\end{equation}
Here $\langle \langle \dots \rangle\rangle$ denotes disorder averaging, and $J$ has dimensions of energy.  We will mostly work in the units where $J =1$ (except in section \ref{subsec: Special Cases of DSSYK} where we will temporarily restore $J$). We sometimes refer to the model eq(\ref{eq: DSSYK Hamiltonian}) as SYK$\!_p$ model since there are $p$ fermion interaction in the Hamiltonian. 

The double scaling limit of the SYK model \cite{Erds2014PhaseTI,Cotler:2016fpe,2018JHEPNarayan,Berkooz:2018jqr} is defined as the large $N$ limit, and $p$ scaling with $N$ as $p \propto\sqrt N$. More precisely, double scaling limit is defined via
   \begin{equation}
      q \equiv  e^{-\lambda},\  \lambda \equiv \frac{p^2}{N}, \hspace{3mm}\textrm{held fixed as } N \rightarrow \infty\ ,
   \end{equation}
where we have defined a parameter $q$ with $ 0 \le q  \le 1$. 

The operators of interest in this model are again fermion products, analogous to the ones that appear in the Hamiltonian but with $\tilde p$ number of fermions. We hence define $O \equiv i^{\tilde p \over 2} \tilde J_{\tilde I} X_{\tilde I}$, where $\tilde I$ is an ordered index set $\{i_1,i_2,\dots i_{\tilde p}\}$ and $X_{\tilde I}$ denotes a fermion product $X_{\tilde I} = \chi_{i_1} \dots \chi_{i_{\tilde p}}$. $\tilde J_{\tilde I}$ is disorder (independent of $J_I$ introduced in eq(\ref{eq: DSSYK Hamiltonian})) satisfying
\[
\langle \langle  \tilde J_{\tilde I} \tilde J_{\tilde K}  \rangle \rangle  = {\delta_{\tilde I \tilde K} \over \binom{N}{\tilde p} } \ \tilde J^2.
\]
We will normalize the operators such that $\tilde J =1$. The double scaled limit of operators are defined as 
\begin{equation}
    \tilde q \equiv e^{-m \lambda},\ m = {\tilde p \over p}, \hspace{3mm}\textrm{held fixed as } N \rightarrow \infty.
\end{equation}
Here we have defined another parameter $\tilde q$ with $ 0 \le \tilde q  \le 1$. 
Throughout this work, we will take the strict $N\rightarrow \infty$ limit and neglect all $1/N$ corrections. To summarize, in the double scaling limit, we get the DSSYK model parametrized by $\lambda, m$ or equivalently $q,\tilde q$. 

In the so called triple scaling limit $\lambda\rightarrow 0$, with low temperatures $\beta \sim \lambda^{-{3 \over 2}}$, the familiar SYK physics governed by the Schwarzian action is recovered \cite{Cotler:2016fpe}. The advantage of DSSYK is that we not only have an extra tunable $\lambda$, but also the ability to obtain closed-form expressions for several observables across \emph{all energy scales} — not just in the low-energy regime as in the solution of original SYK model.

The relevant object of interest for us in this work are the partition function and the two point function defined as follows
\begin{eqnarray}
\label{eq:Def of Z}
Z(\beta) &\equiv& \langle \langle \mbox{Tr}\left( e^{-\beta H} \right) \rangle\rangle,     \\
\label{eq:Def of G}
G(t) &\equiv& {1 \over Z(\beta) }\ \langle \langle \mbox{Tr}\left( O(t) Oe^{-\beta H} \right) \rangle\rangle.
\end{eqnarray}
Here  the trace is normalized such that $\mbox{Tr}({\mathbb I})=1$. 
 Using combinatorial techniques, these observables can be reformulated as sums over chord diagrams \cite{2018JHEPNarayan}, which in turn can be recast in terms of a transfer matrix. The contributions at each order can be explicitly written down, and resummed to yield closed form expressions, Our goal here is not to explain all the steps involved  which are well-described in the original references. Instead, we will be content with
 a brief review with results relevant to our problem.
 
\paragraph{Partition function $Z(\beta)$}
The partition function can be reconstructed  from the disorder averaged moments of the Hamiltonian $\langle \langle \mbox{Tr}H^k\rangle \rangle$. For the moments, we have that 
\begin{equation}
    \langle \langle \mbox{Tr}H^k\rangle \rangle = i^{k p \over 2}\sum_{I_1 ,\dots I_k} \langle \langle  J_{I_1}\dots J_{I_k} \rangle \rangle \mbox{Tr}\left( X_{I_1}\dots X_{I_k} \right).
\end{equation}

The disorder averaging reduces to Wick contractions among the index sets $I_k$, as shown in eq(\ref{eq:Wicks contraction in moments}). Each choice of contraction can be represented by a {\it chord diagram} with $k$ nodes — corresponding to the insertions of $X_I$ — arranged on a circle, reflecting the cyclicity of the trace (see \cite{cordial} for more details). Chords connect pairs of nodes to indicate index contractions. This gives an infinite sum
\begin{equation}\label{eq:Wicks contraction in moments}
     \langle \langle \mbox{Tr}H^k\rangle \rangle =  i^{kp \over 2}  \binom{N}{p}^{-k/2} \sum_{\mbox{contractions}} \ \sum_{I_1,\dots I_{k\over 2}} \mbox{Tr}\wick{( \c1 X_{I_1} \c3 X_{I_2} \c1 X_{I_1} .\c2..\c3...  \c2 X_{I_k}  )}.
\end{equation}

If the chord diagrams are non-crossing — i.e.,  there are no intersections between chords—then the contracted fermion products are adjacent and square to the identity $X_I^2 \propto {\mathbb I}$. In the presence of chord crossings, the contracted fermion pairs can still be brought adjacent by commuting them, at the cost of introducing sign factors, since 
\begin{equation*}
X_I X_J  = (-1)^{|I| \cap|J|} X_J X_I ,
\end{equation*}
 where $|I| \cap |J|$ denotes the number of indices shared between the index sets $I$ and $J$. In the double-scaled limit, $|I| \cap |J|$ can be shown to be Poisson distributed, and this leads to the following result
\begin{equation}\label{eq:Moments via Chord diagram}
     \langle \langle \mbox{Tr}H^k\rangle \rangle  = \sum_{\mbox{chord diagrams with k nodes}}  q^{\mbox{number of intersections}}.
\end{equation}
 A transfer matrix can be constructed for implementing this procedure, which acts on the Hilbert space of open chord states. Denoting the state with $l$ open chords as $| l \rangle$, the moments of the Hamiltonian can be expressed as a simple transition amplitude 
 \begin{equation*}
 \langle \langle \mbox{Tr}H^k\rangle \rangle = \langle 0 | T^k | 0 \rangle,
 \end{equation*}
 where $T$ is a transfer matrix operator explicitly defined in \cite{2018JHEPNarayan}. As a result, the partition function is given by 
 \begin{equation*}
Z(\beta) =  \langle 0 | e^{-\beta T} | 0 \rangle.
 \end{equation*}
 The spectrum of the transfer matrix $T$ can be determined and is found to have finite width.
 
The final result, as derived in \cite{2018JHEPNarayan} by performing the infinite sum is,
\begin{equation}\label{DSSYK Results Partition}
    \begin{split}
        Z(\beta) &= \int\limits_{-E_0}^{E_0} dE \ \rho(E) e^{-\beta E} = \int_0^\pi d\theta\  \Psi(\theta) e^{-\beta E(\theta) }, \\
    \end{split}
\end{equation}
where the width as well as the explicit density of states are given by
\begin{equation*}
E_0=\frac{2}{\sqrt{1-q}} \qquad \text{and} \qquad \rho(E) ={\sqrt{1-q}\ (q;q)_\infty \over 4\pi\sin \theta} (e^{\pm 2i\theta};q)_\infty\ .
\end{equation*} 
The answers are written in terms of the q-Pochammer symbol $(a;q)_\infty \equiv \prod\limits_{k=0}^\infty(1-aq^k)$. We have also used here a convenient notation $(e^{ \pm i  b};q)_k \equiv  (e^{ib};q)_k(e^{-ib};q)_k$. In the second equality above for $Z(\beta)$, we have used a different angular  variable $\theta$ to parameterize the  energy spectrum. It is related to actual energy via $E(\theta) =- E_0 \cos \theta$, and the corresponding density of states is
\begin{equation}\label{eq:Def Psi in DSSYK}
    \Psi(\theta) 
    = {(1-q)^2 (q;q)_\infty^3 \over 2\pi}\ {1 \over |\Gamma_q({2i\theta\over \lambda})|^2 }\ .  
\end{equation}
Here we encounter the q-Gamma function $\Gamma_q(x)  \equiv (1-q)^{1-x} {(q;q)_\infty \over (q^x ; q)_\infty }$. As the name suggests, this is a generalization of the familiar Gamma function which is recovered in the $q\to 1$ limit, i.e., we have $\lim_{ q \rightarrow 1}\Gamma_q(x) = \Gamma(x)$.

\begin{figure}
    \centering
    \includegraphics[width=0.65\linewidth]{ 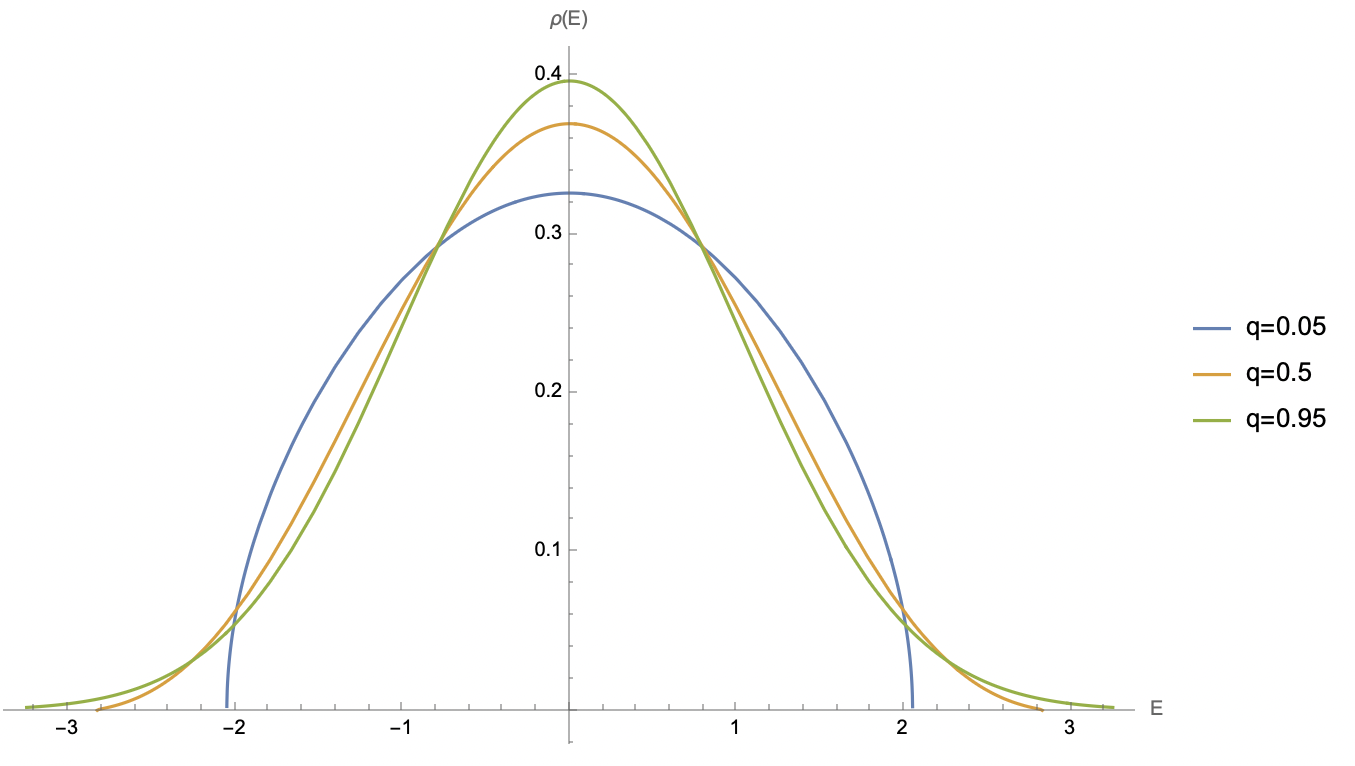}
    \caption{Density of states against energy for systems with different values of q.}
    \label{fig:density of states}
\end{figure}

\begin{figure}
    \centering
    \subfloat[]{%
        \includegraphics[width=0.45\linewidth]{ 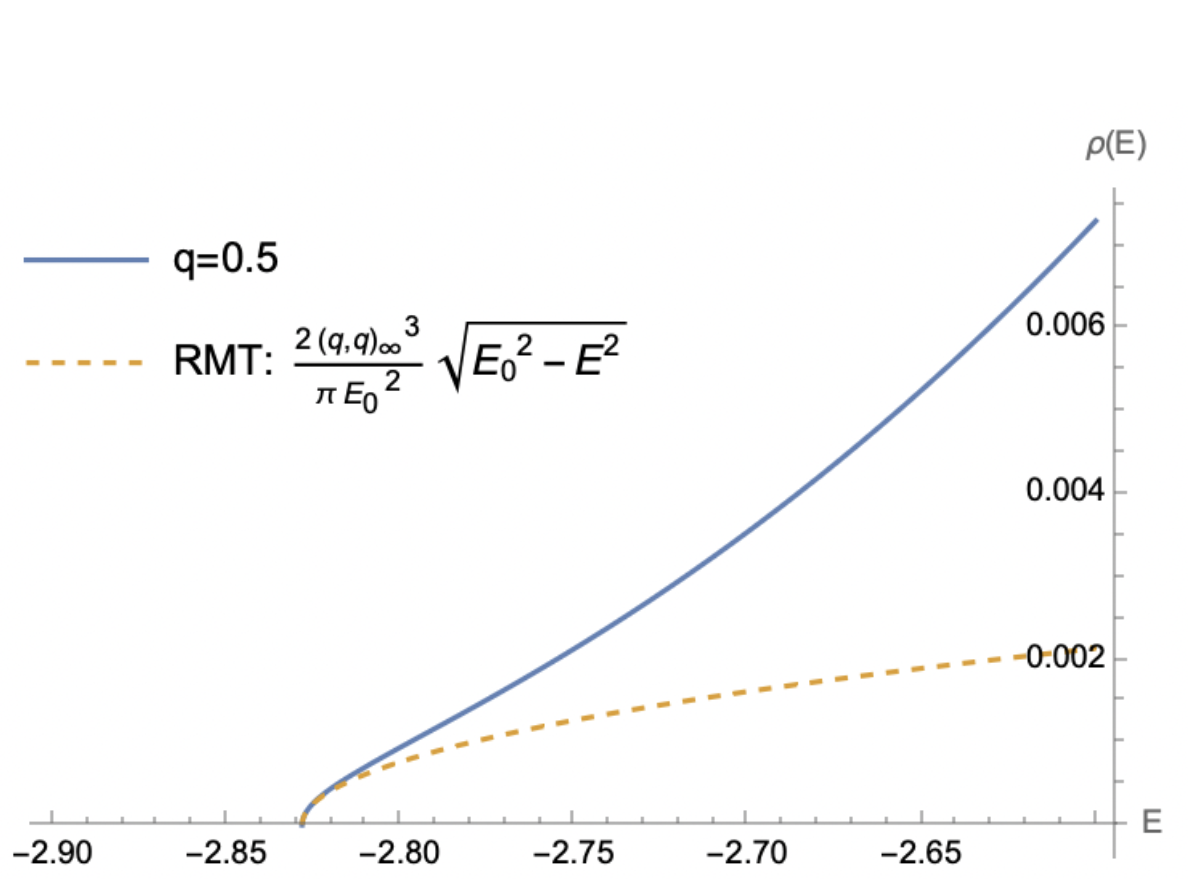}
        \label{fig:edgedistribution}
    }
    \hspace{0.015\linewidth} 
    \subfloat[]{%
        \includegraphics[width=0.45\linewidth]{ 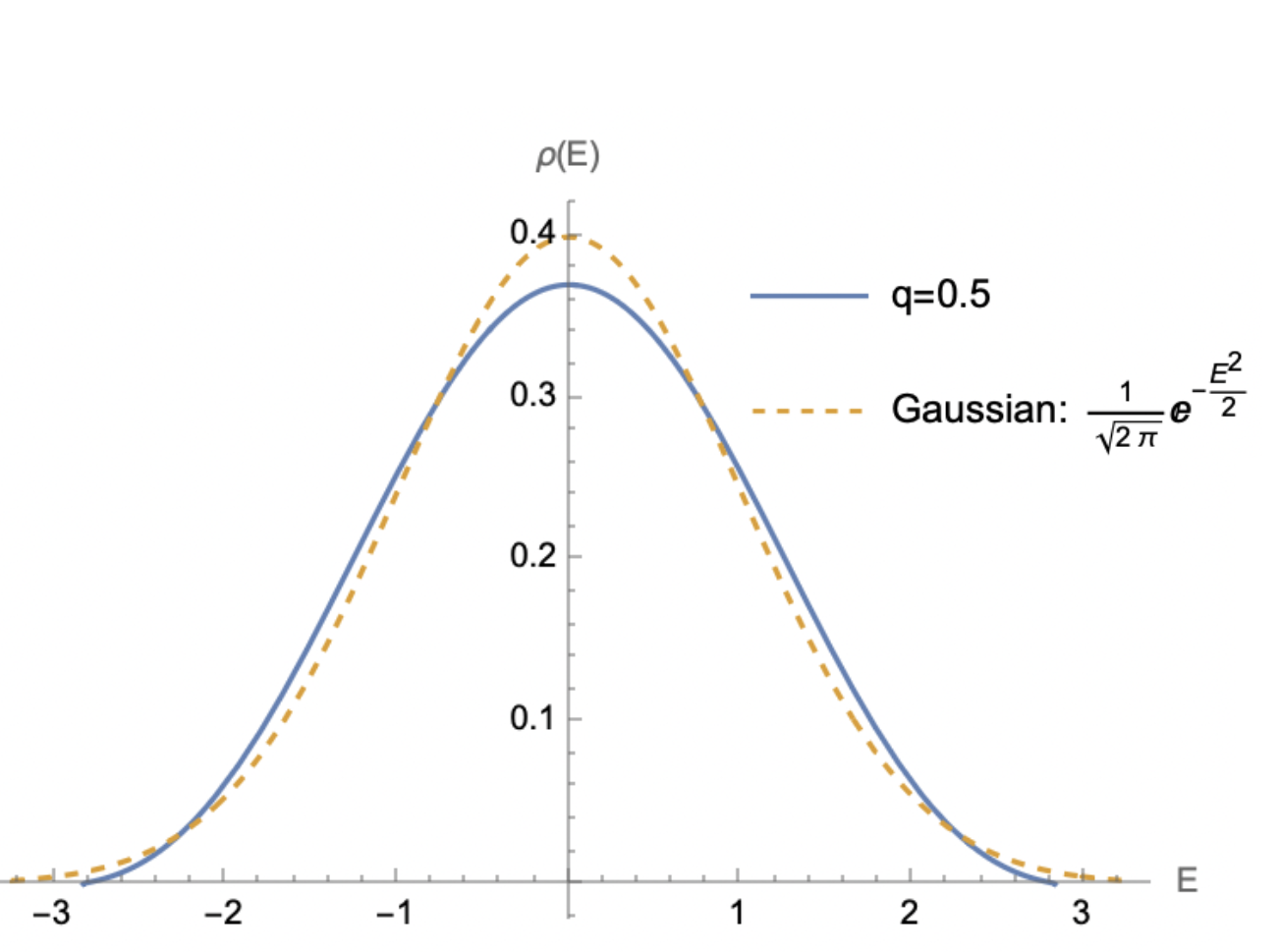}
        \label{fig:q vs gaussian}
    }
    \caption{Density of states for finite $q$ against semicircular and Gaussian density of states}
    \label{fig:limits of dos}
\end{figure}

To gain more insight into the spectrum described by above formulae,  we plot in figure(\ref{fig:density of states}) the density of states $\rho(E)$ for three different values of $q$. 
\begin{itemize}
    \item For $q=0$, the density of states 
$\rho(E)$ follows a semicircular distribution supported on $-2 \le E \le 2$; the case 
$q=0.05$ in figure(\ref{fig:density of states}) closely resembles this expected form. We remind the reader that $q\to 0$ corresponds to the limit where every term in the Hamiltonian describes a huge number of fermions interacting together. 
\item In contrast, for $q=1$, $\rho(E)$ takes a Gaussian form, and the case $q=0.95$ shown in figure(\ref{fig:density of states}) approximates this behavior well. We remind the reader that $q\to 1$ corresponds to the limit where every term in the Hamiltonian describes very few fermions interacting together. 
\item  Notably, for finite $q$ say $q=0.5$, the density of states retains a semicircular shape near the spectral edges—as illustrated in fig(\ref{fig:edgedistribution}), while closely approximating a Gaussian in the bulk, as seen in  fig(\ref{fig:q vs gaussian}).
\end{itemize}
Already  in these asymptotic behaviors, we see the emergence of our toy models. 

While the integral form of the partition function eq (\ref{DSSYK Results Partition}) is useful for understanding the dominant contributions at small $\lambda$, it is often convenient to work with a different representation for explicit calculations. In particular, the partition function admits a useful $q$-series expansion (see \cite{Berkooz:2018jqr})
\begin{equation}\label{eq:Z Bessel expansion}
\begin{split}
    Z(\beta) &= \frac{2}{\beta E_0} \sum_{p=0}^\infty (-1)^p q^{p+{p \choose 2}} (2p+1) I_{2p+1}({\beta E_0}) ,
\end{split}
\end{equation} 
where $I_{2p+1}$ denotes the modified Bessel function.  It will also be convenient to define the object
\begin{equation}\label{eq:Z_k as a integral}
    Z_k(\beta) \equiv \langle k | e^{-\beta T} | 0 \rangle =   \int_0^\pi d\theta \Psi(\theta) e^{2 \beta \cos \theta \over \sqrt{1-q}}\ {H_k(\cos \theta | q) \over \sqrt{(q;q)_k}}, 
\end{equation}
where $H_k(\cos \theta | q)$ are continuous q Hermite polynomials \cite{wiki:Continuous_q-Hermite_polynomials}. We mention some  properties of $Z_k$ in detail in Appendix(\ref{Appendix: $Z_k$ details}).

\paragraph{Two point function $G(t)$ : }
The chord diagram prescription  introduced earlier in the context of partition functions extends naturally to the two point function of $O$ operators we defined before. In the transfer matrix language, it leads to the same expression as in eq(\ref{eq:Def G Two point function}) (see \cite{2018JHEPNarayan}), with the averaged matrix elements being given by 
\begin{equation}\begin{split}\label{eq:Def f in DSSYK result}
    f (E_1,E_2) 
    & =   \sum_k \tilde q^k\  {H_k(-{E_1\over E_0} | q) H_k(-{E_2\over E_0} | q) \over (q;q)_k} \\
    & = {(1-q)^{-2m-3} \over (q;q)_\infty^3 \Gamma_q(2m)} \ (e^{\pm i \theta_1\pm i \theta_2};q)_m |\Gamma_q({i (\theta_1+\theta_2)\over \lambda})\Gamma_q({i (\theta_1-\theta_2)\over \lambda})|^2,
\end{split}\end{equation}
where again $E_i = - {2 \cos \theta_i \over \sqrt{1-q}}$ in the second line above. One can check that to ${\cal O}(\tilde q)$, the  $f(E_1,E_2)$ has the same form as the one assumed in RMT model eq(\ref{eq:f for RMT}).  As with the partition function, a $q$-series expression can be derived for $G(t)$,
\begin{equation}\label{eq:G def as a series}
    G(t) =  {1 \over Z(\beta)}   \sum_{k=0}^\infty \tilde q^k Z_k(it) Z_k(\beta-it) .
\end{equation}  
Since $Z_k(\beta)$ has a $q$ series expansion eq(\ref{eq:Z_k as a sum}), this is a very explicit result. 
Below, we summarize additional useful properties of the DSSYK two-point function, beyond those discussed in section \ref{subsec:Description of the isolated quantum system}. 
\begin{itemize}
    \item \textbf{Identity operator case}: For $\tilde q = 1$, using the orthonormality property described in Appendix(\ref{Appendix: $Z_k$ details}), we obtain $G(t) = 1$.  This corresponds to the trivial case of the identity operator.

    \item \textbf{Short-time behavior of $G(t)$}: The two-point function has the following small time expansion:
    \begin{equation}\label{eq:G t expansion}
        G(t) = 1 - \frac{i t (1 - \tilde q)\, \partial_\beta Z(\beta)}{Z(\beta)} + {\cal O}(t^2).
    \end{equation}
    This follows from the properties of $Z_k$  in Appendix(\ref{Appendix: $Z_k$ details}). The result is also consistent with the normalization $G(t = 0)$ being $\beta$-independent, as was assumed in section \ref{sec:Heat Transport setup}.
\end{itemize}
The key result of this section is the two point function of the DSSYK model given in  eq(\ref{eq:G def as a series}). We conclude this subsection by highlighting a few special parameter choices where the DSSYK model reduces to simpler, analytically tractable models.

\subsubsection{Special Cases}\label{subsec: Special Cases of DSSYK}
The DSSYK model depends on the parameters: $q$ and $\tilde q$. For certain special choices of these parameters, the model simplifies significantly, reducing to much simpler models, as outlined below. 
\begin{itemize}
    \item \textbf{RMT Model:} For $q = \tilde q = 0$, the DSSYK model reduces to the Random Matrix Theory (RMT) model. To see this, note that in this limit, the density of states and matrix elements simplify (eqn(\ref{eq:Def Psi in DSSYK}) and eqn(\ref{eq:Def f in DSSYK result})) to 
    \begin{equation}\label{eq:small q, small tilde q DSSYK is RMT}
         \rho(E) = {\sqrt{4-E^2} \over 2\pi} ,\qquad f(E_1,E_2) = 1 ,
    \end{equation}
   matching the RMT case discussed in section \ref{subsec:RMT model example} with the identification $E_0 = 2$, $\Delta = 0$, and $f_0 = 1$. Consequently, all results for heat transport in this limit coincide with those of the RMT model.\footnote{More generally, for small $\tilde q$, the DSSYK model continues to agree with the RMT model at leading order in $\tilde q$. At ${\cal O}(\tilde q)$, this correspondence holds provided we identify the operator parameters as $\Delta ={4 \tilde q \over 1-q}  , f_0 =1 $.}
    \item \textbf{Gaussian Model}:  In the case of $q=1$ and finite $\tilde q$, the model reduces to Gaussian model. This reduction is not immediately evident from the original DSSYK model expressions but becomes clear in the chord diagram formulation. 
    
    Recall the definition of the Hamiltonian moments in eq(\ref{eq:Moments via Chord diagram}). When $q=1$, the moments $\text{Tr} \langle\langle H^{2k} \rangle\rangle$ simply count the number of Wick contractions among $2k$ nodes, for which the generating function is a Gaussian. More explicitly, we have
\[
 \langle \langle \mbox{Tr} H^{2k} \rangle \rangle \big|_{q=1} = (2k-1)!! = \int_{-\infty}^\infty dE \ {e^{-{E^2 \over 2}} \over \sqrt{2\pi} } E^{2k},
\]
This then gives the density of states to be
\begin{equation}\label{eq:q=1 density of states}
    \rho(E) ={1 \over J \sqrt{2\pi}} e^{-{E^2 \over 2J^2}} ,
\end{equation}
which is the same as the result in section \ref{subsec:Gaussian model example}. Note that we have reinstated the disorder coupling $J$ - previously set to 1 in eq(\ref{eq:disorder average J}). To compute the averaged matrix elements of operator $f(E_1,E_2)$, note that \cite{koekoek2010hypergeometric}
\begin{equation}
    \lim_{q \rightarrow 1, \mbox{ fixed } E}    {H_k(-{E\over E_0} | q) \over \sqrt{(q;q)_k}  } = {H_k(-{E\over J \sqrt 2}) \over \sqrt{2^n n!}}\ , \qquad E_0 = {2 J\over \sqrt{1-q}},
\end{equation}
where $H_k(x)$ is the Hermite polynomial. We can then obtain the function $f(E_1,E_2)$ given in eq(\ref{eq:Def f in DSSYK result}) to be\footnote{ We have used the Mehler's formulae to perform the sum 
\[ 
    \sum_{k=0}^\infty { \tilde q^k H_k(-{E_1 \over \sqrt 2})H_k(-{E_2 \over \sqrt 2})     \over 2^k k!} = {1 \over \sqrt{1-\tilde q^2}} e^{ {\tilde q \over 4(1-\tilde q^2)} ( (1-\tilde q) (E_1+E_2)^2 -  (1+\tilde q) (E_1-E_2)^2) }.
\]
}
\begin{equation}
    f(E_1,E_2) = {1 \over \sqrt{1-\tilde q^2}} e^{ {\tilde q \over 2(1-\tilde q^2)J^2} ( 2 E_1 E_2  -  \tilde q  (E_1^2+E_2^2)) }.
\end{equation}
This is identical to the expression in eq(\ref{eq:f in gaussian model}) with $\gamma = 1,f_0=1$.  Consequently, all results for heat transport in this limit coincide with those of the gaussian model.

\end{itemize}

\subsection{Regimes in the DSSYK model}\label{subsec:Regimes of DSSYK}

The two-point function of the DSSYK model, given in eq(\ref{eq:G def as a series}) and eq(\ref{eq:G tilde omega in DSSYK}), involves several parameters that make a general analysis nontrivial. However, by focusing on particular limits or parameter regimes, the model simplifies and as we will see reduces to well-understood physical systems. In this section, we explore such regimes, highlighting how they offer valuable physical intuition and how simple models can capture complicated dynamics of the DSSYK model.

Since this subsection involves a detailed exploration of a number of limits, we begin with a brief summary of the key regimes that arise in the DSSYK model. 
\begin{itemize}
\item As we will show in section \ref{subsec:DSSYK Long Time regime}, the low-temperature, long-time limit ($\beta \gg 1$) of DSSYK at finite $q$ coincides with the behavior of the Cold RMT models introduced in section \ref{subsec:Cold RMT model example}. We therefore refer to this as the {\it Cold RMT regime}. In this regime, the dynamics of DSSYK model is mostly governed by states near the ground state, where both the density of states and the matrix elements exhibit the same scaling as in random matrix theory. For instance, we will find that, $\rho(E) \propto \sqrt{E-E_0}$ which is the same as the RMT (or cold RMT) spectrum near the edges. 
\item As $q \rightarrow 1$, or equivalently 
$\lambda \rightarrow 0$, the Cold RMT regime continues to exist in the low-temperature limit $\beta\gg \lambda^{-{3\over 2}}$. However, a distinct new regime emerges at higher temperatures, 
$\beta \ll \lambda^{-{3\over 2}}$, which we term the {\it Conformal regime}, detailed in section \ref{subsec:DSSYK conformal regime}. More precisely the conformal regime is  $\lambda \rightarrow 0$ limit of the DSSYK model, with  $\beta  \sqrt \lambda$ held fixed.  In this regime, the density of states is approximately Gaussian in the angular variable $\theta$, and thermodynamics is dominated by states near a saddle-point value of $\theta$ determined by the combination $\beta\sqrt\lambda$. A schematic overview of these regimes is shown in figure(\ref{fig:cartoon beta regimes}). 
\end{itemize}
We now turn to a detailed discussion of how these distinct regimes emerge from the DSSYK model.
The results of this subsection is summarized in figure(\ref{fig:Cartoon relations the DSSYK model and other toy models}).

\subsubsection{Conformal Regime}\label{subsec:DSSYK conformal regime}
The $ \lambda \rightarrow 0 $ limit is the most well-studied regime of the DSSYK model \cite{Mukhametzhanov:2023tcg, Berkooz:2018jqr, 2018JHEPNarayan}, as this is where the low-energy dynamics of the SYK model are recovered. More precisely, the Schwarzian theory governing the infrared behavior of SYK emerges in the so-called \textit{triple-scaling limit}\cite{Cotler:2016fpe}, where  $\lambda \to 0$  with $ \beta \lambda^{3\over 2} $ held fixed. In this limit, both the partition function and the two-point function of the DSSYK model (with appropriately chosen $ m $) match those of the low-energy SYK model. In fact at all energies, the $\lambda\rightarrow 0$ regime is equivalent to the large-$ p $ limit of SYK with $ m = \frac{1}{p} \to 0 $ as shown in \cite{Goel:2023svz}. We will now consider a somewhat different limit, which can be loosely seen as the high temperature version  of the above limit where we take  $\lambda \rightarrow 0$, with $\beta \sqrt \lambda$ held fixed instead. In such a limit, we will soon see that the DSSYK model reduces to the conformal model that we discussed in section \ref{subsec:Conformal model example}.  

To see this, we first note that the density of states takes a simplified form in this limit (details in appendix(\ref{Appendix: Conformal Regime})),
\begin{equation}\label{eq:Small lambda simplification of d.o.s}
   \frac{ \Psi(\theta) }{ E_0 \sin \theta }   =   
   \begin{cases}
      \dfrac{2 \sqrt{2\pi}}{ \lambda } \, e^{-{\pi^2 \over 2 \lambda}} \, \theta  & \text{for } \theta \ll \lambda \\
      \sqrt{\dfrac{2}{\pi}} \, e^{- \frac{2}{\lambda}(\frac{\pi}{2}-\theta)^2} & \text{for } \theta \gg \lambda
   \end{cases}
\end{equation}
 In the low-energy limit $ \theta \ll \lambda $, the energy relation $ E + E_0 \approx  {\theta^2 \over \sqrt{\lambda}} $ implies $ \rho(E) \propto \sqrt{E + E_0} $, which matches the Cold RMT result given in eq(\ref{eq:rho cold RMT}) with $ \alpha = \frac{1}{2} $. At low temperatures $ \beta \gg \lambda^{-{3\over2}} $, it turns out that only such low-energy states contribute, and the model effectively reduces to the Cold RMT regime described in section \ref{subsec:Cold RMT model example} - see Appendix(\ref{Appendix: Conformal Regime}) for more details. However, new physics arises in the opposite limit $ \beta \ll \lambda^{-{3\over2}} $, where contributions from states with $ \theta \gg \lambda $ dominate and will be the main focus of this section. We do the detailed analysis in Appendix(\ref{Appendix: Conformal Regime}), and here we give only the results.

 The main simplification that occurs in the $ \beta \ll \lambda^{-{3\over2}} $ regime is that the partition function turns out to have a saddle at some $\theta$. We will find it more useful to state the results in terms of the variable $\phi \equiv {\pi \over 2}- \theta$. Due to the saddle point, the thermodynamics is entirely dominated by states near $\phi = \phi_0$ \cite{2018JHEPNarayan} where 
 \begin{equation}\label{eq:DSSYK Saddle eqn in phi0}
\begin{split}
    {4\phi_0  \over \beta E_0 \lambda} &= \cos\phi_0 .
    \end{split}
\end{equation}
Note that $\phi_0 \in (0,{\pi \over 2})$. It is now simple to state the result for the two point function in terms of $\phi_0$ which is \cite{Mukhametzhanov:2023tcg}
\begin{equation}\label{eq: Two point function phi conformal}
    \tilde G(\omega) = { 2(2\cos \phi_0)^{2m-1} \over E_0 \lambda  \Gamma(2m)} \ e^{\beta \omega \over 2} \Gamma(m + {i\omega \over \lambda E_0 \cos \phi_0 }) \Gamma(m - {i\omega \over \lambda E_0 \cos \phi_0 }).
\end{equation}
We show in Appendix(\ref{Appendix: Conformal Regime}), the above result agrees qualitatively well with the numerical results even when $\beta \sim \lambda^{-{3\over 2}}$. The two-point function given in eq(\ref{eq: Two point function phi conformal}),  exactly matches the conformal model result eq(\ref{eq: Conformal two point function in frequency}), upon choosing $f_0 = 1$ and hence we term this regime as conformal regime.

It is sometimes useful to split the conformal regime into high and low temperature conformal regimes where  it is possible to obtain analytic expressions for $\phi_0$. We have 
\begin{equation}\label{eq:Def of high and low temp conformal regime}
\begin{split}
      \phi_0 &= \begin{cases}
        {\pi\over 2}  (1-\frac{4}{\beta  E_0 \lambda } ) &\beta E_0 \lambda \gg 1, \mbox{ low temperature conformal regime}\\
        {\beta E_0 \lambda \over 4  } & \beta E_0 \lambda \ll 1, \mbox{ high temperature conformal regime}
    \end{cases}
\end{split}
\end{equation}

\paragraph{Spectral Function}
As an aside, we compute the spectral function $A(\omega)$ of the DSSYK model in the conformal regime via 
\[ A(\omega)=(1+e^{-\beta \omega})\tilde G(\omega). \] 
We get
\[
 A(\omega) ={  \pi \cosh({\beta\omega\over 2}) f_0 \beta (2\cos(\phi_0))^{2m}\over 2 \phi_0 \Gamma(2m)\sin(\pi m) \cosh({\pi \beta\omega\over 4\phi_0})   } \ \mbox{Re}\left[{\Gamma \left(m-\frac{i \beta  \omega }{4\phi_0 }\right)\over \Gamma \left(1-m-\frac{i \beta  \omega }{4\phi_0}\right)}\right].
\]   
In the low temperature conformal limit, $\phi_0 \rightarrow {\pi \over 2} - {2\pi\over\beta E_0 \lambda}$ the above result with $m={1 \over 4}$ matches exactly with the spectral function given for the SYK with four fermion interaction given in eq(8) of \cite{Larzul:2022kri}\footnote{In the DSSYK model, the parameter $m = {\tilde p \over p}$ is the number of fermions in the operator divided by the number of fermions in the Hamiltonian. For the $\mbox{SYK}_4$ Hamiltonian of reference \cite{Larzul:2022kri}, $p=4$ since the Hamiltonian is quartic in fermions and $\tilde p = 1$, since they compute the two point function of  a single fermion. Hence $m ={1 \over 4}$. We also have to take their $J^2 = {\lambda \over 4 \pi}$.}. Note that for more general $m$, the the temperature dependence is $A(\omega) \sim \beta^{1-2m}$.

\subsubsection{Cold RMT regime}\label{subsec:DSSYK Long Time regime}

\begin{figure}
    \centering
    \begin{minipage}[b]{0.45\linewidth}
        \centering
        \includegraphics[width=\linewidth]{ 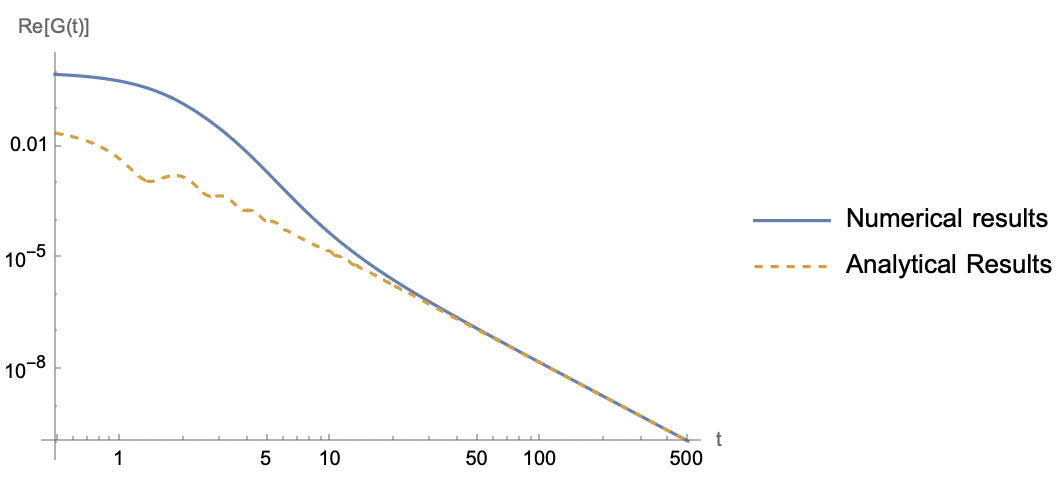}
        \caption{Re$[G(t)]$ vs $t$  for $q=\tilde q=0.5,\ \beta=1$}
        \label{fig:ReG Long Time}
    \end{minipage}
    \hspace{0.05\linewidth} 
    \begin{minipage}[b]{0.45\linewidth}
        \centering
        \includegraphics[width=\linewidth]{ 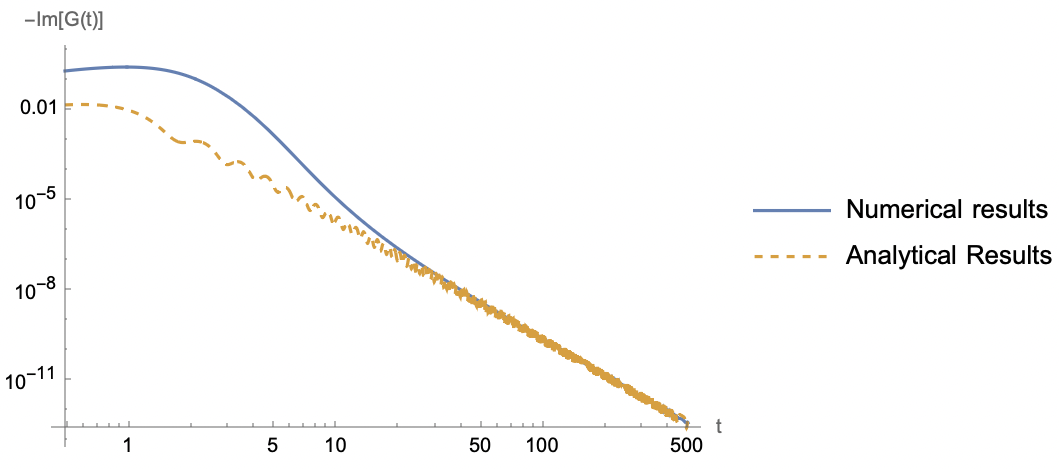}
        \caption{Im$[G(t)]$ vs $t$ a for $q=\tilde q=0.5,\ \beta=1$}
        \label{fig:ImGlongtime}
    \end{minipage}
\end{figure}

\begin{figure}
   \centering
  \includegraphics[width=0.7\linewidth]{ 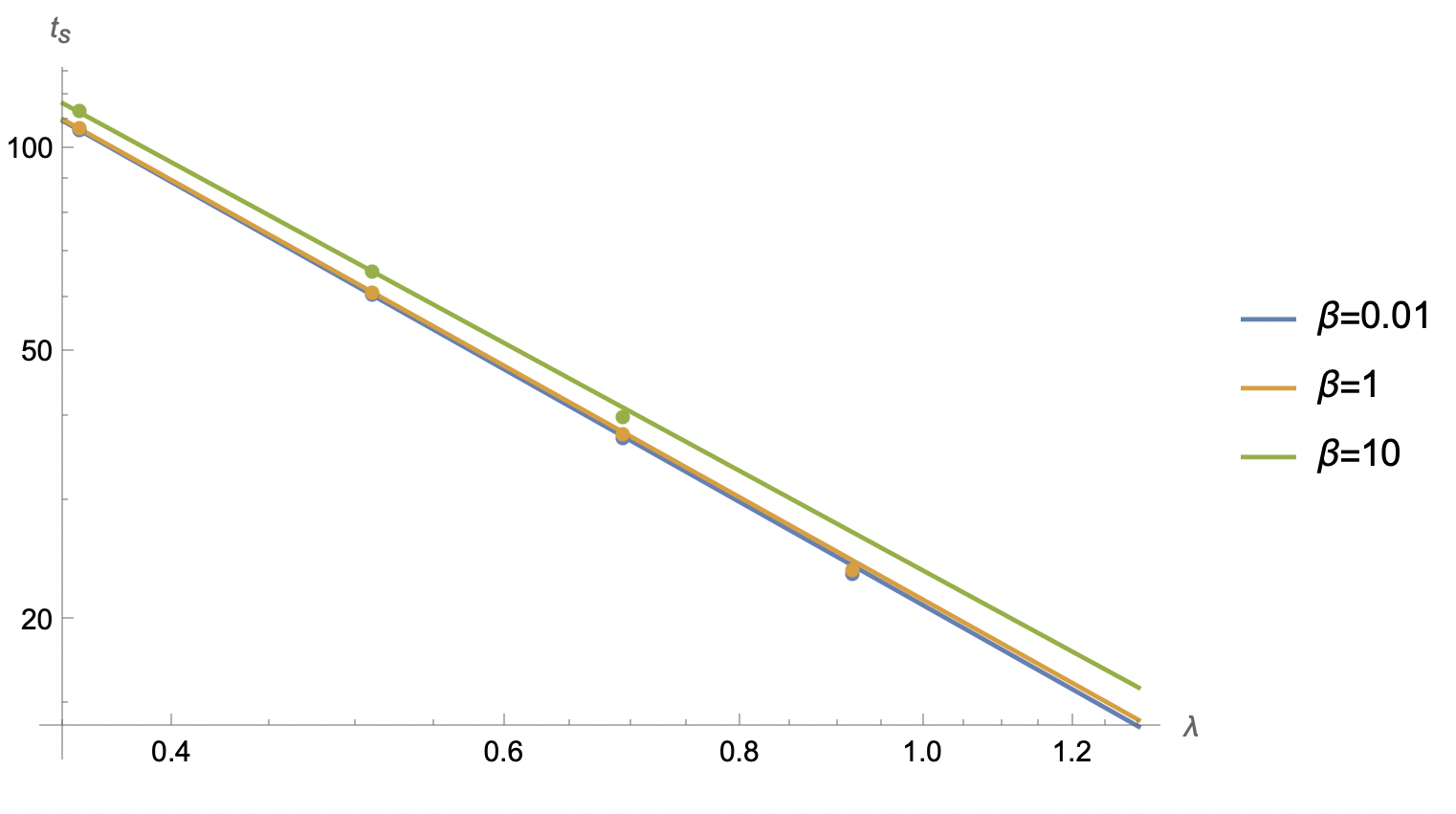}
  \caption{$\lambda$ vs $t_s$ for three $\beta$ values; $t_s\sim{20\over \lambda^{3\over 2}}$}
  \label{fig:lambdavsts}
 \end{figure}

At long times, we can obtain an explicit expression for the two-point function for the DSSYK model using the $q$-series expression given in eq(\ref{eq:G def as a series}). Since $Z_k$ itself is a $q$-series (see eq(\ref{eq:Z_k as a sum})) involving Bessel functions, we employ the asymptotic expansion of the Bessel function at large arguments (see Appendix \ref{sec:Asymptotics} for details) to extract the large-time behavior. The resulting expression for the two-point function is valid for finite $q$ in the regime $t \gg 1$,
\begin{equation}\begin{split}\label{eq:Long time G}
     G(t) & =
 { 4  (q;q)_\infty^6 (\tilde q^2;q)_\infty \over   \pi E_0^3  ( t(t+i \beta) )^{3\over 2}  Z(\beta) }    \left( {\cosh\left({E_0 \beta}\right) \over (\tilde q;q)^4_\infty}-{\sin\left({2E_0t+  i E_0 \beta}\right)  \over (-\tilde q;q)^4_\infty} \right).
   \end{split}
\end{equation}
It is interesting to note that the time dependence of $G(t)$ matches that of the RMT models, as given in eq(\ref{eq:Two point function RMT long time}). Before discussing the physical implications, we verify numerically the validity of the above result. This is illustrated in figures \ref{fig:ReG Long Time} and \ref{fig:ImGlongtime}, where we plot $\mathrm{Re}\, G(t)$ and $\mathrm{Im}\, G(t)$ as functions of $t$ for a specific set of parameters, and observe good agreement with eq(\ref{eq:Long time G}) at large times.
One can also extract a characteristic timescale, the \textit{scrambling time} $t_s$, which we define here as the time beyond which the two-point function is well approximated (within 10\% accuracy) by the expression in eq(\ref{eq:Long time G}).\footnote{Actually we use just the $\mathrm{Re}(G(t))$ to extract $t_s$, since the $t_s$ extracted from the $\mathrm{Im}(G(t))$ is fairly close}  To characterize the scale $t_s$ as a function of $\lambda$, we plot $t_s$ v/s $\lambda$ for a few different values of $\beta$ (fig(\ref{fig:lambdavsts})). It is observed that $t_s$ is well approximated by the relation $t_s \sim \frac{20}{\lambda^{3\over 2}}$ over a wide range of $\beta$, namely $0.01 \leq \beta \leq 10$.
In what follows, we show that the low temperature long time limit of DSSYK ($\beta,t \gg 1$) matches exactly to cold RMT model by comparing the DSSYK partition function and the two point function. 
The matching of partition function is as follows : In the low-temperature limit, the DSSYK model is sensitive only to states near the ground state at $E = -E_0$. From eq(\ref{eq:Def Psi in DSSYK}), we obtain 
\begin{equation*}
\lim_{E \rightarrow -E_0} \rho(E) = \frac{2 (q;q)_\infty^3}{\pi E_0^{2}} \sqrt{E_0^2 - E^2},
\end{equation*}
which matches the  density of states of the cold RMT model given in eq(\ref{eq:rho cold RMT}) with $\alpha = \frac{1}{2}$, upon appropriate choice of the normalization constant ${\cal N}_r$. Furthermore, the two-point function given in eq(\ref{eq:Long time G}), together with the already-matched partition function $Z(\beta)$, agrees exactly with the two-point function of the RMT model\footnote{We have already mentioned in section \ref{subsec:Cold RMT model example} that at low temperatures and long times, the two-point functions of the cold RMT and the RMT models coincide.} given in eq(\ref{eq:Two point function RMT long time}), provided the RMT parameters $f_0$ and $\Delta$ are appropriately identified.

In summary, the DSSYK model with parameters $(q, \tilde{q})$ in the cold RMT regime, namely $\beta, t \gg 1$, matches the cold RMT model described in section \ref{subsec:Cold RMT model example}, characterized by the parameters $f_0$, ${\cal N}_r$, $\Delta$, and $\alpha = \frac{1}{2}$, under the following identification\footnote{This result can also be obtained by matching the matrix elements. The function $f(E_1, E_2)$ in DSSYK, given in eq(\ref{eq:Def f in DSSYK result}), can be expanded near the left edge of the spectrum ($E \approx -E_0$) to quadratic order and compared with the RMT result in eq(\ref{eq:f for RMT}).}
\begin{equation}
    \label{eq:DSSYK long time effective f and Delta}
    \begin{split}
    {\cal N}_r & = (q,q)_\infty^3 ,\\
    {f_0} &=   {1 \over 2}(\tilde q^2;q)_\infty \left[\  (-\tilde q,q)_\infty^{-4} +  (\tilde q,q)_\infty^{-4} \ \right] ,\\
    \frac{1+\Delta}{1-\Delta}&=\frac{(-\tilde q;q)_{\infty }^4}{(\tilde q;q)_{\infty }^4}.
    \end{split}
\end{equation}
One can also explicitly verify that $1>\Delta>-1,f_0 >0$, as anticipated in section \ref{subsec:RMT model example}. The discussion above was for finite $q$. As is clear from the discussion below eq(\ref{eq:Small lambda simplification of d.o.s}), for $q\rightarrow 1$ or $\lambda \rightarrow 0$, the cold RMT regime emerges in the limit $t,\beta \gg \lambda^{-{3 \over 2}}$ with the same matching of parameters as in eq(\ref{eq:DSSYK long time effective f and Delta}).

\subsection{Transport in DSSYK Models}\label{subsec:Transport in DSSYK}
  
\begin{figure}
    \centering
    \includegraphics[width=0.5\linewidth]{ 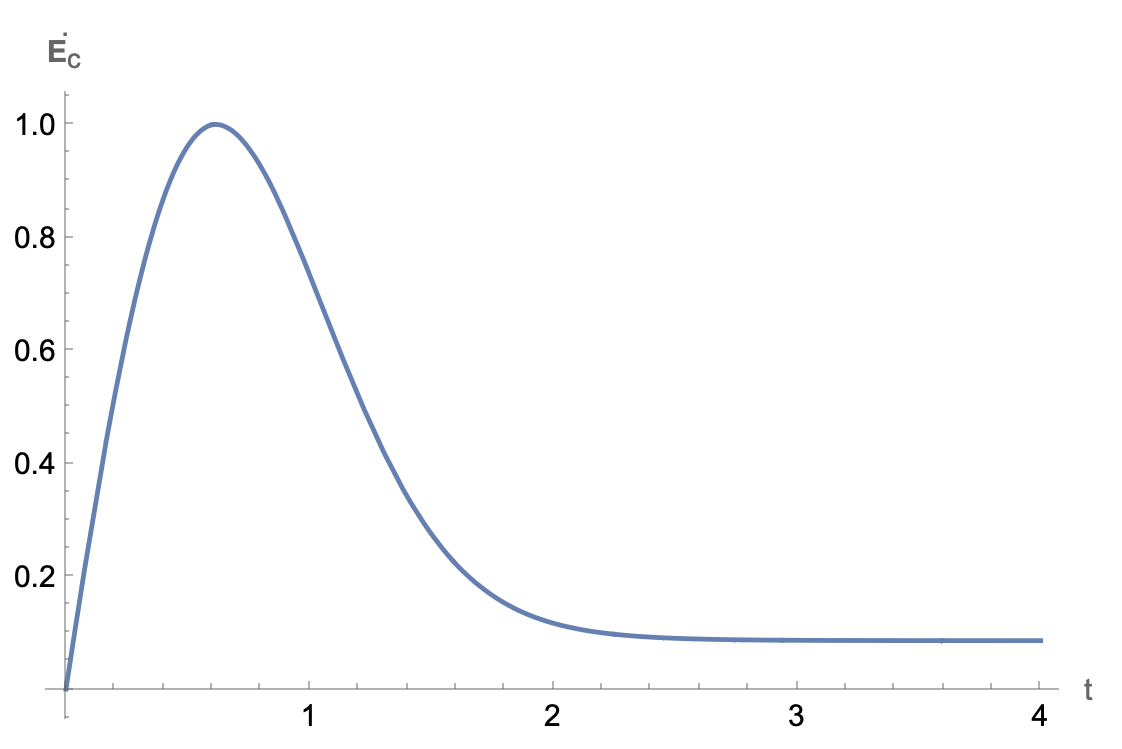}
    \caption{Heat Current vs time for a DSSYK system with $q=0.5, \beta=1$}
    \label{fig:DSSYKHeatcurrent}
\end{figure}

In this section we examine the heat transport in the DSSYK model. As discussed in the introduction the typical heat current plot  given in fig(\ref{fig: typical heat current}) has an initial increase to a peak value and eventually settles down to constant value. The DSSYK model heat current exhibits these  features as expected - the plot of heat current for some typical values of DSSYK parameters are given in fig(\ref{fig:DSSYKHeatcurrent}) which clearly shows the initial increase and eventual NESS regime. For  the case of DSSYK,  the initial increase can be worked out analytically and it turns out to be linear in time. Using the small $t$ expansion of two point function given in eq(\ref{eq:G t expansion}), the heat current at small time is obtained to be
    \begin{equation}\label{eq:Heat current early time DSSYK}
        \dot E_c(t) ={ 2(1-\tilde q)\epsilon^2  \partial_\beta Z(\beta) \over Z(\beta)}\ t  + {\cal O}(t^2).
    \end{equation}
Notice that the RHS above is proportional to the internal energy of the DSSYK system. This
means that, for small times, the fractional change in energy of the DSSYK system is independent of $\beta$.
The eventual saturation in heat current happens since for sufficiently long time, the DSSYK model reduces to an RMT model as discussed in section \ref{subsec:DSSYK Long Time regime}. 

As we saw in the previous section, in specific (not necessarily distinct) limits the  DSSYK model reduces to the simple models introduced in sections \ref{subsec:RMT model example}-\ref{subsec:Gaussian model example} as summarized in figure(\ref{fig:Cartoon relations the DSSYK model and other toy models}). We expect that the heat current for the DSSYK model in those specific limits will be captured well by the  appropriate toy models. We first present the analytical results obtained for the $\lambda\rightarrow 0$ case for conductivities at different temperature limits. In section \ref{subsec: DSSYK numerics}, we give the typical plots of heat currents against RMT and Conformal model results and  provide numerical evidence showing that the behavior away from the \textit{RMT} and \textit{Conformal} regimes smoothly interpolates between these analytical regimes. 

\subsubsection{DSSYK Model in the conformal regime \texorpdfstring{$\lambda \rightarrow 0$}{}}
\begin{figure}
    \centering
        \includegraphics[width=0.65\linewidth]{ 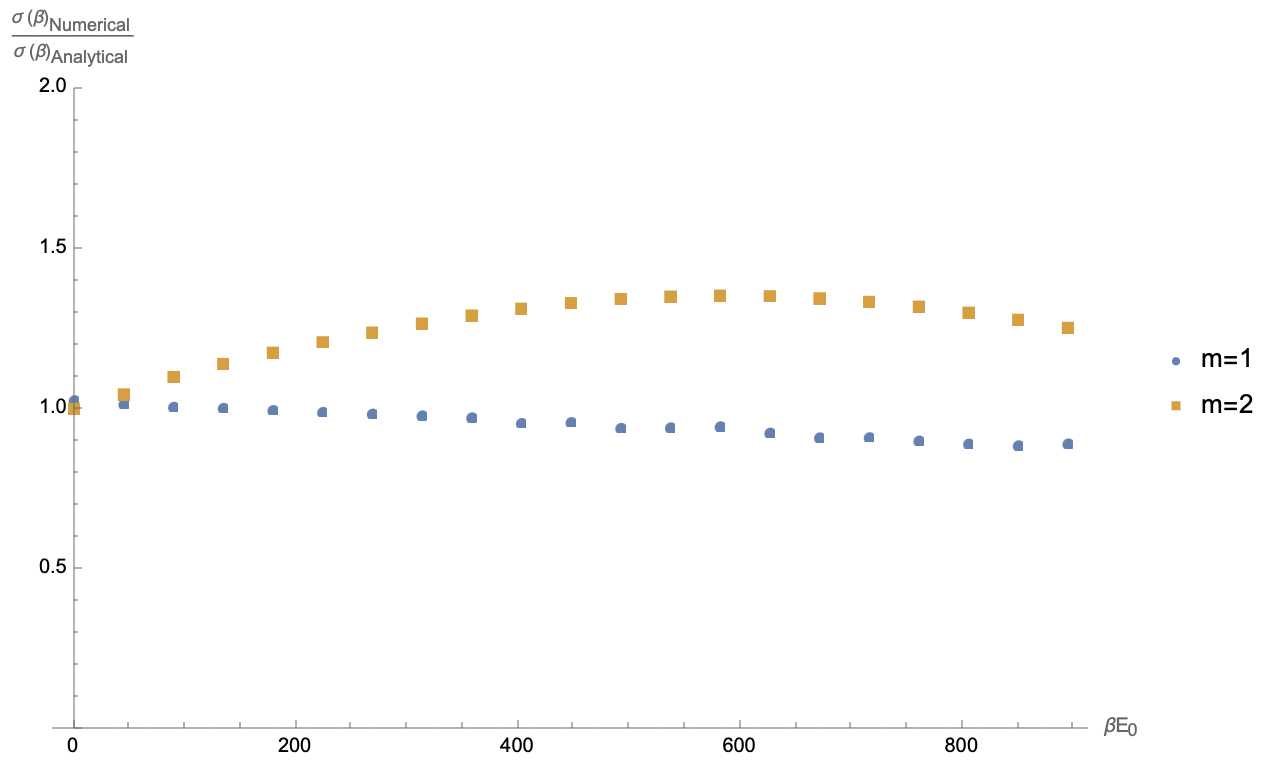}
    
    \caption{Ratio of numerical values of $\sigma(\beta)$ against analytical result for $\lambda=0.05$ and different values of $m$.}
    \label{fig:ratio}
\end{figure}

In the conformal regime, i.e $\lambda \rightarrow 0, \beta \ll \lambda^{-{3 \over 2}}$, as discussed in section \ref{subsec:DSSYK conformal regime}, the two point function agrees with the conformal toy model \ref{subsec:Conformal model example}. Hence the conductivity in this regime can be read off from the results of the conformal model given in eq(\ref{eq:Conductivity conformal model}) with appropriate $\phi_0$. More explicitly, we have the conductivity of the DSSYK model in the conformal regime 
\begin{equation}
    \sigma   = c_m\ \epsilon ^2 f_0^2 \phi_0 \beta \cos ^{4 m}(\phi_0), \hspace{2 cm}  {4\phi_0  \over \beta E_0 \lambda} = \cos\phi_0,
\end{equation}
where $c_m \equiv 2\sqrt\pi m^2 {\Gamma(2m)\over \Gamma(2m+{3\over 2})}$.
We check numerically in figure(\ref{fig:ratio}) that the conductivity for the DSSYK model agrees with the above formulae very well.
Note that even outside the conformal regime, say around $\beta E_0 \sim {1 \over \lambda^2}$ as we see in figure(\ref{fig:ratio}) the  conformal regime results are only a few percentages off from the actual result at least for $m=1$ case.  We end with some explicit results for the conductivity $\sigma$ of the DSSYK model in the low and high temperature conformal regime,
\begin{itemize}
    \item \textbf{Low temperature conformal regime ${1 \over \lambda } \gg \beta \sqrt \lambda \gg 1$ }:
    \begin{equation}
        \sigma \approx  {1\over 2}\epsilon^2 f_0^2 c_m \pi^{1+4m} \lambda^{-2m} \beta^{1-4m} 
    \end{equation}
    \item \textbf{High temperature conformal regime $1 \gg \beta \sqrt \lambda$}  
    \begin{equation}
        \sigma \approx   {\epsilon^2\over 2} f_0^2  c_m  \sqrt \lambda \beta^2 
    \end{equation}   
\end{itemize}

\paragraph{Comparison to results in literature}
In this subsection we obtained results about heat transport between two DSSYK sytems in the conformal regime. As mentioned before, the low temperature conformal regime of DSSYK model with $m={1 \over p}$ matches the low temperature SYK$_p$ models. Hence the results for transport in the DSSYK model in this regime must match with that of the low temperature SYK$\!_{1 \over m}$ model. 

In the work \cite{Larzul:2022kri}, the authors couple two SYK$_4$ systems via a free fermion channel (or equivalently a SYK$_2$ system). While this setup is not exactly the same as ours, we can easily construct a setup similar to theirs by connecting two of our setups. Then their conductivity can be extracted just from the conductivity between the SYK$_4$ system and the channel SYK$_2$ system. In eq(\ref{conductivity conformal mc mh}), we already have worked out the conductivity between two different conformal systems - upon  using $m_c={1 \over 4}, m_h={1 \over 2}$ the conductivity is obtained to be
\begin{equation}
    \sigma \sim \beta^{-{1\over 2}} = \sqrt T,
\end{equation}
which agrees with the weak coupling results of \cite{Larzul:2022kri}. Additionally, when the coupling between the $\mathrm{SYK}_4$ and $\mathrm{SYK}_2$ systems is very strong, the authors of \cite{Larzul:2022kri} find that the conductivity behaves as $\sigma \sim T$. It is interesting to note that this result is also reproduced by the weak coupling calculation discussed above, provided we replace $m_c = \frac{1}{4} \rightarrow \frac{1}{2}$. We leave a detailed exploration of this apparent connection (if any) to future work.


\subsubsection{Numerical results for heat transport in DSSYK for  \texorpdfstring{$t < (1-q)^{-\frac{3}{2}}$ and finite $\lambda$}{tq}}\label{subsec: DSSYK numerics}

In this section, we focus on the large-time behavior of systems with finite $q$ values, specifically analyzing the quantity $\dot E_-(t)$, which was defined in eq(\ref{eq:Def Heat current}). As discussed in the introduction, at late times, $\dot E_-(t)$ becomes the dominant contribution to the heat current, making it the relevant observable in this regime. The intermediate $q$ does not quite reduce to any of our toy models. However, what we see numerically is that there is still an interpolation \emph{in time} between the conformal and RMT regimes.
This suggests that, in some sense, the intermediate $q$ DSSYK can be thought of as RG running from the conformal model at high temperatures/small times to a RMT-like model at low temperatures/large times.

\begin{figure}
    \centering
    \subfloat[$\beta=1$]{%
        \includegraphics[width=0.45\linewidth]{ 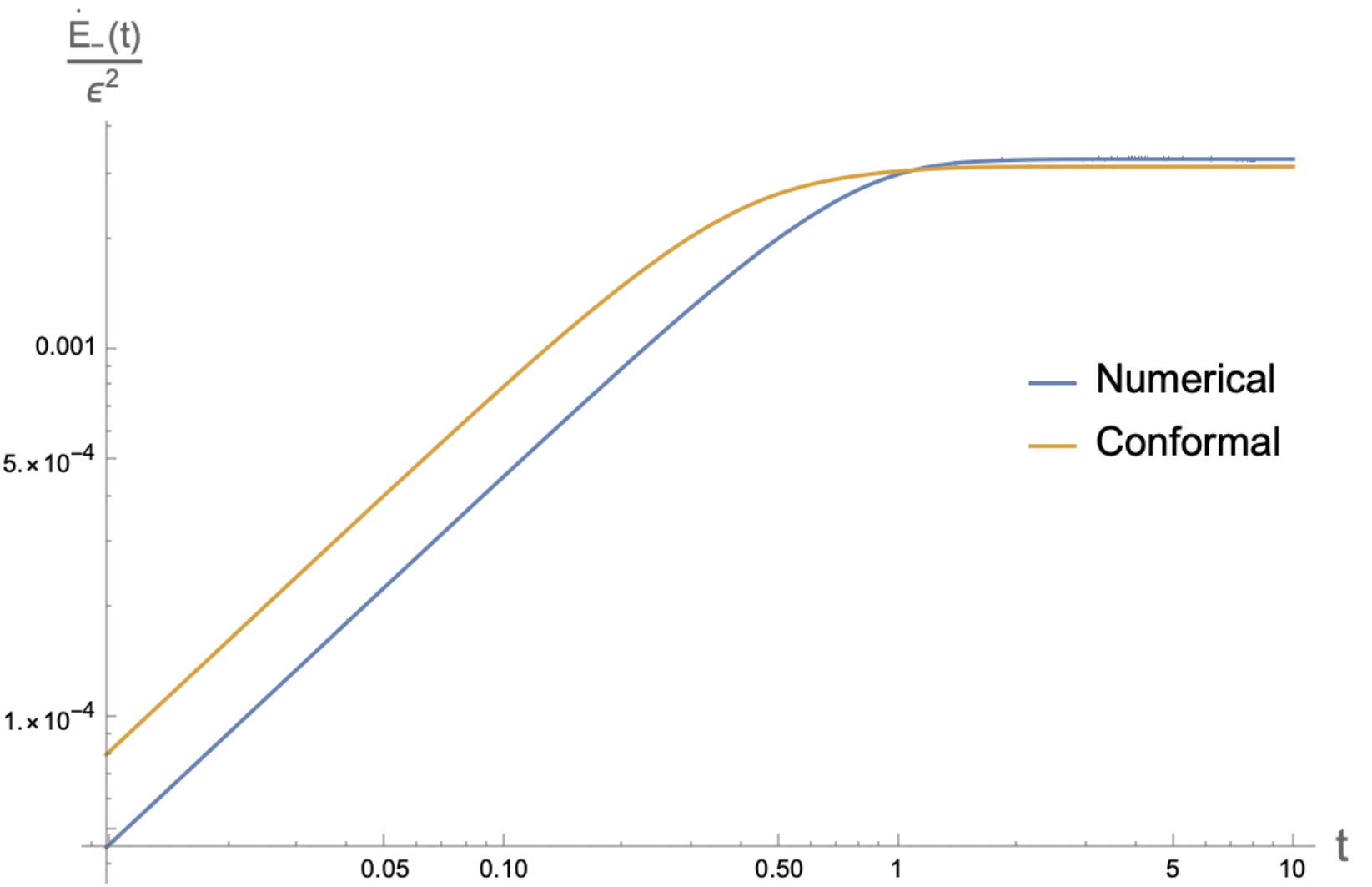}
        \label{fig:Ecdiff 1}
    }
    \hspace{0.015\linewidth} 
    \subfloat[$\beta=100$]{%
        \includegraphics[width=0.45\linewidth]{ 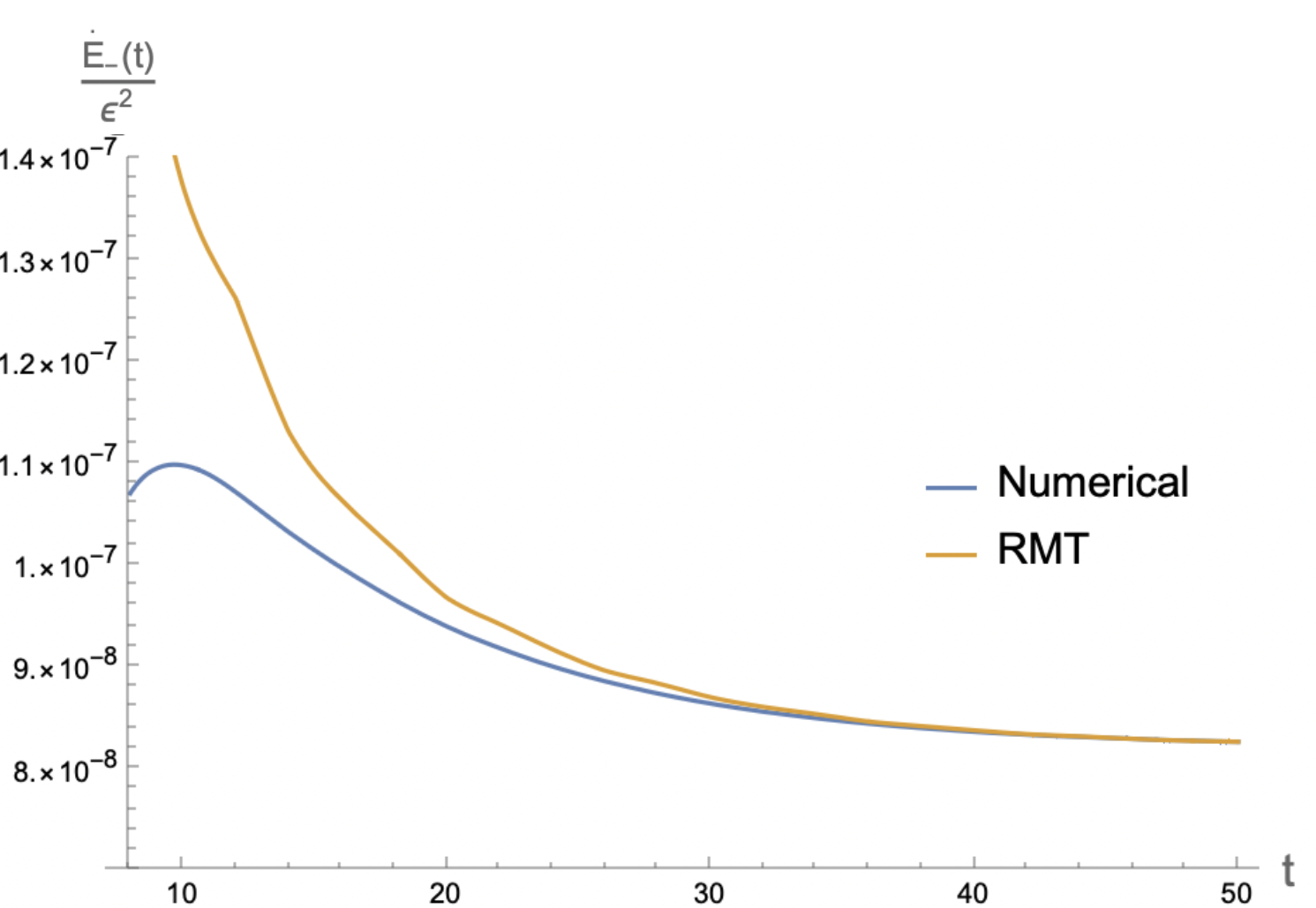}
        \label{fig:Ecdiff 100}
    }
    \caption{$\dot E_-(t)$ results for $q=0.4$ and different values of $\beta$ against corresponding results for conformal model and RMT results.  We have normalized the RMT results so that it agrees with the DSSYK results at $t=50$.}
    \label{fig:Ecdiff}
\end{figure}

To explore this, we compare numerical results for 
$\dot E_-(t)$ with that of the RMT and conformal toy models at two different inverse temperatures, $\beta=1$ and $\beta=100$ in figure(\ref{fig:Ecdiff}).
\begin{itemize}
    \item For $\beta=1$, it is seen that the system reaches non equilibrium steady state (NESS) fairly quickly around $t>1$ and the NESS regime heat current turns out to match that of the  conformal toy model as shown in the figure(\ref{fig:Ecdiff 1}). In computing the heat current for the conformal case in the figure(\ref{fig:Ecdiff 1}), we use the conformal two point function given in eq(\ref{eq:G for conformal model preserving KMS}) with $\phi_0$ found by solving eq(\ref{eq:DSSYK Saddle eqn in phi0}). 
    \item  At lower temperatures say $\beta =100$,  around time $t\gtrsim 50$, 
    does not exactly match RMT. But, we see that, with an appropriate normalization, RMT does get the fall off right - see figure(\ref{fig:Ecdiff 100}). In computing the heat current for the RMT case in the figure(\ref{fig:Ecdiff 100}), we use the two point function given in eq(\ref{eq:Two point function RMT long time}) with parameters as given in eq(\ref{eq:DSSYK long time effective f and Delta}).
\end{itemize}  We thus observe numerically that  the crossover between RMT-like and conformal like behavior figure(\ref{fig:cartoon beta regimes}), previously observed in the $q\rightarrow 1$ regime, persists even at finite $q$ values. 

\begin{figure}
    \centering
    \subfloat[$m=1$]{%
        \includegraphics[width=0.6\linewidth]{ 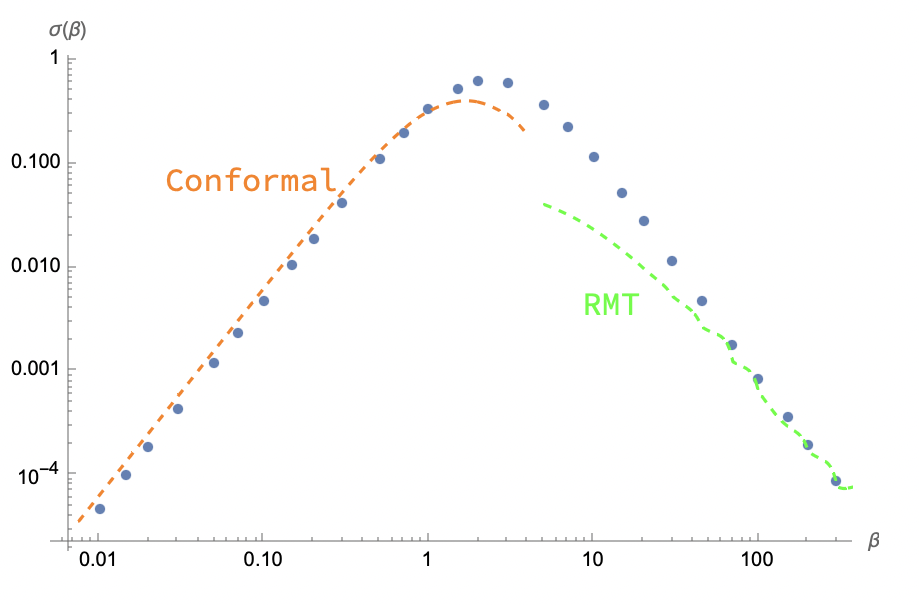}
        \label{fig:m1}
    }
    \hspace{0.015\linewidth} 
    \subfloat[$m=2$]{%
        \includegraphics[width=0.62
    \linewidth]{ 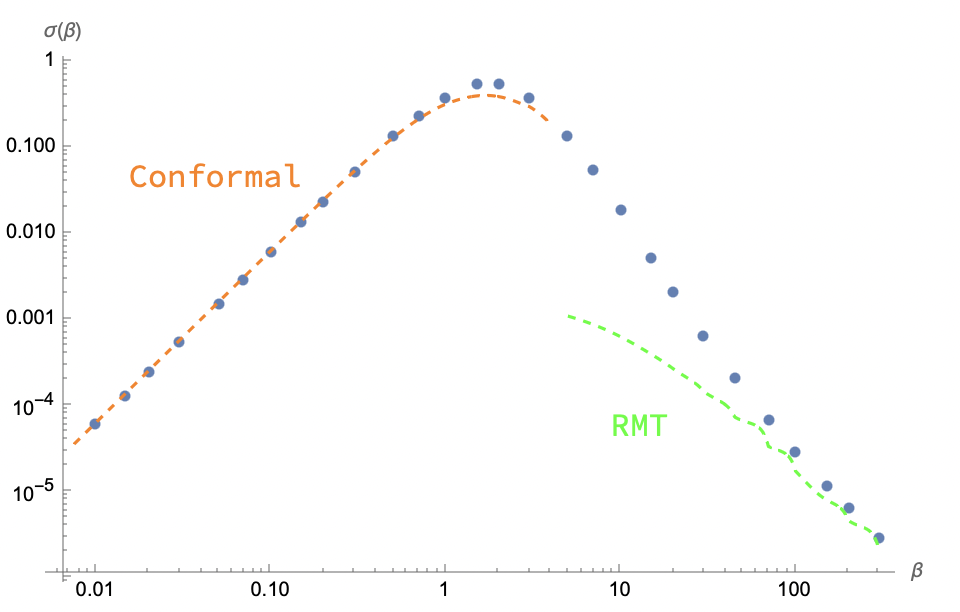}
        \label{fig:m2}
    }
    \caption{Numerical values of $\sigma(\beta)$ vs Conformal model and Cold RMT results for $q=0.4$ and $m=1,m=2$.}
    \label{fig:rationew}
\end{figure}

\paragraph{Finite $q$ conductivity:}
In the above discussion, we observed that the NESS heat current of the DSSYK model matches the results of the conformal and RMT toy models in the high- and low-temperature limits, respectively. This behavior is also reflected in the thermal conductivity. In figure(\ref{fig:rationew}), we plot the thermal conductivity $\sigma$ as a function of temperature for the DSSYK model, RMT toy model, and conformal toy model for a few different values of $m$. As expected, we find good agreement between the DSSYK model and the conformal model at high temperatures ($\beta < \beta_d$ with $\beta_d \sim 1$), and with the RMT toy model at low temperatures ($\beta > \beta_s$ with $\beta_s \sim 70$).

\begin{figure}
    \centering
    \subfloat[]{%
        \includegraphics[width=0.46\linewidth]{ 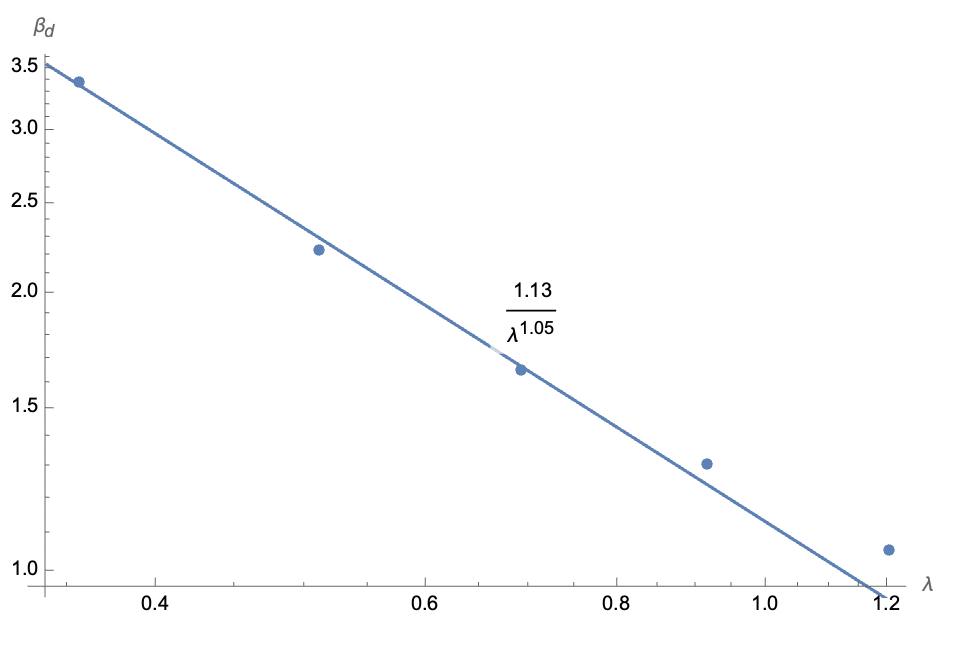}
        \label{fig: betad}
    }
    \hspace{0.015\linewidth} 
    \subfloat[]{%
        \includegraphics[width=0.46\linewidth]{ 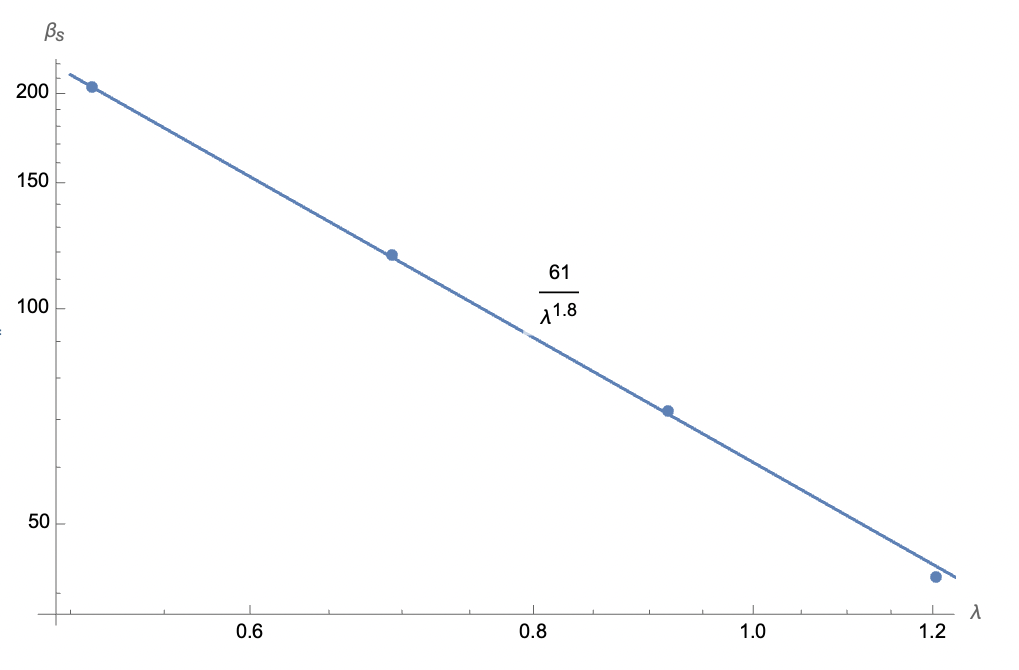}
        \label{fig: betas}
    }
    \caption{$\beta_s$ and $\beta_d$ plotted against corresponding values of $\lambda$.}
    \label{fig: betas and betad}
\end{figure}

In fact, we can make this observation more quantitative by defining $\beta_d^{-1}$ as the temperature above which the DSSYK conductivity deviates from the RMT result by more than $50\%$, and $\beta_s^{-1}$ as the temperature below which the DSSYK conductivity deviates from the conformal result by more than $20\%$. In figure(\ref{fig: betas and betad}), we plot $\beta_s$ and $\beta_d$ as functions of $\lambda$. At finite $\lambda$, the transition between conformal vs RMT regimes is not as sharp in temperature as suggested by fig(\ref{fig:cartoon beta regimes}). Numerics suggests an intermediate region between the temperatures $\beta_s^{-1}$ and $\beta_d^{-1}$, that separates a low temperature RMT-like behavior from a high temperature conformal-like behavior. 


\section{Conclusions and Outlook}\label{sec:Conlusions and Outlook}

In this work, we have analyzed the heat transport  between two quantum systems using a set of exactly solvable models. We demonstrated that even relatively simple toy models - motivated by random matrix theory and conformal symmetry - are sufficient to capture a number of essential features of thermal transport, including the approach to non-equilibrium steady state. We also worked out the transport in the case of solvable but nontrivial case of DSSYK models. We showed that the behavior of heat transport in the DSSYK systems interpolate smoothly between the conformal and the RMT toy model results models with the transition temperature $\sim \lambda^{-{3\over 2}}$. The behavior of the DSSYK model appears suggestive of an RG flow between conformal and RMT regimes.

We end with some comments on interesting future directions. An obvious question is to investigate higher order corrections in coupling strength $\epsilon$ between the systems. At higher order, the heat current will depend on higher-point correlation functions. In the context of DSSYK model, this question is accessible - although more complicated than the two point function results and may reveal new regimes of behavior. It would be interesting to see if there is a new scaling regime as happens in the SYK case \cite{Larzul:2022kri}.

Another interesting direction is to investigate transport in other solvable versions of the DSSYK model \cite{Narayan:2023wlk,Berkooz2020ComplexSM,Berkooz:2020xne} with more symmetries, where an even richer behavior is expected. Apart from introducing extra parameters, one could also study charge transport rather than just the heat transport. A different kind of extension is to consider a chain made of the kind of toy models we consider here, as well as a DSSYK chain.\footnote{See \cite{Gu_2017,bucca2024} for related construction.} This would allow us to see how our statements extend to higher dimensional heat transport.

In this work, we have refrained from commenting on the possible holographic interpretation of our results. There are many existing proposals for the gravitational duals for DSSYK\cite{Blommaert:2024ydx,Susskind:2022bia}. It would be interesting to give a gravitational picture of various regimes we see in this work.

We conclude with the hope that the exact results obtained here  will lead to deeper insights into quantum transport.

\acknowledgments

We would like to acknowledge discussions with Subhro Bhattacharjee, Abhishek Dhar, Manas Kulkarni, Gautam Mandal, Shiraz Minwalla, Sandip Trivedi and Spenta Wadia.
This work was presented at the GautamFest Conference at TIFR Mumbai in January 2025, and at the ICTS-TIFR string seminar in March 2025. RL thanks
the organizers of both for giving him the opportunity to present this work.
RL acknowledges the support of the Department of Atomic Energy,
 Government of India, under project no. RTI4001, and would also like to acknowledge his debt to the
 people of India for their steady and generous support of research in the basic sciences. PN would like to acknowledge SERB grant MTR/2021/000145.

\appendix

\section{Expression for energy current }\label{sec: Appendix energy transport}
We compute the expectation value of $E_c$ corresponding to the coupled quantum system discussed in section \ref{sec:Heat Transport setup}. Alternately, we could have  started with a finite temperature state of the two systems (same on both sides) and turn on the interaction Hamiltonian (at time $t=0$ say) and ask about the quenching of the system. More precisely we will be interested in the object,\footnote{For SYK the equation involves disorder averaging, hence $E_c(t) \equiv {1 \over Z } \langle \langle Tr \left[ H_c(t) \rho_0 \right] \rangle \rangle$}
\begin{equation}
   E_c(t) \equiv {1 \over Z }  \mbox{Tr} \left[ H_c(t) \rho_0 \right] , \qquad \rho_0 \equiv e^{-\beta_c H_c} e^{-\beta_h H_h} ,  \qquad Z(\beta) =   \mbox{Tr} \left[  \rho_0 \right]\ .
\end{equation}
We compute the above quantity to $\mathcal{O}(\epsilon^2)$. To evaluate this, we can go to the interaction picture by defining ${\cal H} = H_0 + H_I$ in the Hamiltonian eq(\ref{eq:Setup of the model}) with $H= H_c + H_h$ and $H_{I} = \epsilon O_c O_h \theta(t)$. Denoting the interaction picture operators and states by the superscript $\ ^{(0)}$, for instance $H_I^{(0)}(t) = e^{i H_0 t} H_I e^{-i H_0 t}$ or $\rho_0^{(0)}(t) = U(t) \rho_0 U^\dagger(t)$ where $U(t) \equiv  e^{i H_0 t} e^{-i H t }$ is the interaction picture evolution operator. The energy expectation value is thus given by,
\begin{equation}
  Z  E_c(t) =  \mbox{Tr}\left[  H_c^{(0)}(t)  \rho_0^{(0)}(t)  \right]  =  Tr\left[ U^\dagger(t)  H_c U(t) \rho_0  \right] 
\end{equation}
We now expand $U(t)$ perturbatively in $\epsilon$. We thus have the following result where ,
\begin{equation}\label{energy expectation value}
    \begin{split}
       \dot E_c(t) & = {1\over Z}(-i\epsilon)^2\ \int_0^{t_1} dt_2\   Tr\left[   [ H_I^{(0)}(t_1), [ H_I^{(0)}(t_2) ,  H_c]]\  \rho_0  \right] 
    \end{split}
\end{equation}
Now we factorize the Hilbert space\footnote{ \label{footnote reference} For the SYK systems considered in this paper, the intermediate steps are slightly different since we need to perform disorder averages, but the final result eq(\ref{general Ex}) is the same. To see this, note that the heat current in the SYK case can be written as,
\[\dot E_c(t)=- \epsilon^2 {i^{p_{\tilde c} + p_{\tilde h}}\over Z_c Z_h} \int_0^{t_1} dt_2\  \langle \langle\ \mbox{Tr}\left(  [ J_{\tilde I,\tilde A} X_{\tilde I}^{(0)}(t_1)Y_{\tilde A}^{(0)}(t_1)  , [ J_{\tilde J,\tilde B} X_{\tilde J}^{(0)}(t_2)Y_{\tilde B}^{(0)}(t_2) , H_c] ]\ \rho_0 \right)   \rangle \rangle\]
where $\langle\langle \cdots \rangle\rangle$ denotes the disorder average. We can further use the factorization property of the disorder average, $\langle \langle  J_{\tilde I, \tilde A} J_{\tilde J,\tilde B}  \rangle \rangle = \langle \langle J_{\tilde I} J_{\tilde J} \rangle \rangle  \langle \langle J_{\tilde A} J_{\tilde B} \rangle \rangle$ to simplify the expression leading the second line of  eq(\ref{general Ex}).   },
\begin{equation}\label{general Ex}
    \begin{split}
       \dot  E_c(t) & = -{\epsilon^2 \over Z}    \ \int_0^{t_1} dt_2\ \bigg(  Tr\left[ O_c(t_1) [O_c(t_2),  H_c] e^{-\beta_c H_c} \right]    Tr\left[ O_h(t_1) O_h(t_2) e^{-\beta_h H_h} \right]  \\
         & \hspace{20mm} - Tr\left[  [O_c(t_2),  H_c]O_c(t_1) e^{-\beta_c H_c} \right]    Tr\left[ O_h(t_2) O_h(t_1) e^{-\beta_h H_h} \right] \bigg) \\
&= - \epsilon^2 \  \int_0^{t_1} dt_2\  \bigg( i\left(\partial_{t_2}  G_c(t_1-t_2)\right) G_h(t_1-t_2) -i\left(\partial_{t_2} G_c(t_2-t_1) \right) G_h(t_2-t_1) \bigg)\\
& = - 2 \epsilon^2  \int_0^{t_1} d\tilde t\  \mbox{ Im} \left(\partial_{\tilde t}   G_c(\tilde t) \ G_h (\tilde t) \right)  \\
& = - \epsilon^2 \ \mbox{Im} \  \left(  G_c(t_1) G_h(t_1)    +  \int_0^{t_1} d\tilde t \left[  \partial_{\tilde t}   G_c(\tilde t)  G_h(\tilde t)-     G_c(\tilde t) \partial_{\tilde t} G_h(\tilde t) \right]  \right)
\end{split}
\end{equation}
In going to the third line, we have used $G^*(-t) = G(t)$. Since $G(t=0)=1$ is real, we have also dropped $G_c(0) G_h(0)$  term in the last expression. We thus get eq(\ref{eq:Def Heat current}).

 \section{Properties of \texorpdfstring{$Z_k$}{Zk}}\label{Appendix: $Z_k$ details}
In this appendix, we elaborate on the properties of the quantities $Z_k(\beta)$, defined in eq(\ref{eq:Z_k as a integral}), which play an important role in the analysis of the DSSYK model. We begin by listing the key properties of $Z_k$ , followed by a detailed discussion of their asymptotic behavior, which ultimately leads to the result in eq(\ref{eq:Long time G}). A particularly useful representation of $Z_k(\beta)$ is given by the following $q$-series expansion, analogous to that of the partition function:
\begin{eqnarray}\label{eq:Z_k as a sum}
    Z_k(\beta) = \frac{2}{\beta E_0} \sum_{p=0}^\infty \frac{(-1)^p q^{p + \binom{p}{2}} (k + 2p + 1) (q;q)_{k+p}}{(q;q)_p \sqrt{(q;q)_k}} I_{k + 2p + 1}(\beta E_0)
\end{eqnarray}
The quantity $Z_k(\beta)$ has many useful structural properties which will be used throughout the main analysis ,

\begin{itemize}
    \item \textbf{Orthonormality:} 
    \[
        \sum_{k=0}^\infty Z_k(\beta_1) Z_k(\beta_2) = Z(\beta_1+\beta_2).
    \]
    This follows by inserting \eqref{eq:Z_k as a integral} for both $Z_k$’s, using the completeness relation
    \[
        \sum_{k=0}^\infty \frac{H_k(\cos\theta_1\,|\,q) H_k(\cos\theta_2\,|\,q)}{(q;q)_k}
        = \frac{\delta(\theta_1 - \theta_2)}{\Psi(\theta_1)},
    \]
    which collapses one integration and yields $Z(\beta_1+\beta_2)$.

    \item \textbf{Recursion relation:} 
    \[
        \sqrt{1-q} \,\partial_\beta Z_k(\beta)
        = \sqrt{1 - q^{k+1}} \,Z_{k+1}(\beta) + \sqrt{1 - q^{k}} \,Z_{k-1}(\beta).
    \]
    This is obtained by differentiating \eqref{eq:Z_k as a integral} with respect to $\beta$ and using the recurrence
    \[
        2\cos\theta\, H_k(\cos\theta) 
        = \sqrt{1 - q^{k+1}}\, H_{k+1} + \sqrt{1 - q^{k}}\, H_{k-1},
    \]
    then recognising the resulting integrals as $Z_{k\pm 1}(\beta)$.

    \item \textbf{$\beta=0$ case:} 
    \[
        Z_k(0) = \delta_{k,0}.
    \]
    Setting $\beta=0$ makes the exponential unity; orthogonality of $H_k$ then gives $\delta_{k,0}$.

    \item \textbf{$k=0$ case:} 
    \[
        Z_0(\beta) = Z(\beta).
    \]
    This follows directly from $H_0=1$ and $(q;q)_0=1$ in \eqref{eq:Z_k as a integral}.
\end{itemize}

\subsection{Long-Time Limit of the Two-Point Function  \texorpdfstring{$Z_k(\beta)$}{Zkbeta}}\label{sec:Asymptotics}
In this subsection, we determine the large-$t$ behavior of $Z_k(it)$ and $Z_k(\beta - it)$, since the two-point function is expressed in terms of these in eq(\ref{eq:G def as a series}). To study the large-$t$ limit, we note that $Z_k(\beta)$ admits a $q$-series representation given in eq(\ref{eq:Z_k as a sum}), which involves modified Bessel functions of the first kind. This structure allows us to use the known asymptotic forms of the Bessel functions to evaluate $Z_k(\beta)$ in the appropriate limit. Consequently, we obtain the long-time behavior of the two-point function, as summarized in eq(\ref{eq:Long time G}). The relevant asymptotics of the Bessel function $I_\nu(z)$ for integer $\nu$ is as follows (see Section 8.451 of \cite{GradshteynRyzhik}),

\begin{itemize}
    \item For $\tau$ away from the positive imaginary axis 
\begin{equation}\label{eq: Leading Bessel away from +ve i axis}
      I_\nu(\tau) \approx {e^{\tau} - i (-1)^\nu e^{-\tau}  \over \sqrt{2 \pi \tau}} ,\hspace{20mm} |\tau| \rightarrow \infty,  -{ 3\pi \over 2}< \arg \tau < {\pi \over 2}
\end{equation}
\item For $\tau$ away from negative imaginary axis we have,
\begin{equation}\label{eq: Leading Bessel away from -ve i axis}
I_\nu(\tau) \approx {e^{\tau} + i (-1)^\nu e^{-\tau}  \over \sqrt{2 \pi \tau}} ,\hspace{20mm} |\tau| \rightarrow \infty,  -{ \pi \over 2}< \arg z < {3\pi \over 2}
\end{equation}
\begin{itemize}
    \item As a special case of eq(\ref{eq: Leading Bessel away from -ve i axis}), we also have that 
\begin{equation}\label{eq: Bessel function im t}
   \lim_{t \rightarrow \infty} I_{\nu}(i t)  =  \sqrt{2 \over \pi t}  \begin{cases}
       \sin \left( {\pi \over 4}+t \right) & \mbox{ if } \nu \in \mbox{ even} \\
      -i  \cos\left( {\pi \over 4}+t \right) & \mbox{ if } \nu \in \mbox{ odd} 
   \end{cases} 
\end{equation}
\end{itemize}
\end{itemize}

\paragraph{$Z_k(\beta-it)$ at large $t$}: 
We will first compute $Z_k(\beta- i t)$ at large $t$. Since this the argument is away from the positive imaginary axis, we can use eq(\ref{eq: Leading Bessel away from +ve i axis}). We then get 

\begin{equation}\label{eq: Zk intermediate}
   \lim_{|\beta- it| \rightarrow \infty} Z_k(\beta - it ) = {(1-q)^{3\over 4}\over 2\sqrt{\pi}} {C_{q,k} \over (\beta-it)^{3\over 2}}  (e^{2(\beta-it)\over \sqrt{1-q}}  +i (-1)^k e^{-2(\beta-it)\over \sqrt{1-q}})
\end{equation}
where $C_{q,k} \equiv \sum_{p=0}^\infty {(-1)^p q^{p+\binom{p}{2}} (k+2p+1) (q;q)_{k+p} \over (q;q)_{p} \sqrt{(q;q)_k}} = {(q;q)_\infty^3 H_k(1|q) \over \sqrt{(q;q)_k}}$\footnote{The last equality can be checked numerically. We can also obtain the same result by evaluating the $Z_k$ given as an integral in eq(\ref{eq:Z_k as a integral}) in the long time regime. In this regime, the integral is expected to be dominated by contributions near the edges of the distribution, $\theta = 0$ and $\theta = \pi$. It can thus be split into two parts, with each integrand expanded around its respective edge. After a change of variables near $\theta = \pi$, we obtain:
\begin{equation}
\begin{split}
    Z_k(\beta - i t)& = {2(q,q)_\infty^3\over \pi \sqrt{(q;q)_k}}H_k\left(1|q\right)\left( e^{\frac{2(\beta- i t) }{\sqrt{1-q}}}\int\limits_0^{\pi \over 2}  d\theta \theta^2 e^{-\frac{(\beta-i t) \theta^2}{\sqrt{1-q}}} + (-1)^k e^{\frac{-2(\beta- i t) }{\sqrt{1-q}}}\int\limits_0^{\pi \over 2} d\theta \theta^2 e^{(\beta- i t) \theta^2\over \sqrt{1-q}}\right) \\
    &= {(q;q)_\infty^3 (1-q)^{3\over 4}H_k(1|q)\over 2\sqrt(\pi)(\beta-i t)^{3\over 2}}\left( {e^{2(\beta-it)\over \sqrt{1-q}}}+i (-1)^k  {e^{-2(\beta-it)\over \sqrt{1-q}}}\right) \\
\end{split}
\end{equation}
where the upper limit is changed to $\infty$ as the integral gets contribution only from $\theta\approx 0$. Comparing this result against \ref{eq: Zk intermediate},  $C_{q,k}$ can be read off to be $ {(q;q)_\infty^3 H_k(1|q) \over \sqrt{(q;q)_k} }$}.

\paragraph{$Z_k(it)$ at Long Time}: Now we compute the next ingredient in eq(\ref{eq:G def as a series}), namely $Z_k(it)$ at large $t$. Plugging in the result eq(\ref{eq: Bessel function im t}), we get 
\begin{equation}\label{eq:large time Z_k(it)}
   \lim_{t \rightarrow \infty} Z_k(it) = - {(1-q)^{3 \over 4} (q;q)_\infty^3 H_k(1|q)\over \sqrt{\pi t^3}\sqrt{(q;q)_k} }\begin{cases}
       i \sin\left( {\pi \over 4}+{2t\over \sqrt{1-q}} \right) & \mbox{ if } k \in \mbox{ odd} \\
      \cos\left( {\pi \over 4}+{2t\over \sqrt{1-q}} \right)   & \mbox{ if } k \in \mbox{ even} \\
    \end{cases} 
\end{equation}
With both ingredients in eq(\ref{eq:G def as a series}) now determined, we can proceed to assemble them and obtain the two point function as given in eq(\ref{eq:Long time G}). \footnote{Using the n-orthogonality of $q$ Hermite polynomials, note the identity,
\begin{eqnarray*}
    \sum_{k =0}^\infty (\pm \tilde q)^k { H^2_k(1|q) \over (q;q)_k } & = { (\tilde q^2 ; q)_\infty \over(\pm\tilde q ; q)_\infty^4 }\ . 
 \end{eqnarray*}}

\section{Additional details of DSSYK in conformal Regime}\label{Appendix: Conformal Regime}
In this appendix, we elaborate on the details of the calculations in the conformal regime. We first discuss the simplifications that occur in the limit $\lambda \rightarrow 0$, where the model becomes analytically tractable. Following this, we present explicit expressions for the partition function and the two-point function in the high-temperature limit $\beta \ll \lambda^{-{3\over 2}}$.

\paragraph{Density of states: }
The key simplification of the density of states (eqn(\ref{eq:Def Psi in DSSYK})) in this limit arises from the behavior of the $ \Gamma_q $ function, which enters both the density of states eq(\ref{eq:Def Psi in DSSYK}) and the averaged matrix elements eq(\ref{eq:Def f in DSSYK result}). In the small-$ \lambda $ limit (up to non-perturbative corrections), one has \cite{Berkooz:2018jqr}
\begin{equation}\label{eq:Small lambda Gamma_q}
    \lim_{\lambda \rightarrow 0} \frac{ (1-q)^2 (q;q)_\infty^3 q^{1\over 8} }{ |\Gamma_q({2 i \theta \over \lambda})|^2 } = 
    8 \sqrt{\frac{2 \pi}{ \lambda }} \, e^{ -\frac{2\pi^2}{ \lambda} -\frac{2(\frac{\pi}{2}-\theta)^2}{ \lambda}} 
    \sin (\theta ) \sinh \left( \frac{2 \pi  \theta }{\lambda } \right) \sinh \left( \frac{2 \pi (\pi- \theta) }{\lambda } \right).
\end{equation}
From this, the density of states $ \rho(E) = \frac{\Psi(\theta)}{E_0 \sin \theta} $ (see eq(\ref{DSSYK Results Partition})) takes a simplified form in two distinct limits
\footnote{We use $ \lim_{\lambda \rightarrow 0}(q,q)_\infty^3 = \left( \frac{2 \pi}{\lambda} \right)^{3\over 2} e^{-\pi^2 / (2 \lambda)} $ with $ q = e^{-\lambda} $. Notably, in the limit $ \lambda \to 0 $, $ E - E_0 \sim \lambda^{3\over 2} $, the density of states reduces to the well-known $ \sinh $ form characteristic of the Schwarzian theory.} given in the main text which we reproduce below,
\[ 
   \frac{ \Psi(\theta) }{ E_0 \sin \theta }   =   
   \begin{cases}
      \dfrac{2 \sqrt{2\pi}}{ \lambda } \, e^{-{\pi^2 \over 2 \lambda}} \, \theta  & \text{for } \theta \ll \lambda \\
      \sqrt{\dfrac{2}{\pi}} \, e^{- \frac{2}{\lambda}(\frac{\pi}{2}-\theta)^2} & \text{for } \theta \gg \lambda
   \end{cases}
\]
Given the simple expression for density of states, we can now turn to computing physical observables such as partition function and two point function.

\paragraph{Partition function}
The partition function eq(\ref{eq:Def Z}) can be studied in two distinct regimes,  $\beta \ll \lambda^{-3/2}$ and $\beta \gg \lambda^{-3/2}$, corresponding to high and low temperatures, respectively. These regimes are characterized by the dominance of different regions of the spectrum. In particular, when $\beta \gg \lambda^{-3/2}$, the contribution to the partition function (and hence the two-point function) is dominated by states with $\theta \ll \lambda$. This is because, in this limit, the thermal factor \(e^{-\beta E(\theta)}\) strongly suppresses contributions from higher-energy states, and for small \(\theta\) one has \(E(\theta) \approx -E_0 + \frac{E_0}{2}\theta^2\), so the weight behaves as \(e^{\beta E_0} e^{-\frac{\beta E_0}{2}\theta^2}\), effectively cutting off the integral at \(\theta \sim (\beta E_0)^{-1/2} \ll \lambda\), ensuring that only states with \(\theta \ll \lambda\) contribute significantly.
 Note that the density of states in this region then matches that of the cold random matrix theory (RMT), implying that the resulting physics in this regime is identical to that of the cold RMT case discussed in Section \ref{subsec:Cold RMT model example}.

In the remainder of this appendix, we turn our attention to the opposite regime,  $\beta \ll \lambda^{-3/2}$, where the physics is governed by a different part of the spectrum. To analyze this limit systematically, we consider $\lambda \to 0$ while keeping $\beta \sqrt{\lambda}$ fixed. In this regime, it is self-consistent to assume that the dominant contribution to the partition function arises from the region  $\theta \gg \lambda$. This allows us to use the simplified form of the density of states given in eq(\ref{eq:Small lambda simplification of d.o.s}), leading to the following expression for the partition function:

\begin{equation}\label{eq: Conformal Partition fn}
\begin{split}
    Z(\beta)
    & \approx  \sqrt{2 \over \pi \lambda} \int_0^\pi d\theta \sin \theta \     e^{{1 \over \lambda} \left( -2({\pi \over 2}-\theta)^2 + \beta E_0 \lambda \cos \theta \right)  } 
\end{split}
\end{equation}

Since $\lambda \rightarrow 0$, the above integral has a saddle at $\theta = \theta_0$, where $\theta_0$ is given by \cite{2018JHEPNarayan}, 
\begin{equation}\label{eq:Saddle point in Z for SYK like}
        {\pi \over 2} - \theta_0 = {\beta E_0 \lambda \over 4} \sin \theta_0
\end{equation}
It is also convenient to re-express the saddle point equation in terms of another variable  $\phi$, defined by $\phi + \theta = \frac{\pi}{2}$ as given in eq(\ref{eq:DSSYK Saddle eqn in phi0}). It is to be noted that $\theta_0$ only depends on combination $\beta E_0  \lambda \sim \beta \sqrt \lambda$ which is held fixed in the $\lambda \rightarrow 0$. To be self consistent with the assumption that the partition function gets contribution only from  $\theta \gg \lambda $ region, we need to ensure $\theta_0 \gg \lambda$, i.e., $\beta E_0 \lambda \ll \lambda^{-1}$, i.e., $\beta \ll \lambda^{-{3 \over 2}}$ which is obeyed since $\beta \sim {1 \over \sqrt \lambda}$. Note that for certain limits,  the saddle equation can be solved explicitly as below
\begin{equation}
\theta_0 = 
    \begin{cases}
         {2\pi \over \beta E_0 \lambda }  &  \beta E_0 \lambda \gg 1 \\
         {\pi \over 2}  - {\beta E_0 \lambda \over 4}   &  \beta E_0 \lambda \ll 1 
    \end{cases}
\end{equation}
Since $E_0 \sim {1 \over \sqrt \lambda}$, we will refer to the DSSYK regime with ${1 \over \lambda^{3 \over 2} } \gg \beta \gg {1 \over \sqrt \lambda }$ regime as the low temperature conformal regime and DSSYK regime with $ \beta \ll {1 \over \sqrt \lambda }$ regime as the high temperature conformal regime. The low temp conformal regime has overlaps with  the high temperature Schwarzian theory obtained in the tripled scaling limit mentioned earlier since the region ${1 \over \lambda^{3 \over 2}} \gg \beta \gg {1 \over \lambda^{1 \over 2}}$ is accessible from both sides.

\begin{figure}
    \centering
    \subfloat[$m=1$]{%
        \includegraphics[width=0.65\linewidth]{ 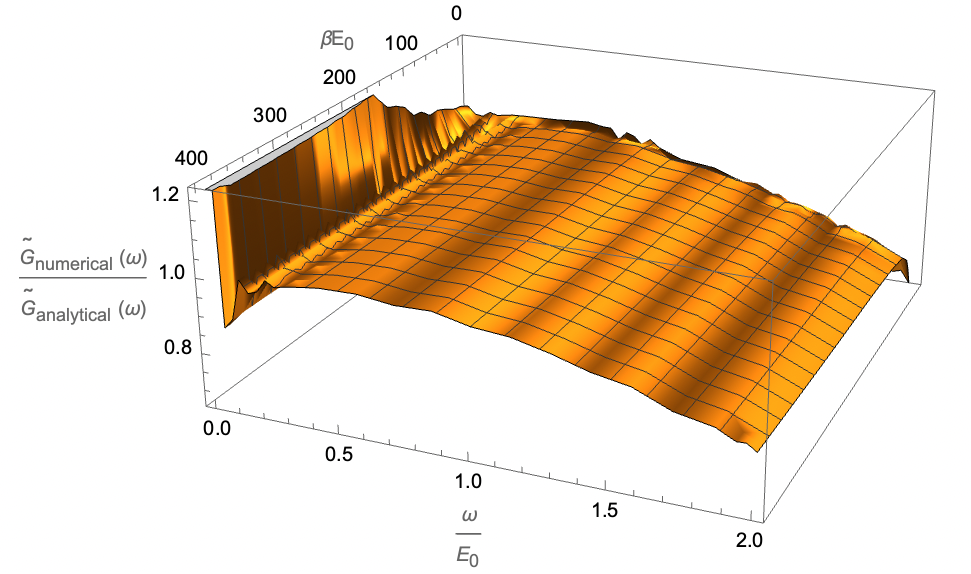}
        \label{fig:ratio_m1}
    }
    \hspace{0.015\linewidth} 
    \subfloat[$m=2$]{%
        \includegraphics[width=0.65\linewidth]{ 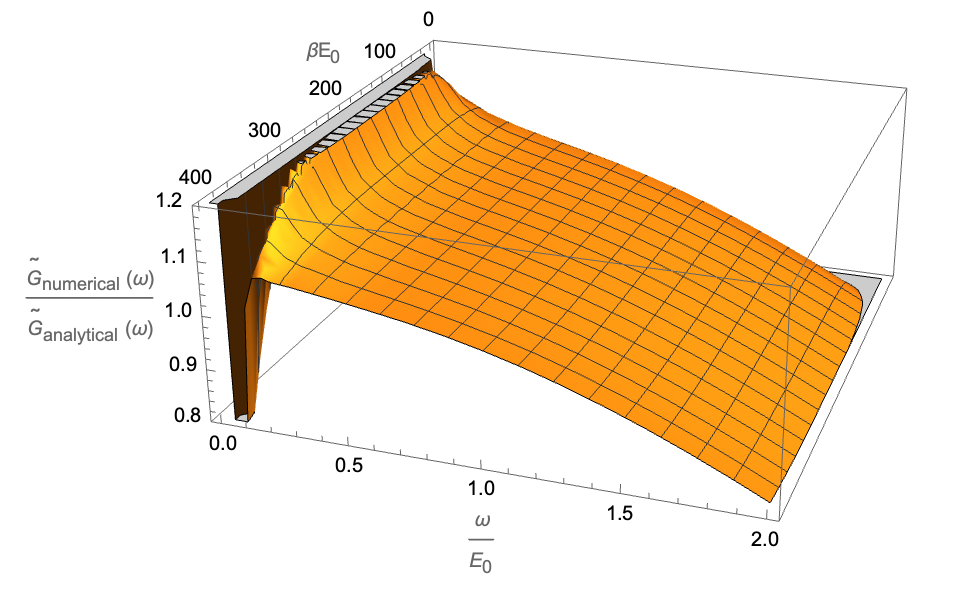}
        \label{fig:ratio_m2}
    }
    \caption{Ratio of numerical values of $\tilde G(\omega)$ against analytical result(eq(\ref{eq: Two point function omega conformal})) for $\lambda=0.05$ and different values of $m$.}
    \label{fig:G ratio}
\end{figure}

\paragraph{Two point function}  
Before turning to solving the two point function in this limit we first note the frequency-space two-point function $\tilde G(\omega)$ (using eqn(\ref{eq:Def tilde G})), represented as an integral over $\theta$
\begin{equation}\label{eq:G tilde omega in DSSYK}
    \tilde G(\omega) = { 1  \over Z(\beta)} \int d\theta \Psi(\theta)e^{\beta E_0 \cos \theta} g(\theta, \theta_\omega)
\end{equation}
where $\theta_\omega= \arccos\left( \cos \theta -{\omega \over E_0} \right)-\theta$ and the insertion factor $g(\theta, \theta_\omega)$ is as follows
\begin{equation}
    \label{eq:rho f in DSSYK}
      g(\theta, \theta_\omega) \equiv    \ {|(e^{i(2\theta+\theta_\omega)},q)_m \Gamma_q( { i(2\theta+\theta_\omega) \over \lambda } )\Gamma_q(m + {i\theta_\omega \over \lambda } )|^2
    \over E_0(1-q) \sin \theta \Gamma_q(2m) \ |\Gamma_q({2i(\theta+\theta_\omega) \over \lambda})|^2 }  
\end{equation}
 Having established the general form of the frequency-space two-point function $\tilde G(\omega)$ in eq(\ref{eq:G tilde omega in DSSYK}), we now focus on its behavior in the $\lambda \to 0$ limit, which is relevant in the regime $\beta \ll \lambda^{-3/2}$ discussed earlier. To simplify the analysis, we first consider the case $\omega = 0$, which already captures the essential features of the small-$\omega$ physics. For $\omega =0$ (i.e., $\theta_\omega=0$), since the partition function already has a saddle at $\theta_0$\footnote{Since there are no exponential in $\theta$ pieces in the insertion factor $g$ given in eq(\ref{eq:rho f in DSSYK}),  the saddle for the two point function eq(\ref{eq:G tilde omega in DSSYK}) does not shift from the saddle of partition function $\theta_0$ given in eq(\ref{eq:Saddle point in Z for SYK like}).  This argument continues to hold in the case of $\omega \ne 0$. } given by eq(\ref{eq:Saddle point in Z for SYK like}),  the insertion factor $g$ given in  eq(\ref{eq:rho f in DSSYK}) evaluates to a constant and we have the result\footnote{Recall that $\lim_{q \rightarrow 1}\Gamma_q(m) = \Gamma(m)$ for finite $m$.}
\begin{equation}\label{eq:Two point function Gtilde small lambda with omega eq 0}
    \tilde G(\omega=0) = { |(e^{2i\theta_0},q)_m|^2\Gamma(m)|^2 \over E_0 \lambda \sin\theta_0 \Gamma(2m)}={ 2(2 \sin \theta_0)^{2m-1}\Gamma(m)^2 \over E_0 \lambda \Gamma(2m)}
\end{equation}
where we have used $|(e^{2i\theta},q)_m|^2 = (2 \sin\theta)^{2m}$ for $\theta \gg \lambda$ at small $\lambda$. We will now evaluate $\tilde G(\omega)$ for small  $\omega$ with ${\omega \over E_0\lambda}$ held fixed (or equivalently $\omega \beta$ held fixed) as $\lambda \rightarrow 0$. In this case, one can solve for $\theta_\omega = {\omega \over E_0 \sin\theta}$ - note that $\theta_\omega \sim {\cal O}(\lambda)$. As seen before, the saddle does not shift from $\theta_0$. As for the $\omega=0$ case, the insertion factor $g$ given in eq(\ref{eq:rho f in DSSYK}) evaluates to a constant and we have the result\footnote{Apart from the explicit factor $\Gamma(m)^2 \rightarrow |\Gamma(m+{i\theta_\omega \over \lambda})|^2$ the
 extra factor compared to $\omega=0$ case is 
\[
{  |(e^{i(2\theta_0+\theta_\omega)},q)_m|^2  |\Gamma_q( { i(2\theta_0+\theta_\omega\over \lambda } )|^2   \over |(e^{2i\theta_0},q)_m|^2 |\Gamma_q( { i(2\theta_0+2\theta_\omega\over \lambda } )|^2 } \approx {\sin^{2m-1} (\theta_0+{\theta_\omega \over 2}) \sin (\theta_0+\theta_\omega) \over \sin^{2m} (\theta_0) }\ e^{\theta_\omega (\pi - 2 \theta_0) \over \lambda} e^{-{3 \theta_\omega^2 \over 2\lambda}}
\]
where we have used eq(\ref{eq:Small lambda Gamma_q}). The above expression evaluates to $e^{\beta \omega \over 2}$ for small $\theta_\omega \sim {\cal O}(\lambda)$ upon using the saddle equation eq(\ref{eq:Saddle point in Z for SYK like}). 
}
\begin{equation}\label{eq: Two point function omega conformal}
    \tilde G(\omega) = { 2(2\sin \theta_0)^{2m-1} \over E_0 \lambda  \Gamma(2m)} \ e^{\beta \omega \over 2} \Gamma(m + {i\omega \over \lambda E_0 \sin \theta_0 }) \Gamma(m - {i\omega \over \lambda E_0 \sin \theta_0 })
\end{equation}
Note that eq(\ref{eq: Two point function omega conformal}) is  consistent with \cite{Mukhametzhanov:2023tcg}\footnote{With the identification $v = 1 - \frac{2}{\pi} \theta$, the Fourier transform of their result for $G(t)$  exactly matches eq(\ref{eq: Two point function omega conformal}).}. In terms of the variable $\phi$ introduced earlier, the two-point function can be written as shown in eq(\ref{eq: Two point function phi conformal}). We can confirm that the above analytical result is consistent with numerical results even for slightly large $\lambda$ and a wide range of $\beta$ beyond the expected regime $\beta E_0 \sim {1 \over \lambda}$. In fig(\ref{fig:G ratio}), we observe that the analytical expression agrees with the numerical data across a broad range of $\omega$, with an accuracy of approximately $20\%$. 

\newpage

\bibliographystyle{ieeetr}
\bibliography{draft.bib}

\begin{thebibliography}{10}

\bibitem{zwanzig2001nonequilibrium}
R.~Zwanzig, {\em Nonequilibrium Statistical Mechanics}.
\newblock OUP USA, 2001.

\bibitem{Kapusta:2006pm}
J.~I. Kapusta and C.~Gale, {\em {Finite-temperature field theory: Principles and applications}}.
\newblock Cambridge Monographs on Mathematical Physics, Cambridge University Press, 2011.

\bibitem{kamenev2011field}
A.~Kamenev, {\em Field Theory of Non-Equilibrium Systems}.
\newblock Cambridge University Press, 2011.

\bibitem{balakrishnan2020elements}
V.~Balakrishnan, {\em Elements of Nonequilibrium Statistical Mechanics}.
\newblock Springer International Publishing, 2020.

\bibitem{Bernard:2012je}
D.~Bernard and B.~Doyon, ``{Energy flow in non-equilibrium conformal field theory},'' {\em J. Phys. A}, vol.~45, p.~362001, 2012.

\bibitem{Almheiri:2019jqq}
A.~Almheiri, A.~Milekhin, and B.~Swingle, ``Universal constraints on energy flow and syk thermalization,'' {\em JHEP}, vol.~04, p.~245, 2020.

\bibitem{Bernard:2016nci}
D.~Bernard and B.~Doyon, ``{Conformal field theory out of equilibrium: a review},'' {\em J. Stat. Mech.}, vol.~1606, no.~6, p.~064005, 2016.

\bibitem{Essler:2014qza}
F.~H.~L. Essler, G.~Mussardo, and M.~Panfil, ``{Generalized Gibbs Ensembles for Quantum Field Theories},'' {\em Phys. Rev. A}, vol.~91, no.~5, p.~051602, 2015.

\bibitem{2012PhRvE..85a1126D}
A.~Dhar, K.~Saito, and P.~H{\"a}nggi, ``Nonequilibrium density-matrix description of steady-state quantum transport,'' {\em Phys. Rev. E}, vol.~85, p.~011126, January 2012.

\bibitem{Doyon:2012bg}
B.~Doyon, ``Nonequilibrium density matrix for thermal transport in quantum field theory,'' {\em Communications in Mathematical Physics}, vol.~351, no.~1, pp.~155--200, 2017.

\bibitem{PhysRevB.88.195129}
C.~Karrasch, R.~Ilan, and J.~E. Moore, ``Nonequilibrium thermal transport and its relation to linear response,'' {\em Phys. Rev. B}, vol.~88, p.~195129, Nov 2013.

\bibitem{PhysRevB.90.161101}
A.~De~Luca, J.~Viti, L.~Mazza, and D.~Rossini, ``Energy transport in heisenberg chains beyond the luttinger liquid paradigm,'' {\em Phys. Rev. B}, vol.~90, p.~161101, Oct 2014.

\bibitem{PhysRevB.93.205121}
A.~Biella, A.~De~Luca, J.~Viti, D.~Rossini, L.~Mazza, and R.~Fazio, ``Energy transport between two integrable spin chains,'' {\em Phys. Rev. B}, vol.~93, p.~205121, May 2016.

\bibitem{Biella:2019pxg}
A.~Biella, M.~Collura, D.~Rossini, A.~De~Luca, and L.~Mazza, ``{Ballistic transport and boundary resistances in inhomogeneous quantum spin chains},'' {\em Nature Commun.}, vol.~10, no.~1, p.~4820, 2019.

\bibitem{RevModPhys.93.025003}
B.~Bertini, F.~Heidrich-Meisner, C.~Karrasch, T.~Prosen, R.~Steinigeweg, and M.~\ifmmode \check{Z}\else \v{Z}\fi{}nidari\ifmmode~\check{c}\else \v{c}\fi{}, ``Finite-temperature transport in one-dimensional quantum lattice models,'' {\em Rev. Mod. Phys.}, vol.~93, p.~025003, May 2021.

\bibitem{hartnoll2018holographic}
S.~A. Hartnoll, A.~Lucas, and S.~Sachdev, {\em Holographic quantum matter}.
\newblock MIT press, 2018.

\bibitem{KitaevTalk1}
A.Kitaev, ``A simple model of quantum holography.'' \url{http://online.kitp.ucsb.edu/online/entangled15/kitaev/}, 2015.

\bibitem{PhysRevLett.70.3339}
S.~Sachdev and J.~Ye, ``Gapless spin-fluid ground state in a random quantum heisenberg magnet,'' {\em Phys. Rev. Lett.}, vol.~70, pp.~3339--3342, May 1993.

\bibitem{PhysRevD.94.106002}
J.~Maldacena and D.~Stanford, ``Remarks on the sachdev-ye-kitaev model,'' {\em Phys. Rev. D}, vol.~94, p.~106002, Nov 2016.

\bibitem{Polchinski2016TheSI}
J.~Polchinski and V.~Rosenhaus, ``The spectrum in the sachdev-ye-kitaev model,'' {\em Journal of High Energy Physics}, vol.~2016, pp.~1--25, 2016.

\bibitem{Sachdev:2010uj}
S.~Sachdev, ``{Strange metals and the AdS/CFT correspondence},'' {\em J. Stat. Mech.}, vol.~1011, p.~P11022, 2010.

\bibitem{Maldacena:2015waa}
J.~Maldacena, S.~H. Shenker, and D.~Stanford, ``{A bound on chaos},'' {\em JHEP}, vol.~08, p.~106, 2016.

\bibitem{Maldacena:2016upp}
J.~Maldacena, D.~Stanford, and Z.~Yang, ``{Conformal symmetry and its breaking in two dimensional Nearly Anti-de-Sitter space},'' {\em PTEP}, vol.~2016, no.~12, p.~12C104, 2016.

\bibitem{Engelsoy:2016xyb}
J.~Engels{\"o}y, T.~G. Mertens, and H.~Verlinde, ``{An investigation of AdS$_{2}$ backreaction and holography},'' {\em JHEP}, vol.~07, p.~139, 2016.

\bibitem{PhysRevX.5.041025}
S.~Sachdev, ``Bekenstein-hawking entropy and strange metals,'' {\em Phys. Rev. X}, vol.~5, p.~041025, Nov 2015.

\bibitem{Gross:2016kjj}
D.~J. Gross and V.~Rosenhaus, ``{A Generalization of Sachdev-Ye-Kitaev},'' {\em JHEP}, vol.~02, p.~093, 2017.

\bibitem{Yoon:2017nig}
J.~Yoon, ``{SYK Models and SYK-like Tensor Models with Global Symmetry},'' {\em JHEP}, vol.~10, p.~183, 2017.

\bibitem{Narayan2018SupersymmetricSM}
P.~Narayan and J.~Yoon, ``Supersymmetric syk model with global symmetry,'' {\em Journal of High Energy Physics}, vol.~2018, pp.~1--60, 2018.

\bibitem{Gu2020NotesOT}
Y.~Gu, A.~Y. Kitaev, S.~Sachdev, and G.~M. Tarnopolsky, ``Notes on the complex sachdev-ye-kitaev model,'' {\em Journal of High Energy Physics}, vol.~2020, pp.~1--74, 2020.

\bibitem{Bhattacharya2017SYKMC}
R.~Bhattacharya, S.~Chakrabarti, D.~P. Jatkar, and A.~Kundu, ``Syk model, chaos and conserved charge,'' {\em Journal of High Energy Physics}, vol.~2017, pp.~1--16, 2017.

\bibitem{Liu:2019niv}
J.~Liu and Y.~Zhou, ``{Note on global symmetry and SYK model},'' {\em JHEP}, vol.~05, p.~099, 2019.

\bibitem{Wenbo}
W.~Fu, D.~Gaiotto, J.~Maldacena, and S.~Sachdev, ``Supersymmetric sachdev-ye-kitaev models,'' {\em Phys. Rev. D}, vol.~95, p.~026009, Jan 2017.

\bibitem{Yoon2017SupersymmetricSM}
J.~Yoon, ``Supersymmetric syk model: bi-local collective superfield/supermatrix formulation,'' {\em Journal of High Energy Physics}, vol.~2017, pp.~1--48, 2017.

\bibitem{Peng:2017spg}
C.~Peng, M.~Spradlin, and A.~Volovich, ``{Correlators in the $\mathcal{N}=2$ Supersymmetric SYK Model},'' {\em JHEP}, vol.~10, p.~202, 2017.

\bibitem{Larzul:2022kri}
A.~Larzul and M.~Schir\`o, ``Energy transport across two interacting quantum baths without quasiparticles,'' {\em Phys. Rev. B}, vol.~108, p.~115120, Sep 2023.

\bibitem{Larzul:2022yss}
A.~Larzul, S.~J. Thomson, and M.~Schir{\`o}, ``Are fast scramblers good thermal baths?,'' Apr. 2022.

\bibitem{Erds2014PhaseTI}
L.~Erdős and D.~Schr{\"o}der, ``Phase transition in the density of states of quantum spin glasses,'' {\em Mathematical Physics, Analysis and Geometry}, vol.~17, pp.~441--464, 2014.

\bibitem{Cotler:2016fpe}
J.~S. Cotler, G.~Gur-Ari, M.~Hanada, J.~Polchinski, P.~Saad, S.~H. Shenker, D.~Stanford, A.~Streicher, and M.~Tezuka, ``{Black Holes and Random Matrices},'' {\em JHEP}, vol.~05, p.~118, 2017.
\newblock [Erratum: JHEP 09, 002 (2018)].

\bibitem{2018JHEPNarayan}
M.~{Berkooz}, P.~{Narayan}, and J.~{Sim{\'o}n}, ``{Chord diagrams, exact correlators in spin glasses and black hole bulk reconstruction},'' {\em Journal of High Energy Physics}, vol.~2018, p.~192, Aug. 2018.

\bibitem{Berkooz:2018jqr}
M.~Berkooz, M.~Isachenkov, V.~Narovlansky, and G.~Torrents, ``\href{https://arxiv.org/pdf/1811.02584.pdf}{Towards a full solution of the large N double-scaled SYK model},'' {\em JHEP}, vol.~03, p.~079, 2019.

\bibitem{Berkooz:2020fvm}
M.~Berkooz, N.~Brukner, V.~Narovlansky, and A.~Raz, ``{Multi-trace correlators in the SYK model and non-geometric wormholes},'' {\em JHEP}, vol.~21, p.~196, 2020.

\bibitem{Lin:2022rbf}
H.~W. Lin, ``{The bulk Hilbert space of double scaled SYK},'' {\em JHEP}, vol.~11, p.~060, 2022.

\bibitem{Berkooz:2022mfk}
M.~Berkooz, M.~Isachenkov, M.~Isachenkov, P.~Narayan, and V.~Narovlansky, ``{Quantum groups, non-commutative AdS$_{2}$, and chords in the double-scaled SYK model},'' {\em JHEP}, vol.~08, p.~076, 2023.

\bibitem{Mertens:2022irh}
T.~G. Mertens and G.~J. Turiaci, ``{Solvable models of quantum black holes: a review on Jackiw{\textendash}Teitelboim gravity},'' {\em Living Rev. Rel.}, vol.~26, no.~1, p.~4, 2023.

\bibitem{Susskind:2021esx}
L.~Susskind, ``{Entanglement and Chaos in De Sitter Space Holography: An SYK Example},'' {\em JHAP}, vol.~1, no.~1, pp.~1--22, 2021.

\bibitem{Susskind:2022bia}
L.~Susskind, ``De sitter space, double-scaled syk, and the separation of scales in the semiclassical limit,'' {\em Journal of Holography Applications in Physics}, vol.~5, no.~1, pp.~1--30, 2025.

\bibitem{Susskind:2022dfz}
L.~Susskind, ``Scrambling in double-scaled syk and de sitter space,'' May 2022.

\bibitem{Blommaert:2024ydx}
A.~Blommaert, T.~G. Mertens, and J.~Papalini, ``{The dilaton gravity hologram of double-scaled SYK},'' {\em JHEP}, vol.~06, p.~050, 2025.

\bibitem{Narayan:2023wlk}
P.~Narayan and S.~T. S, ``{SYK Model with global symmetries in the double scaling limit},'' {\em JHEP}, vol.~05, p.~083, 2023.

\bibitem{Berkooz2020ComplexSM}
M.~Berkooz, V.~Narovlansky, and H.~Raj, ``Complex sachdev-ye-kitaev model in the double scaling limit,'' {\em arXiv: High Energy Physics - Theory}, 2020.

\bibitem{Berkooz:2020xne}
M.~Berkooz, N.~Brukner, V.~Narovlansky, and A.~Raz, ``{The double scaled limit of Super--Symmetric SYK models},'' {\em JHEP}, vol.~12, p.~110, 2020.

\bibitem{Maldacena:2016hyu}
J.~Maldacena and D.~Stanford, ``{Remarks on the Sachdev-Ye-Kitaev model},'' {\em Phys. Rev. D}, vol.~94, no.~10, p.~106002, 2016.

\bibitem{chakrabarti2012remarkable}
D.~Chakrabarti and G.~K. Srinivasan, ``On a remarkable formula of ramanujan,'' {\em Archiv der Mathematik}, vol.~99, pp.~125--135, Aug. 2012.

\bibitem{Dodelson:2024atp}
M.~Dodelson, ``Ringdown in the syk model,'' Aug. 2024.
\newblock CERN-TH-2024-135.

\bibitem{cordial}
M.~Berkooz and O.~Mamroud, ``A cordial introduction to double scaled syk,'' {\em arXiv.org}, July 2024.

\bibitem{wiki:Continuous_q-Hermite_polynomials}
Wikipedia, ``{Continuous q-Hermite polynomials} --- {W}ikipedia{,} the free encyclopedia.'' \url{http://en.wikipedia.org/w/index.php?title=Continuous\%20q-Hermite\%20polynomials&oldid=1121109966}.

\bibitem{koekoek2010hypergeometric}
R.~Koekoek, T.~Koornwinder, P.~Lesky, and R.~Swarttouw, {\em Hypergeometric Orthogonal Polynomials and Their q-Analogues}.
\newblock Springer Monographs in Mathematics, Springer Berlin Heidelberg, 2010.

\bibitem{Mukhametzhanov:2023tcg}
B.~Mukhametzhanov, ``{Large p SYK from chord diagrams},'' {\em JHEP}, vol.~09, p.~154, 2023.

\bibitem{Goel:2023svz}
A.~Goel, V.~Narovlansky, and H.~Verlinde, ``{Semiclassical geometry in double-scaled SYK},'' {\em JHEP}, vol.~11, p.~093, 2023.

\bibitem{Gu_2017}
Y.~Gu, X.-L. Qi, and D.~Stanford, ``Local criticality, diffusion and chaos in generalized sachdev-ye-kitaev models,'' {\em Journal of High Energy Physics}, vol.~2017, May 2017.

\bibitem{bucca2024}
M.~Bucca and M.~Mezei, ``Nonlinear soft mode action for the large-$p$ syk model,'' 2024.

\bibitem{GradshteynRyzhik}
I.~S. Gradshteyn and I.~M. Ryzhik, {\em Table of integrals, series, and products}.
\newblock Elsevier/Academic Press, Amsterdam, seventh~ed., 2007.

\end{thebibliography}

\end{document}